\documentclass[useAMS,fleqn,usenatbib]{mnras}

\usepackage[T1]{fontenc}
\usepackage{ae,aecompl}

\usepackage{amsmath}
\usepackage{amsopn}
\usepackage[british]{babel}
\usepackage[varg]{txfonts}
\usepackage{biblio} 
\usepackage{natbib}
\usepackage{color}

\usepackage{graphicx}
\usepackage{epstopdf} 

\usepackage{microtype}

\definecolor{orange}{rgb}{1.0,0.5,0.}

\DeclareMathOperator{\sech}{sech}

\def\MDM{\ifmmode{\>M_{\textnormal{\sc dm}}}\else{$$M_{\textnormal{\sc dm}}}\fi}

\def\XH{\ifmmode{\>X_{\textnormal{\sc h}}} \else{$X_{\textnormal{\sc h}}$}\fi}
\def\nH{\ifmmode{\>n_{\textnormal{\sc h}}} \else{$n_{\textnormal{\sc h}}$}\fi}

\def\maspyr{\ifmmode{\>\textnormal{mas~yr}^{-1}}\else{mas~yr$^{-1}$}\fi}

\def\mG{\ifmmode{\>\mu\mathrm{G}}\else{$\mu$G}\fi}
\def\erg{\ifmmode{\> {\rm erg}}\else{erg}\fi}
\def\keV{\ifmmode{\> {\rm keV}}\else{keV}\fi}

\def\deg{\ifmmode{\>^{\circ}}\else{$^{\circ}$}\fi}
\def\onedeg{\ifmmode{\>1^{\circ}}\else{$1^{\circ}$}\fi}

\def\xvir{\ifmmode{\>\!x_{vir}}\else{$x_{vir}$}\fi}
\def\Mvir{\ifmmode{\>\!M_{vir} }\else{$M_{vir} $}\fi}
\def\rvir{\ifmmode{\>\!r_{vir}}\else{$r_{vir}$}\fi}
\def\vvir{\ifmmode{\>\!v_{vir}}\else{$v_{vir}$}\fi}
\def\Vvir{\ifmmode{\>\!V_{vir} }\else{$V_{vir} $}\fi}

\def\tratio{\ifmmode{\>\tau}\else{$\tau$}\fi}

\def\rms{\ifmmode{\>r_{\textnormal{\sc ms}}}\else{$r_{\textnormal{\sc ms}}$}\fi}

\def\Mpc{\ifmmode{\>\!{\rm Mpc}} \else{Mpc}\fi}
\def\kpc{\ifmmode{\>\!{\rm kpc}} \else{kpc}\fi}
\def\pc{\ifmmode{\>\!{\rm pc}} \else{pc}\fi}

\def\Gyr{\ifmmode{\>\!{\rm Gyr}} \else{Gyr}\fi}
\def\Myr{\ifmmode{\>\!{\rm Myr}} \else{Myr}\fi}
\def\yr{\ifmmode{\>\!{\rm yr}} \else{yr}\fi}
\def\pyr{\ifmmode{\>\!{\rm yr}^{-1}}\else{yr $^{-1}$} \fi}
\def\s{\ifmmode{\>\!{\rm s}}\else{s}\fi}
\def\ps{\ifmmode{\>\!{\rm s}^{-1}}\else{s$^{-1}$}\fi}
\def\Hz{\ifmmode{\>\!{\rm Hz}}\else{Hz}\fi}

\def\kms{\ifmmode{\>\!{\rm km\,s}^{-1}}\else{km~s$^{-1}$}\fi}

\def\K{\ifmmode{\>\!{\rm K}}\else{K}\fi}

\def\sr{\ifmmode{\>\!{\rm sr}}\else{sr}\fi}
\def\psr{\ifmmode{\>\!{\rm sr}^{-1}}\else{sr$^{-1}$}\fi}
\def\arcs{\ifmmode{\>\!{\rm arcsec}}\else{arcsec}\fi}
\def\parcs{\ifmmode{\>\!{\rm arcsec}^{-1}}\else{arcsec${-1}$}\fi}
\def\parcss{\ifmmode{\>\!{\rm arcsec}^{-2}}\else{arcsec${-2}$}\fi}

\def\cm{\ifmmode{\>\!{\rm cm}}\else{cm}\fi}
\def\cc{\ifmmode{\>\!{\rm cm}^{3}}\else{cm$^{3}$}\fi}
\def\sqc{\ifmmode{\>\!{\rm cm}^{2}}\else{cm$^{2}$}\fi}
\def\pcc{\ifmmode{\>\!{\rm cm}^{-3}}\else{cm$^{-3}$}\fi}
\def\psc{\ifmmode{\>\!{\rm cm}^{-2}}\else{cm$^{-2}$}\fi}

\def\g{\ifmmode{\>\!{\rm g}}\else{g}\fi}
\def\Msun{\ifmmode{\>\!{\rm M}_{\odot}}\else{M$_{\odot}$}\fi}
\def\hMsun{\ifmmode{\> h^{-1}{\rm M}_{\odot}}\else{$h^{-1}$M$_{\odot}$}\fi}

\def\Zsun{\ifmmode{\>\!{\rm Z}_{\odot}}\else{Z$_{\odot}$}\fi}

\def\Lsun{\ifmmode{\>\!{\rm L}_{\odot}}\else{L$_{\odot}$}\fi}

\def\rayl{\ifmmode{\>\!{\rm R}}\else{R}\fi}
\def\mR{\ifmmode{\>\!{\rm mR}}\else{mR}\fi}

\renewcommand{\ion}[2]{\hbox{#1\,{\sc #2}}}

\def\lya{\ifmmode{\>\!{\rm Ly}\alpha}\else{Ly$\alpha$}\fi}

\def\Ha{\ifmmode{\>\!{\rm H}\alpha}\else{H$\alpha$}\fi}
\def\Hb{\ifmmode{\>\!{\rm H}\beta}\else{H$\beta$}\fi}

\def\HI{\ifmmode{\> \textnormal{\ion{H}{i}}} \else{\ion{H}{i}}\fi}
\def\HII{\ifmmode{\> \textnormal{\ion{H}{ii}}} \else{\ion{H}{ii}}\fi}
\def\CIV{\ifmmode{\> \textnormal{\ion{C}{iv}}} \else{\ion{C}{iv}}\fi}
\def\SiIV{\ifmmode{\> \textnormal{\ion{S}{iv}}} \else{\ion{Si}{iv}}\fi}

\def\NH{\ifmmode{\> {\rm N}_{\rm H}} \else{N$_{\rm H}$}\fi}
\def\Ng{\ifmmode{\> {\rm N}_{\rm gas}} \else{N$_{\rm gas}$}\fi}
\def\NHI{\ifmmode{\> {\rm N}_{\HI}} \else{N$_{\HI}$}\fi}
\def\MHI{\ifmmode{\> {\rm M}_{ \HI}} \else{M$_{\HI}$}\fi}

\def\mua{\ifmmode{\>\mu_{ \textnormal{\Ha}}}\else{$\mu_{ \textnormal{\Ha}}$}\fi}
\def\alphabha{\ifmmode{\>\alpha_{B}^{(\textnormal{\Ha})}}\else{$\alpha_{B}^{(\textnormal{\Ha})}$}\fi}

\newcommand{\myemail}{tepper@physics.usyd.edu.au}
\newcommand{\ramses}{{\sc Ramses}}
\newcommand{\agama}{{\small AGAMA}}

\newcommand{\gaia}{{\em Gaia}}

\newcommand{\nexus}{{\sc Nexus}}
\newcommand{\vp}{V_\phi}
\newcommand{\vz}{V_{z}}
\newcommand{\vR}{V_{R}}

\defcitealias{bla21e}{BT21}
\defcitealias{tep22x}{TGBF22}
\defcitealias{gra19a}{GRAVITY Collaboration}

\title[Phase spirals within a clumpy, turbulent ISM]{Galactic seismology: can the \gaia\ `phase spiral' co-exist with a clumpy, turbulent interstellar medium?}

\author[Tepper-Garc\'\i{}a et al.]{%
Thor Tepper-Garc\'\i{}a,$^{1}$\thanks{\myemail} Joss Bland-Hawthorn,$^{1}$ 
Timothy R. Bedding,$^{1}$
Christoph Federrath,$^{2}$ and \newauthor
Oscar Agertz$^{3}$
\\
$^{1}$Sydney Institute for Astronomy, School of Physics, University of Sydney, NSW 2006, Australia\\
$^{2}$Research School of Astronomy and Astrophysics, Australian National University, Canberra, ACT 2611, Australia\\
$^{3}$Lund Observatory, Division of Astrophysics, Department of Physics, Lund University, Box 43, SE-221 00 Lund, Sweden
}

\date{Accepted ---. Received ---; in original form ---}

\pubyear{\date{year}}

\begin{document}
\label{firstpage}
\pagerange{\pageref{firstpage}--\pageref{lastpage}}
\maketitle

\pdfminorversion=5

\begin{abstract}
The \gaia\ satellite revealed a remarkable spiral pattern (`phase spiral', PS) in the $z-\vz$ phase-plane throughout the solar neighbourhood, where $z$ and $\vz$ are the displacement and velocity of a star perpendicular to the Galactic plane. As demonstrated by many groups, the kinematic signature reflects the Galactic stellar disc's response to a dynamical disturbance some $0.3-3$~Gyr ago. However, previous controlled simulations did not consider the impact of the multi-phase interstellar medium (ISM) on the existence of the PS. This is crucial because it has been suggested that this weak signal is highly susceptible to scattering by small-scale density fluctuations typical of the ISM. This has motivated us to explore the formation and fate of the PS in a suite of high-resolution, N-body/hydrodynamical simulations of an idealised Galaxy analogue bearing a realistic ISM that interacts impulsively with a massive perturber. In our models, high gas surface densities within the disc encourage vigorous star formation, which in turn couples with the gas via feedback to drive turbulence.
We find that the PS is almost non-existent if
the disc potential is too strong or
the ISM is highly structured on sub-kiloparsec scales. This can happen in the absence of stellar feedback when the gas is allowed to cool. In the presence of turbulent gas maintained by stellar feedback, the PS has a patchy spatial distribution and a high degree of intermittency on kiloparsec scales.
We anticipate that future studies of the phase-spiral behaviour on all scales will provide crucial information on star-gas dynamics.

\end{abstract}


\section{Introduction} \label{sec:intro}

An unexpected result from the European Space Agency (ESA) \gaia\ astrometric survey was the detection of an unusual phase-space signature in the local stellar disc \citep{ant18b}. 
In Galactic cylindrical coordinates ($R,\phi,z$), individual stars have velocities ($\vR$, $\vp$, $\vz$). In a volume element
defined by ($\Delta R, R\,\Delta \phi, \Delta z$) = ($\pm 0.1$, $\pm 0.1$, $\pm 1$) kpc$^3$
centred on the Sun, the \gaia\ team uncovered a coherent spiral pattern in the $z-\vz$ phase plane, the so-called `phase spiral'.

The phase spiral is most evident when each point of the $z-\vz$ phase plane is represented by either $\langle \vR \rangle$ or $\langle \vp \rangle$, averaged over the local volume. Thereafter, \citet{kha19a} showed the phase spiral emerges more clearly when encoded by $\langle L_z\rangle$,
\footnote{The vertical frequency ($\nu$) of stars depends on the vertical amplitude ($A_z$) of the orbit and the guiding centre radius (or $L_z$); the phase spiral is produced because the vertical frequency depends monotonically on the vertical amplitude of the orbit, and is clearer if either dependence ($A_z$ or $L_z$) is removed.}
which measures stellar angular momentum about the Galaxy's spin axis.
This signature is indicative of a system that is settling from a mildly disturbed state to a stationary configuration through the process of phase mixing \citep{lyn67a}. A recent review of the origins of the phase spiral is provided by \citet{hun25a}.

Since its discovery, the phase-spiral phenomenon has been extensively studied using both analytic methods \citep[e.g.,][]{bin18a,dar19a,ali23a} and numerical simulations \citep[e.g.,][]{lap19a,bla19a,hun21t,asa25a}. In an attempt to bridge the two approaches, \citet[][henceforth \citetalias{bla21e}]{bla21e} carried out a high-resolution, N-body simulation ($N = 10^8$ particles) that closely follows \citet{bin18a}'s analytic model. 
\citet*[][henceforth \citetalias{tep22x}]{tep22x} improved upon this early simulation by adding a gaseous component in the form of a rotationally supported, isothermal, cold disc in vertical hydrostatic equilibrium. Although the last study did not focus on the emergence of the phase spiral, a comparative analysis between the pure N-body simulation and its hydrodynamical extension revealed that the addition of gas, even if `inert' (isothermal, non star-forming), results in significant damping of the vertical perturbation induced onto the disc by a massive external perturber, which is intimately connected to the emergence of the phase spiral.

In the same year, \citet{gar22c} presented an extensive analysis of the emergence and properties of the phase-spiral phenomenon in a synthetic Milky-Way (MW) analogue found in a `zoom-in' cosmological simulation. This study --- the first of its kind --- showed that phase spirals qualitatively similar to the \gaia\ phase spiral are probably ubiquitous in simulated MW analogues evolved within a full cosmological context, accounting for dynamical, hydrodynamical, and basic astrophysical, baryonic process (e.g., gas cooling and heating, star formation, stellar feedback and enrichment). 

Later, \citet{tre23a} demonstrated using an analytic model that a phase spiral can be triggered without an external perturbation by the cumulative effect of many stochastic `kicks' on to the disc stars, resulting from a clumpy background made of dark-matter (DM), giant molecular clouds (GMCs), or a combination of these.\footnote{See \citet{kho19a} and \citet{gra23a} for alternative scenarios.} In a subsequent study, \citet{gil25a} argued that DM substructure within the Galaxy's halo is unlikely to generate a perturbation that is strong enough.

The findings in these recent studies, most notably \citet{tre23a}, are intriguing and have motivated us to extend our earlier study \citepalias{tep22x} on gas-phase evolution in the Galaxy, where we investigated controlled simulations of an idealised MW interacting with a massive perturber with a focus on the emergence and evolution of disc corrugations linked to the phase spiral phenomenon.
The new models presented here improve upon earlier work in at least three key aspects: 1) by considering the presence of gas that is not inert, but instead is allowed to cool and heat, and to form stars, accounting for their feedback and enrichment, in a way that captures the multi-phase nature of the interstellar medium (ISM) and the clumpy structure of the star-formation process in the Galaxy;
2) by avoiding the complexity of a full cosmological simulation and the obfuscating effect of several interactions; and 3) by considering a full-scale, three-dimensional (3D) Galaxy model, accounting for processes such as the disc's self-gravity and its differential rotation, which have been shown to be crucial in this context \citep{dar19a,ban22a}.\\

This paper is organised as follows. In Sec.~\ref{sec:setup} we introduce our new simulation suite. The analysis of the simulation outputs is described in Sec.~\ref{sec:ana}, and the results from this analysis are presented in Sec.~\ref{sec:result}. Our interpretation of the results, their implications and caveats are discussed in Sec.~\ref{sec:summ}. We wrap up with some final thoughts in Sec.~\ref{sec:fin}.

\section{Setup} \label{sec:setup}

Our working model is built around the paradigm that the dynamical perturbation of the Galaxy's disc has been primarily driven by the last significant disc crossing of the Sagittarius (Sgr) dwarf galaxy some 0.5~-~1 Gyr ago \citepalias[][see also \citealt{bin18a}]{bla21e}. Within this paradigm, we simulate the response of the stellar disc adopting different Galaxy models and evolution conditions, with a focus in the formation and evolution of the phase spiral (PS).

In all our models, the Galaxy is approximated by a multi-component system consisting of a host dark matter (DM) halo, a pre-assembled stellar bulge, a pre-assembled stellar disc and, if applicable, a gaseous disc. The DM halo, bulge, and stellar disc have masses \mbox{$M_{\rm \textsc{dm}} \approx 1.45\times10^{12}$~\Msun}, \mbox{$M_{\rm b} \approx 1.5\times10^{10}$~\Msun}, and \mbox{$M_{\rm d} \approx 3.4\times10^{10}$~\Msun}, respectively, sampled with $2\times10^6$,  $10^5$, and  $5\times10^7$ particles. The mass of the gas disc, if present, is different for different models (see below). It is worth noting that all galaxy components are reactive (i.e. `live'). The Sgr dwarf is approximated by a point-mass with \mbox{$M = 2\times10^{10}$~\Msun} (if applicable).

Initially, the DM halo and the bulge follow roughly a \citet{nav97a} profile with a scale radius $r_{\rm s} \approx 15$~kpc, and a \citet{her90a} profile with $r_{\rm s} \approx 0,6$~kpc, respectively; the disc has a radial exponential profile with a scale length $R_{\rm d} \approx 3$~kpc, and a vertical $\sech^2$ profile with an average scale height $h_{\rm s} \approx 0.3$~kpc. The DM halo structure and mass yields a density consistent with the local estimate \citep[\mbox{$\rho_\textnormal{\sc dm} \approx 0.015$ \Msun pc$^{-3}$};][]{guo24a}. This is relevant to a realistic modelling of the total Galactic potential around the solar circle.\\

Specifically, we consider the following suite of simulation runs (q.v. Tab.~\ref{tab:runs}):
\begin{enumerate}
	\item[\bf fg00:] Pure N-body simulation (i.e. no gas), in which the synthetic Galaxy consists of a host DM halo, a stellar bulge and a stellar disc, all reactive and made up of collisionless particles, and is subject to an impulsive interaction with the Sgr proxy.
	\item[\bf fg10\_nsf:] N-body/hydrodynamical simulation similar to fg00, but in which the synthetic Galaxy additionally features an exponential gas disc with a scalelength\footnote{A gas disc with a scalelength twice that of the stellar disc is a common feature of self-consistent dynamical models of the Galaxy \citep[e.g.,][]{bin15a,bla16a}. See Tab.~\ref{tab:runs}. } \mbox{$R_{\rm d} \approx 6$~kpc}; the gas disc is in vertical hydrostatic equilibrium, with a mass \mbox{$M_{\rm gas} \approx 4\times10^9$~\Msun}, corresponding to a gas fraction \mbox{$f_{\rm gas} = 0.1$}, i.e. 10\% the total disc mass (stars and gas); the compound system is evolved adopting a strict isothermal equation of state (EoS) adopting a temperature 
    \mbox{$T = 10^3$~K}; the gas is thus referred to as `inert'.
	\item[\bf fg20\_nsf:] Similar to fg10\_nsf, but  the gas disc has a mass equal to 20\% ($f_{\rm gas} = 0.2$) of the total disc mass (\mbox{$M_{\rm gas} \approx 9\times10^9$~\Msun}), and a temperature 
    \mbox{$T = 2\times10^3$~K}
	\item[\bf fg20\_sf:] Identical initial conditions to fg20\_nsf, but during the system's evolution the gas is allowed to cool and heat, and to form stars, accounting for their feedback and enrichment.
\end{enumerate}
Finally, we add the following two `control' simulations to our suite:
\begin{enumerate}
	\item[{\bf fg00\_iso}, {\bf fg20\_sf\_iso:}] Identical to fg00 and fg20\_sf, respectively, but the synthetic Galaxy is evolved in isolation, i.e. there is {\em no} interaction with the perturber.
\end{enumerate}

\begin{table}
\begin{center}
\caption{ Overview of simulations. Common to all is a synthetic Galaxy, approximated by a multi-component system consisting of a host dark matter (DM) halo, a pre-assembled stellar bulge, and a pre-assembled stellar disc. See text for details.
}
\label{tab:runs}
\begin{tabular}{lccl}
Identifier & $f_{\rm gas}$$^{\;a}$ & Sgr & Remarks  \\
\hline
 fg00 &  0.0 & Yes & -- \\
 fg00\_iso & 0.2 & No & -- \\
 fg10\_nsf & 0.1 & Yes & inert, isothermal$^{\;b}$ \\
 fg20\_nsf & 0.2 & Yes & inert, isothermal$^{\;c}$ \\
 fg20\_sf & 0.2 & Yes & star-forming, multi-phase \\
 fg20\_sf\_iso & 0.2 & No & star-forming, multi-phase \\
\hline
\end{tabular}
\end{center}
\begin{list}{}{}
\item $^a$ {\em Initial} gas fraction, which decreases over time as a result of the production of stars. The gas is included in the form of an exponential gas disc with a scalelength $R_{\rm d} = 6$~kpc in vertical hydrostatic equilibrium. $^{b,c}$ The gas is treated with a strict isothermal EoS adopting a temperature $T = 10^3$~K and $T = 2\times10^3$~K in fg10\_nsf and fg20\_nsf, respectively.
\end{list}
\end{table}

The pure N-body simulation (fg00) and its hydrodynamical counterpart (fg10\_nsf) are adopted from our earlier studies, \citetalias{bla21e} and \citetalias{tep22x}, respectively. The other models are introduced in this study for the first time.

fg20\_sf is regarded as our more `realistic' simulation, and it is the first of its kind to be published. We note that the increased gas fraction in fg20\_sf with respect to fg10\_nsf is necessary in order to trigger the formation of stars from the outset, given that the values of the relevant parameters (notably the star-formation density threshold and efficiency) are fixed based on other constraints \citep[for a discussion see][]{age13a,gri17b}. Gas fractions in the range 10-20\% are consistent with estimates of the total gas mass in the Galactic disc based on dynamical models \citep[][]{kal09a}, as well as consistent with the mean, total baryonic mass of $\sim10^{12}$\Msun\ haloes \citep[][]{bon25a}.

Simulation fg20\_nsf is intended as a benchmark to assess change in the disc response to an increased potential close to the plane relative to fg10\_nsf \citep[q.v.][]{tep22x}, while fg00\_iso and fg20\_nsf\_iso are intended to study the perturbation induced in the disc by dynamical or numerical `noise'. Run fg20\_nsf is motivated by our earlier findings that a stronger disc potential affects the amplitude and lifetime of the vertical corrugations induced in the disc as a result of a direct interaction with a massive satellite \citepalias{tep22x}. Runs fg00\_iso and fg20\_nsf\_iso are motivated by the claims that, even in the absence of an external perturbation, a clumpy environment (be it of baryonic matter or dark matter) may induce perturbations that trigger a phase spiral \citep[][]{tre23a,gil25a}. 
Note that the gas temperature adopted in fg20\_nsf is higher than fg10\_nsf (\mbox{$T = 2\times10^3$~K} rather than $10^3$~K), imposed by the higher mass of the disc and our requirement that \citet{too64a}'s $Q$ satisfies \mbox{$\textnormal{min}(Q) \sim 1$} which, in turn, is motivated by the desire to have a disc that is reasonably stable against fragmentation from the outset.

It is worth noting that in all models, the pre-assembled stellar disc is sampled with $5\times10^7$ particles, which is significantly higher than the resolution of cosmological (zoom-in) simulations \citep[e.g.,][]{gar22c}, at least a factor 10 higher than the corresponding value adopted in most comparable controlled simulations \citep[e.g.,][]{lap19a}, and only a factor $\sim4$ lower than the number of particles in the highest resolution discs simulated to date \citep[e.g.,][]{hun21t,asa25a}. It must be stressed, however, that none of the earlier work -- with the notable exception of \citet[][]{gar22c} -- accounted for the presence of gas, let alone the effect of gas cooling/heating and star formation processes, as we do here. More details about component masses, particle resolution, etc., for each of our runs are provided in Tab.~\ref{tab:runs}\\

All simulations were set up and run with our \nexus\ framework \citep{tep24a}.\footnote{Runs fg00, fg00\_iso, and fg10\_nsf pre-date the publication of our framework, which was however in place by the time we conducted these.}  In brief, the initial conditions (ICs) were created with an extended version of the {\em Action-based Galaxy Modelling Architecture} (\agama) stellar-dynamics library \citep{vas19a}, that allows the treatment of gas components in addition to collisionless components. The ICs were evolved using a modified version of the Adaptive Mesh Refinement (AMR), N-body/hydrodynamical code \ramses, last described by \citet{tey02a}. The gas was treated by adopting either a strict isothermal EoS (fg10\_nsf and fg20\_nsf), or the `sub-grid` galaxy formation module (fg20\_sf and fg20\_sf\_iso) developed by \citet[][see also references therein]{age21l}. We kindly refer the reader to \citet{tep24a} for more details about the \nexus\ simulation framework, in particular the treatment of the gas in fg20\_sf and fg20\_sf\_iso,

All models were run for a total simulation timespan of roughly \mbox{2~Gyr} in a cubic box of size \mbox{600~kpc} per side, adopting a maximum and minimum AMR levels \mbox{$l_{\rm max} = 14$} and \mbox{$l_{\rm min} = 8$}, respectively, implying a limiting spatial resolution of \mbox{$\delta x = 600 / 2^{14} \approx 37$~pc}.
The Sgr proxy (if included) moved along a hyperbolic orbit which intersects the synthetic Galactic plane {\em once} at $R \approx 18$~kpc at $t \approx 100$~Myr. After the first crossing, the perturber's mass was exponentially decreased with a time scale of \mbox{30~Myr}, to reduce its gravitational influence on the disc, thus yielding a clean, one-time impulse onto the synthetic Galaxy. While this is clearly not a realistic approximation of the actual Sgr orbit, it does allow us to study in a clean fashion the dynamical effect on the stellar disc induced by the single transit of a massive perturber (\citealt{bin18a}; \citetalias{bla21e,tep22x}). In addition, as discussed in \citetalias{bla21e}, we believe that it was Sgr's last significant crossing (i.e. when it was massive enough) some \mbox{0.5~-~1~Gyr} ago that triggered the perturbation of the Galaxy's disc that we observe today.

\begin{figure}
\centering
\includegraphics[width=\columnwidth]{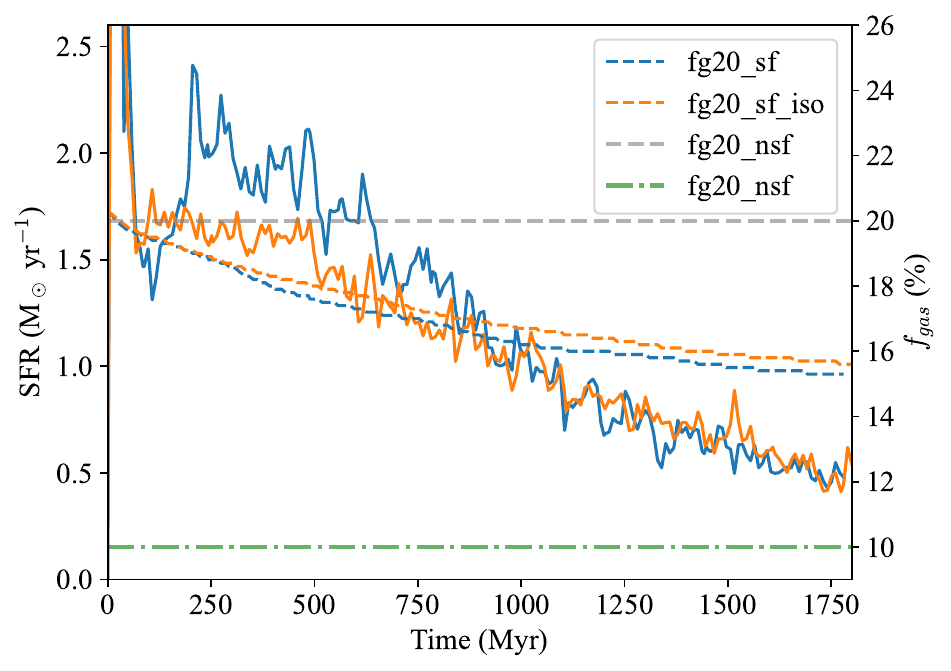}
\caption[  ]{ Star formation rate (continuous curves; scale on left $y$-axis) and total gas fraction (broken curves; scale on right $y$-axis) in our hydrodynamical runs. Blue and orange curves correspond, respectively, to the interacting and isolated star-forming Galaxy model. Note that the inert-gas simulations (horizontal grey and green lines) have no star formation and do not experience gas depletion. }
\label{fig:sfh_fg}
\end{figure}

Fig.~\ref{fig:sfh_fg} provides an overview of the star formation history (SFH) and gas depletion (if applicable) in our hydrodynamical runs. Inert gas simulations have no star formation by design, and thus they are not affected by gas depletion. In fg20\_sf and fg20\_sf\_iso, we observe a similar SFH and gas depletion history, although fg20\_sf displays a slight enhanced SFR in the time span \mbox{$200 \lesssim t~/~\Myr \lesssim 1000$}, and a correspondingly higher gas depletion, probably as a result of the interaction with Sgr. Note that the initial star formation `burst' in both runs is a common feature of this type of simulation, where an initially exponential gas disc is allowed to form stars from the outset. A way around this is to evolve the synthetic galaxy under adiabatic conditions for a few hundred million years to allow the disc to settle (effectively, to lose the exponential cusp). However, the long-term evolution is not affected by this \citep[q.v.][]{bla24a}.\\

For ease of discussion, we will refer to the stars already present at the start of the simulation (bulge or disc) as `pre-existing', and to the stars that form out of the gas during the course of the simulation as `newly formed'. Note that the latter is adopted regardless of the actual age of the stars. Pre-existing stars are immutable (or inert), i.e. they do not evolve. In contrast, newly formed stars age and lose mass and provide feedback to the environment in the form of energy, momentum, and chemically enriched gas as a result of their evolution. We will focus the remainder of our paper on the kinematic study of the pre-existing disc stars, and will address the newly formed stars in a future study. The reason behind this choice is to allow for a like-for-like comparison between the various simulations, as well as between our study and earlier work, regardless of whether or not star-formation is accounted for.

\section{Analysis} \label{sec:ana}

Our main goal is to study the dynamical perturbation of the Galactic disc by the interaction with Sgr under different sets of conditions, using each of the runs introduced above. Our focus is on the formation and evolution of the phase-spiral phenomenon within the context of a `realistic' Galaxy model that accounts for all the dynamically important components (DM, stars, gas), as well as for the relevant astrophysical processes and components (star formation, turbulence, multi-phase ISM). More specifically, we seek to quantify the incidence rate, i.e. when and where the phase spiral appears during the evolution of each of our synthetic Galaxy models.

\begin{figure}
\centering
\includegraphics[width=\columnwidth]{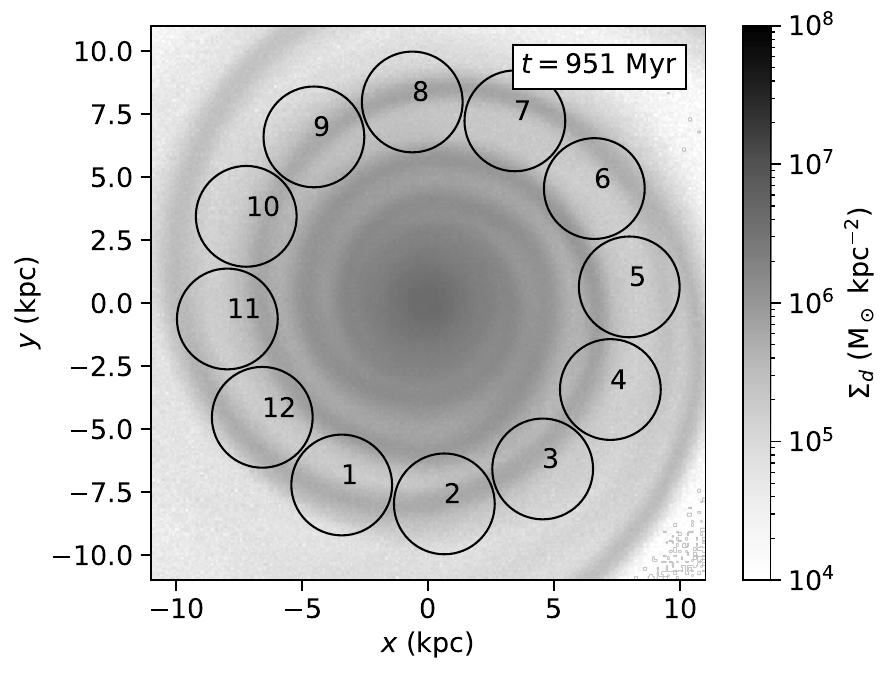}
\vspace{-10pt}
\caption[  ]{ Projected stellar density in fg00 at $t \approx 1$~Gyr. The circles and the inscribed numbers indicate, respectively, the projected sampling volumes and their associated ID along the solar circle ($R_\odot = 8.2$~kpc). Note that the volumes roughly co-rotate with the disc (in anti-clockwise direction). }
\label{fig:vols}
\end{figure}

To this end, we followed part of the methodology presented in our earlier work \citepalias{bla21e}. In brief, we segmented the distribution of pre-existing disc stars around the solar circle \citepalias[\mbox{$R_\odot \approx 8.2$};][]{gra19a} with twelve identical, spherical volumes with a diameter of roughly 4 kpc each (see Fig.~\ref{fig:vols}). The volumes are numbered 1 to 12 in an anticlockwise direction. In each volume, we calculated the distribution of stars in vertical phase space \mbox{$z - \vz$}, weighted either by azimuthal velocity $\vp$, or by radial velocity, $\vR$. We repeated this procedure at different times from start ($t = 0$~Gyr) to finish ($t \approx 2$~Gyr) of the simulation, adopting a time step \mbox{$\delta t \approx 10$~Myr}. 

Note that the volumes are fixed into a rotating frame of reference that approximately follows the Galactic rotation (counter-clockwise) at the solar radius; volume 1 is roughly aligned in azimuth with Sgr's disc crossing location, \mbox{$(x,y) \approx (-18,0)$~kpc}, at the impact epoch ($t \approx 100$~Myr). The intent behind this approach is to sample at all times roughly the same disc region (although not necessarily the same stars, since these will move in and out of the volume as a result of their different orbits).
As discussed in \citetalias{bla21e}, and found by several independent studies \citep[e.g.,][]{lap19a,gar22c}, the precise method used to sample the disc (i.e. co-moving frame vs. fixed frame) does not significantly affect the results.

Note that, prior to calculating the phase space distribution at each time step, the synthetic Galaxy is centred on the centre of mass of the stellar disc using the `shrinking sphere' approach \citep{pow03o}. The angular momentum of the disc is subsequently aligned with the $z$-axis of the simulation box. These are both important steps required to properly quantify dynamical quantities.

We visualised the results by arranging the phase-space distribution at each volume along a row, for a given time step, and stack the resulting rows in ascending order of simulation time $t$. This yields what we refer to as a phase-space 'chronogram'. This is a very useful way to visualise the spatiotemporal evolution of the phase spiral, as we have shown earlier \citepalias[][]{bla21e}, and it has been adopted by more recent studies \citep[][]{gar22c,asa25a}.

Following this approach we end up with two different chronograms: a $\vp$-chronogram,  and a $\vR$-chronogram, each spanning 12 columns (corresponding to the spherical sampling volumes) and 21 rows (with a time step of roughly 10 Myr, from $t=0$ to $t \approx 2$~Gyr). We refer to each volume-timestep pair on a chronogram as a `cell'.

We calculated $\vp$- and $\vR$-chronograms for each of our runs, thus allowing for a like-for-like comparison of the incidence rate of the phase spiral across simulations. For the sake of brevity, we do not include examples of chronograms in this study, but instead refer the reader to \citetalias[]{bla21e} (their fig.~12), and to the more recent work by \citet[][their fig.~2]{gar22c} and \citet[][their fig.~12]{asa25a}.

\subsection{Phase-spiral finder}

We seek to identify the epochs and the corresponding locations across the stellar disc (chronogram cells) where the phase spiral is apparent (not necessarily a perfect match to the \gaia\ phase spiral, it must be noted). Rather than performing a visual analysis as we did in \citetalias{bla21e}, we opted for an automated procedure. To this end, we adopted the approach put forward by \citet{gar22c}, who have successfully applied it in the systematic identification of phase spiral in a zoom-in, cosmological simulation.

Their method, which is based on the Fourier decomposition of the stellar distribution in vertical phase space, can be summarised as follows: The cell (either of a $\vp$- or a $\vR$-chronogram) is partitioned into $N$ concentric annuli of linearly increasing  phase-space radius \mbox{$\tilde{R} \equiv (J_z)^{1/2}$}, where $J_z$ is the vertical action. Next, a Fourier analysis is applied to the distribution of stars within the cell for each annulus, which yields the amplitudes $A_m$ and phases $\phi_m$ of modes $m=0$ to $m=6$ as a function of $\tilde{R}$; effectively a radial profile.

A dipole-like signature is implied by a $A_1/A_0$ profile that is significantly higher than the profiles of the amplitudes of higher modes; a true one-arm spiral signature also requires a monotonically increasing\footnote{For a spiral wrapping up in a clockwise sense, as is the case of the \gaia\ phase spiral. } (roughly linearly) phase $\phi_1$ profile.
Similarly, a two-arm spiral is revealed by a dominant \mbox{$m=2$} amplitude and monotonically increasing $\phi_2$ profile, etc. In this work, we will concentrate on $m=1$ and will defer the analysis of $m=2$ modes \citep[cf.][]{hun22a,asa25a} to a forthcoming study.

In practice, we deem a dipole-like signal significant if the average value of the $A_1/A_0$ profile (i.e. the unweighted arithmetic mean of $A_1/A_0$ over $\tilde{R}$) is strictly higher than the average of the other modes, $\langle A_1/A_0 \rangle > \langle A_m/A_0 \rangle$  for \mbox{$m = 2, 3, \ldots, 6$}. The signal is further considered a strong indication of the presence of a one-arm spiral if the average slope, $\nabla \phi_1$, of the phase profile over $\tilde{R}$ (effectively, the slope of a linear fit to the $\phi_1$ profile) is higher than a given threshold, $(\nabla\phi)_{\rm th}$, and its scatter around a linear fit is below some threshold,  $\sigma(\nabla\phi)_{\rm th}$. The significance (or `strength') of the signal was estimated by the median value of the $A_1/A_0$ profile, which we denote as \mbox{`med$\left[A_1/A_0\right]$'}.

The free parameters of the method, and the values we adopted, are then: $N = 21$, $(\nabla\phi)_{\rm th} = 0.1$, and $\sigma(\nabla\phi)_{\rm th} = 0.3$. We found these values empirically after extensively testing the method with synthetic data (see Appendix \ref{app:val}). In general, we found that the method works remarkably well under idealised conditions (see top row in Fig.~\ref{fig:ps_toy}), and it showed  satisfactory performance when noise was added to the data (see second to last rows in Fig.~\ref{fig:ps_toy}), which is the case of the data retrieved from our simulations. It is worth noting that the noise induced by the variability in the simulation data is what makes the introduction of $(\nabla\phi)_{\rm th}$ and $\sigma(\phi)_{\rm th}$ necessary. Indeed, under ideal conditions, $(\nabla\phi)_{\rm th}$ is a constant, i.e. $\sigma(\nabla\phi)_{\rm th} \equiv 0$ (see top row in Fig.~\ref{fig:ps_toy}).

\begin{figure*}
\centering
\includegraphics[width=0.49\textwidth]{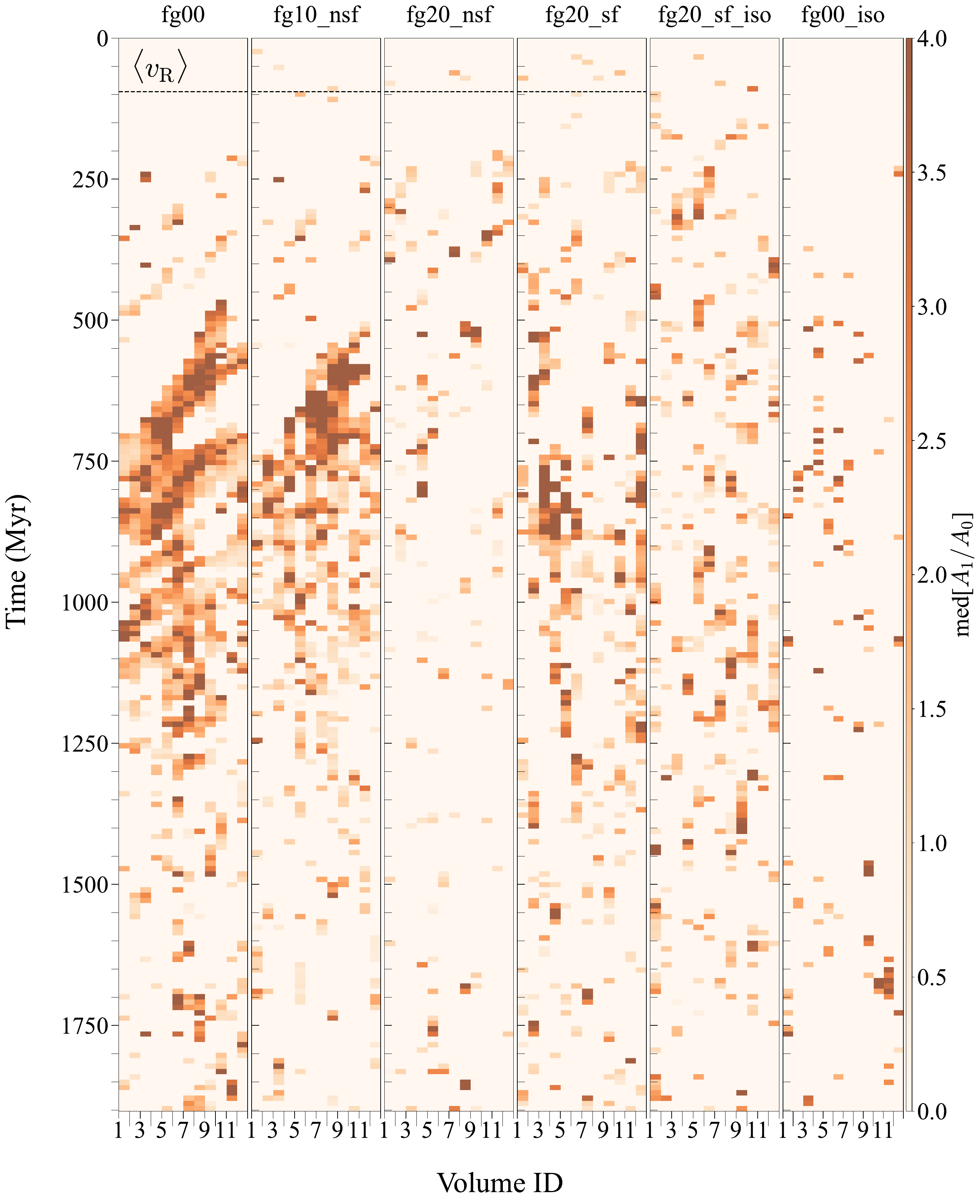}
\hfill
\includegraphics[width=0.49\textwidth]{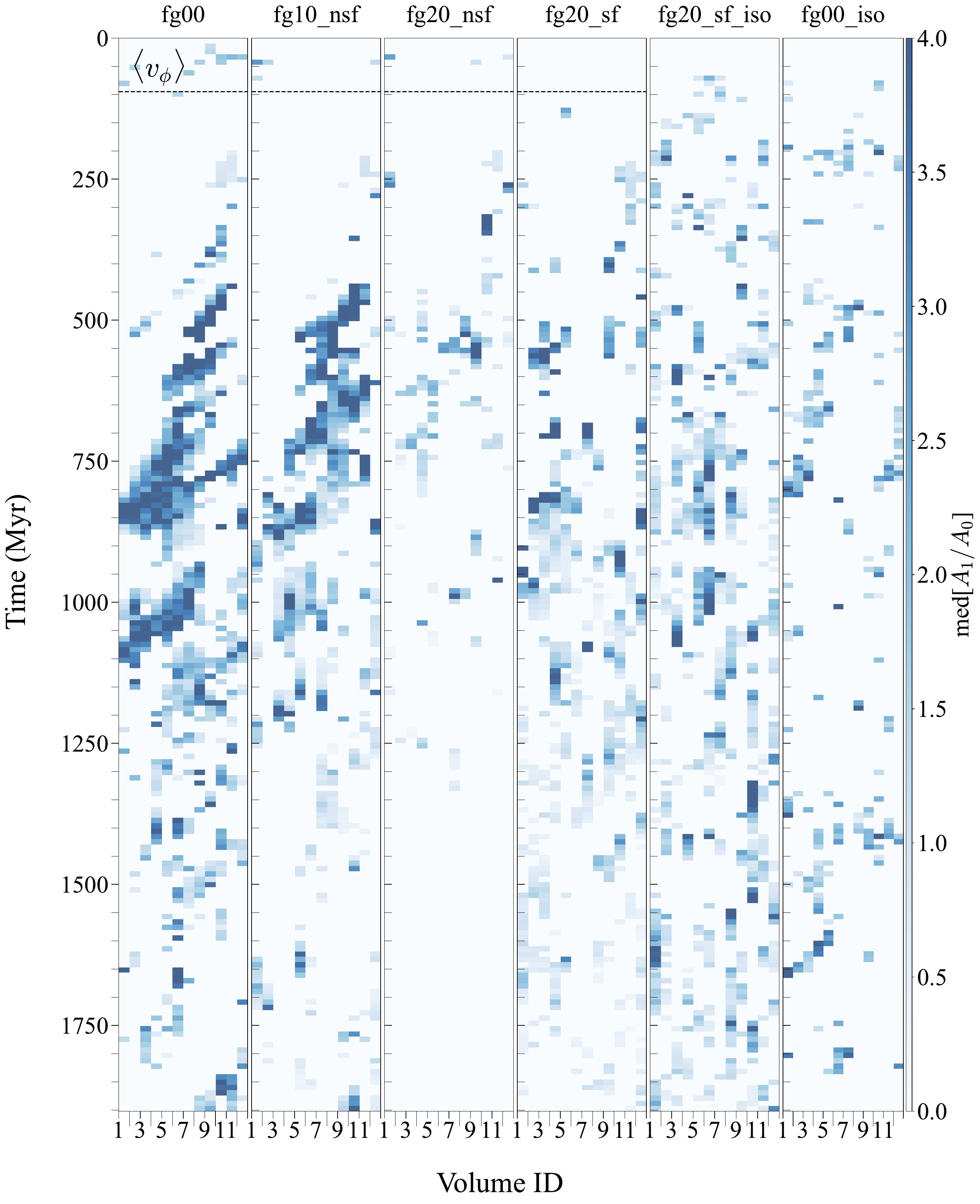}
\caption[  ]{
Incidence maps, which display the presence of the phase spiral and its strength (see Sec.~\ref{sec:ana}), weighted by $\vR$ (left; orange hues) and $\vp$ (right; blue hues), over roughly two billion years of evolution along the solar circle of the synthetic galaxy. For each of $\vR$ and $\vp$, each column corresponds to one of our simulations, as indicated by the column header (cf. Tab.~\ref{tab:runs}). Within each column, a cell flags a particular location and time as indicated by the label on the horizontal axis and the vertical axis, respectively; each location corresponds to one of the co-rotating volumes along the solar circle (cf. Fig.~\ref{fig:vols}), and consecutive vertical cells are separated with a time step of $\delta t \approx 10$~Myr. The horizontal, dashed line in each column (if applicable) indicates the approximate disc crossing epoch ($t \sim 100$ Myr). In each map, a non-empty cell indicates a positive detection, i.e. the presence of a one-arm PS; a darker hue indicates a stronger (i.e. with a higher contrast with respect to the background) signal.
}
\label{fig:incidence_map}
\end{figure*}

\section{Results} \label{sec:result}

\subsection{Incidence rates across space and time} \label{sec:chrono}

Applying the phase-spiral finder algorithm introduced above, we transformed each of the $\vp$- and $\vR$--chronograms described in Sec.~\ref{sec:ana} into an `incidence' map. This is a matrix of $12\times21$ cells, each displaying the value of \mbox{med$\left[A_1/A_0\right]$}, in which non-empty cells indicate that the corresponding location (column) and time (row) in the synthetic disc hosts a phase spiral. The result of applying this to each simulation yields the set of incidence maps displayed in Fig.~\ref{fig:incidence_map}. To enhance the visual appearance of the maps, we have normalised the detection strength to a maximum value of 4, as indicated by the colour-bar next to each incidence map. Note that there is no obvious quantitative meaning attached to the value of \mbox{med$\left[A_1/A_0\right]$}. But since we are not after a match to the \gaia\ phase spiral, our choice has no impact on our analysis; what matters is the differential comparison between the various models.

A few comments preceding the discussion of the results are appropriate. During the testing phase of the phase-spiral finder algorithm, we found that the $\vR$- and $\vp$-incidence maps must be treated differently. This is due to the fact that the range of $\vR$ extends to both negative and positive values, and thus the corresponding phase-space map  displays both negative and positive values across a roughly symmetric range. In contrast, the $\vp$-weighted phase-space distribution is always positive, and appears on top of an underlying (unperturbed) distribution. The Fourier decomposition approach appears to work better in the case of the $\vR$-weighted maps, so it seems reasonable to apply a suitable transformation to the $\vp$-weighted maps to obtain a distribution with negative and positive values.

\citet[][]{gar22c} shifted the $\vp$-weighted map by the mean value $\langle \vp \rangle$ across the map, i.e. effectively displaying the distribution of stars weighted by $\vp - \langle \vp \rangle$.
We followed a similar approach, but we normalised the map by unperturbed distribution prior to estimate and subtract the mean value of the map. The unperturbed distribution is straightforward to get for any of our simulations, and it is roughly given by a 2D Gaussian centred on the origin of the $z - \vz$ plane.\footnote{In practice, we chose for this purpose volume = 1, which is statistically equivalent to any other volume. {\bf See \citetalias{bla21e}, their fig.~1, and the top row of their fig. 12., for visual examples of the unperturbed distribution.}}

We note that the PS-finding algorithm can lead to false positives and false negatives (of order a few percent; see below) but, overall, it usually delivers satisfactory results. We have visually inspected randomly selected volumes in our $\vR$- and $\vp$-incidence maps, and generally confirmed the results of the automated procedure. Therefore, and notwithstanding the caveats of the method, we favour it over a pure visual inspection. In this way, all simulations are put on equal footing, and the comparative analysis is rendered robust against subjectivity.

\begin{figure}
\centering
\includegraphics[width=0.49\columnwidth]{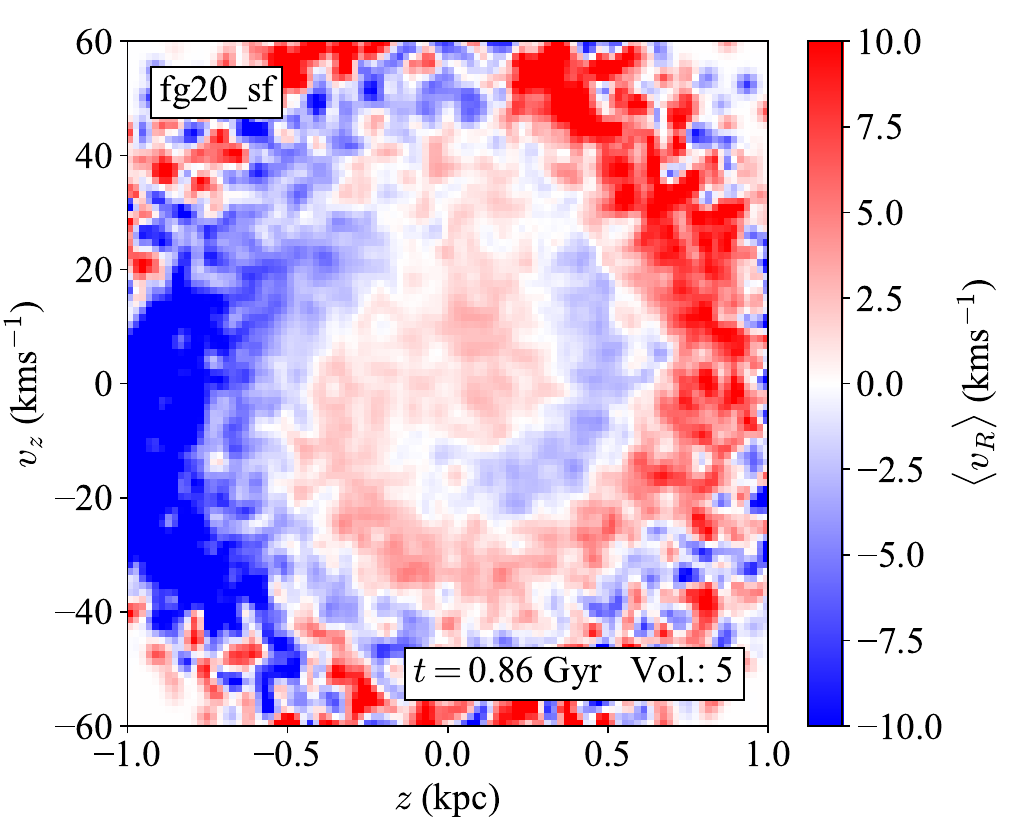}
\includegraphics[width=0.49\columnwidth]{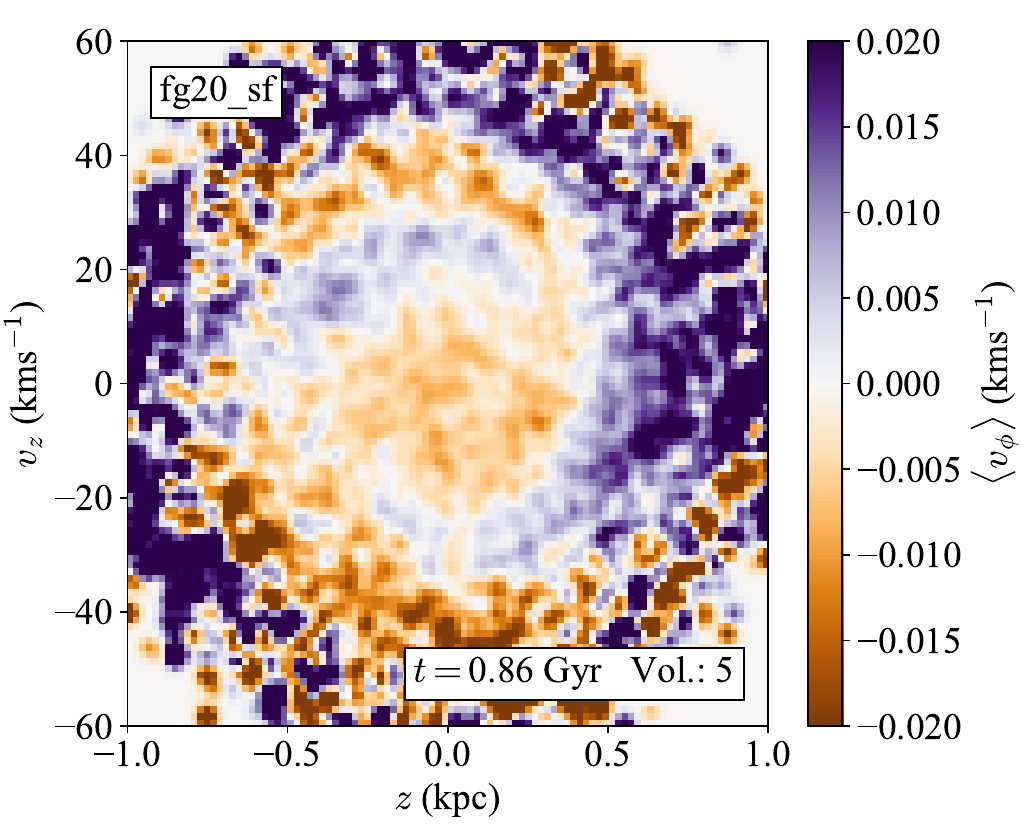}\\
\rule[1ex]{0.9\columnwidth}{0.5pt}\\
\includegraphics[width=0.49\columnwidth]{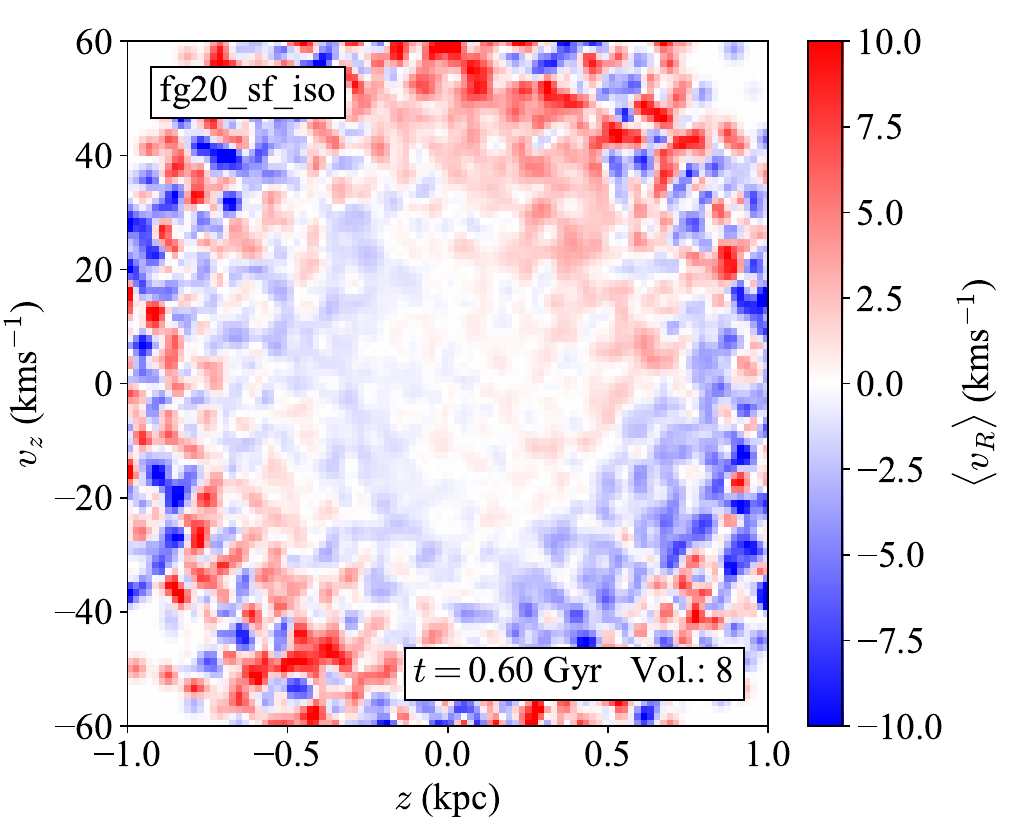}
\includegraphics[width=0.49\columnwidth]{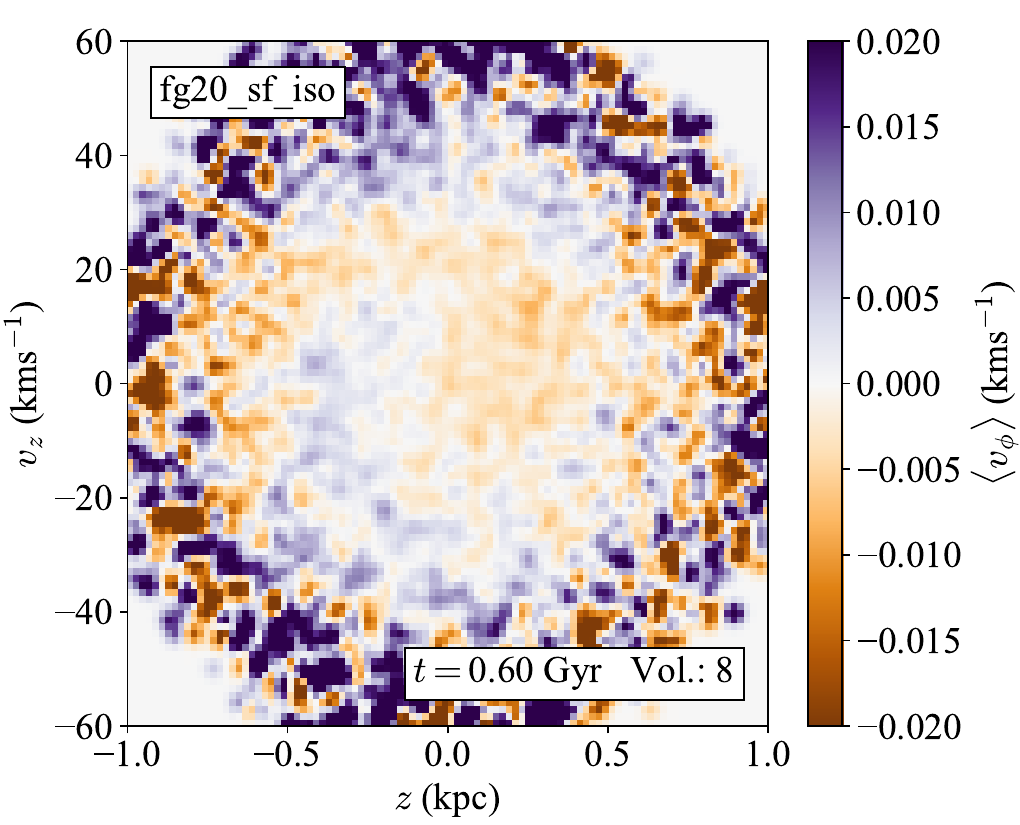}
\includegraphics[width=0.49\columnwidth]{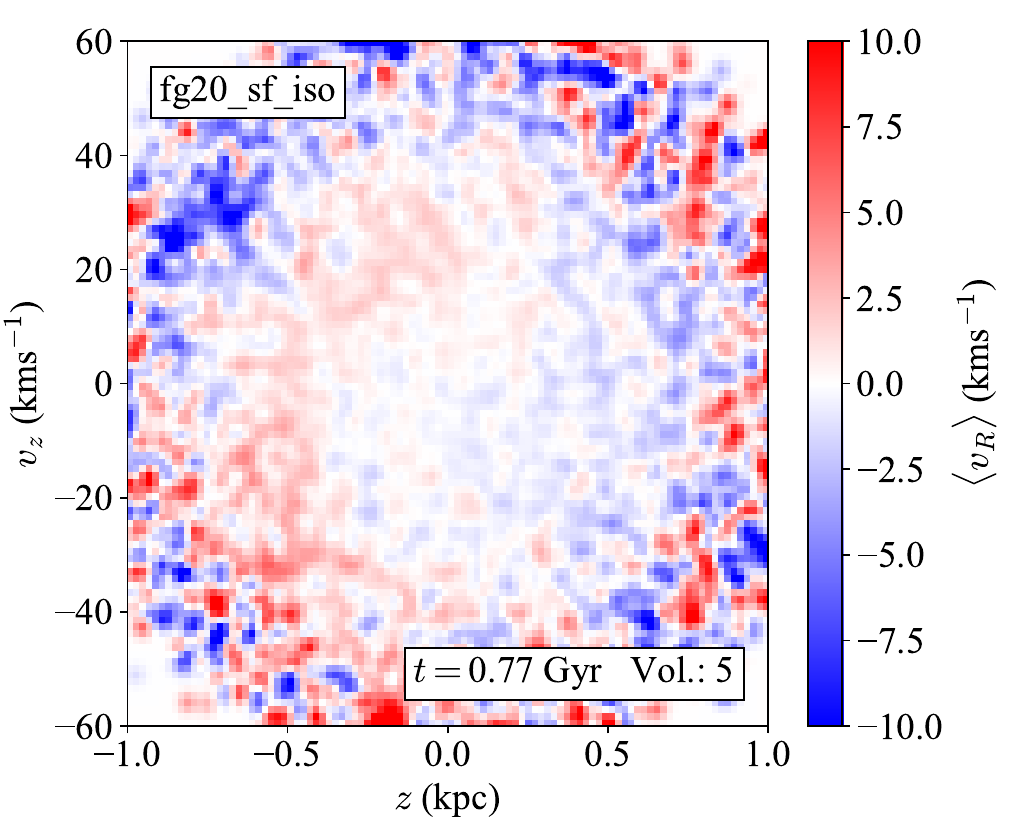}
\includegraphics[width=0.49\columnwidth]{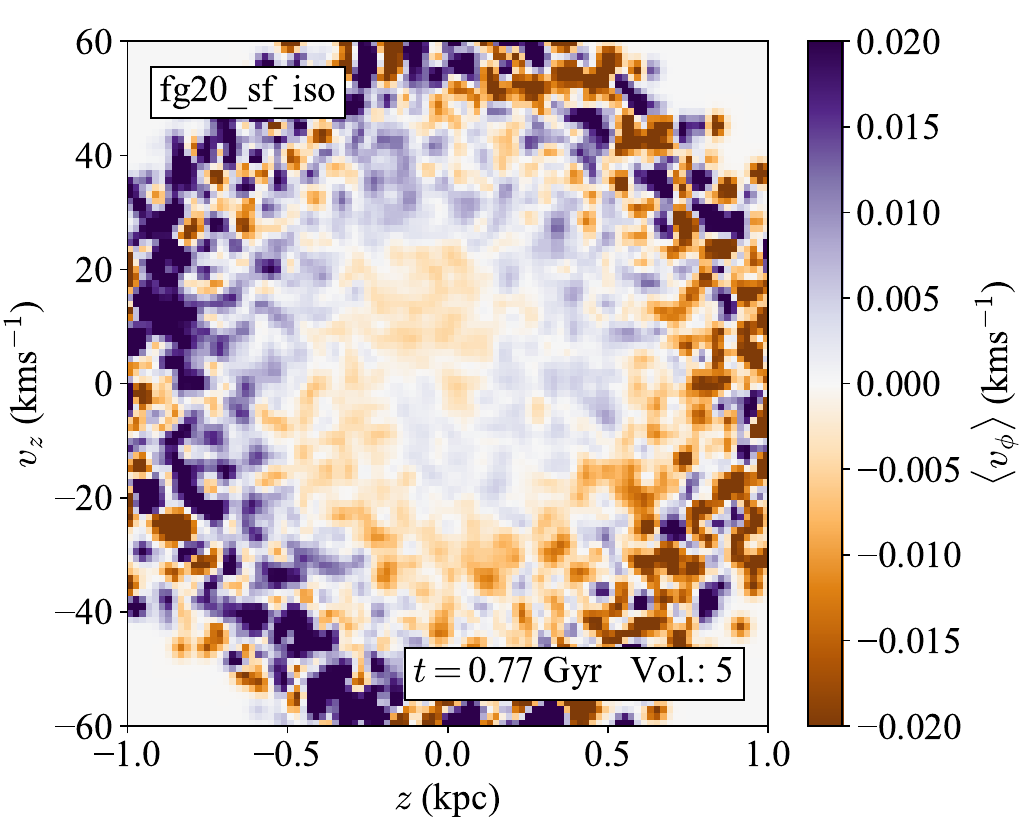}
\includegraphics[width=0.49\columnwidth]{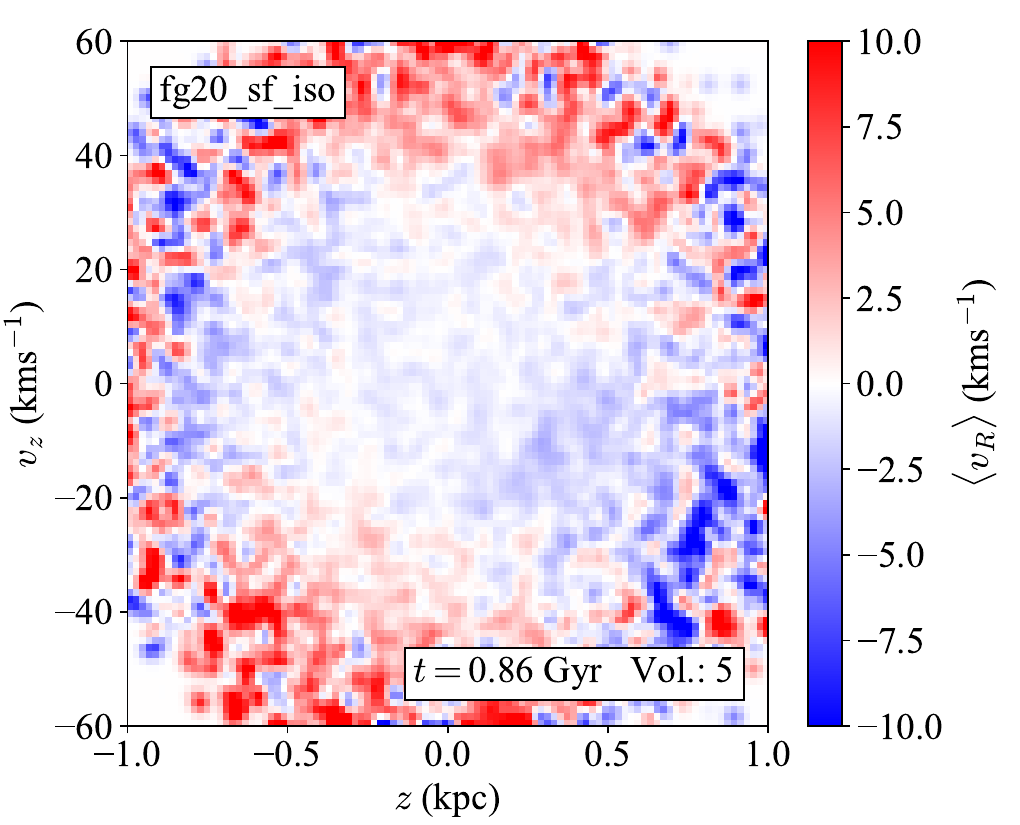}
\includegraphics[width=0.49\columnwidth]{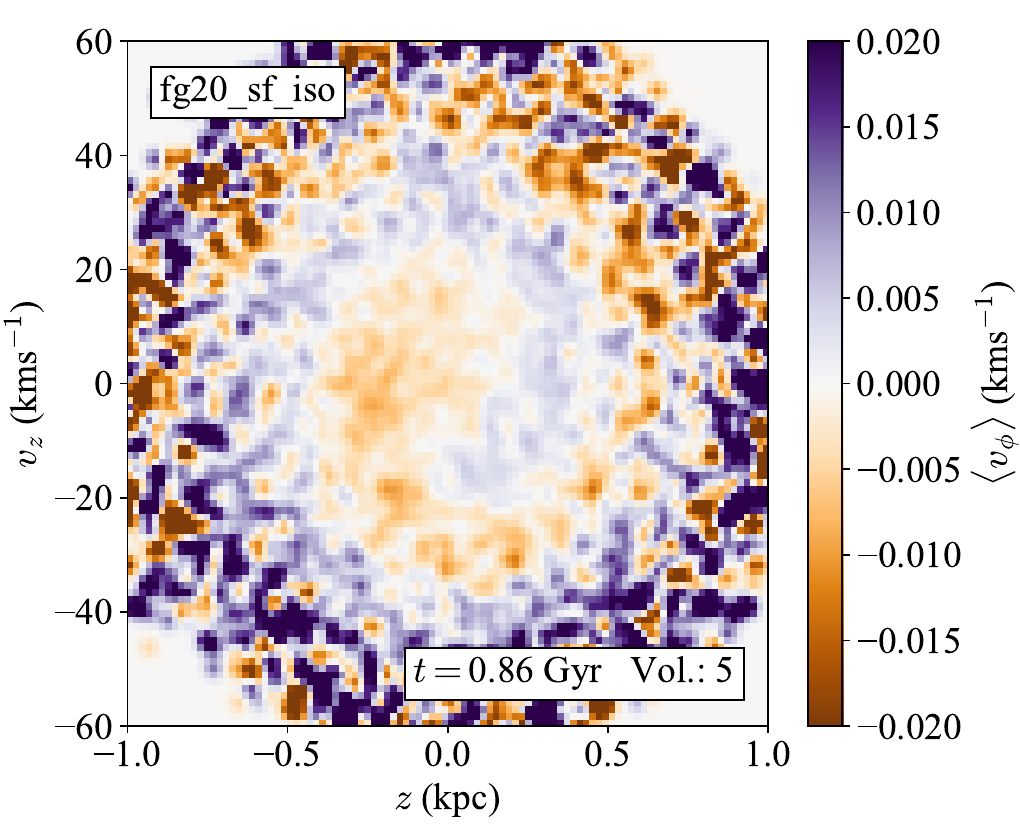}
\includegraphics[width=0.49\columnwidth]{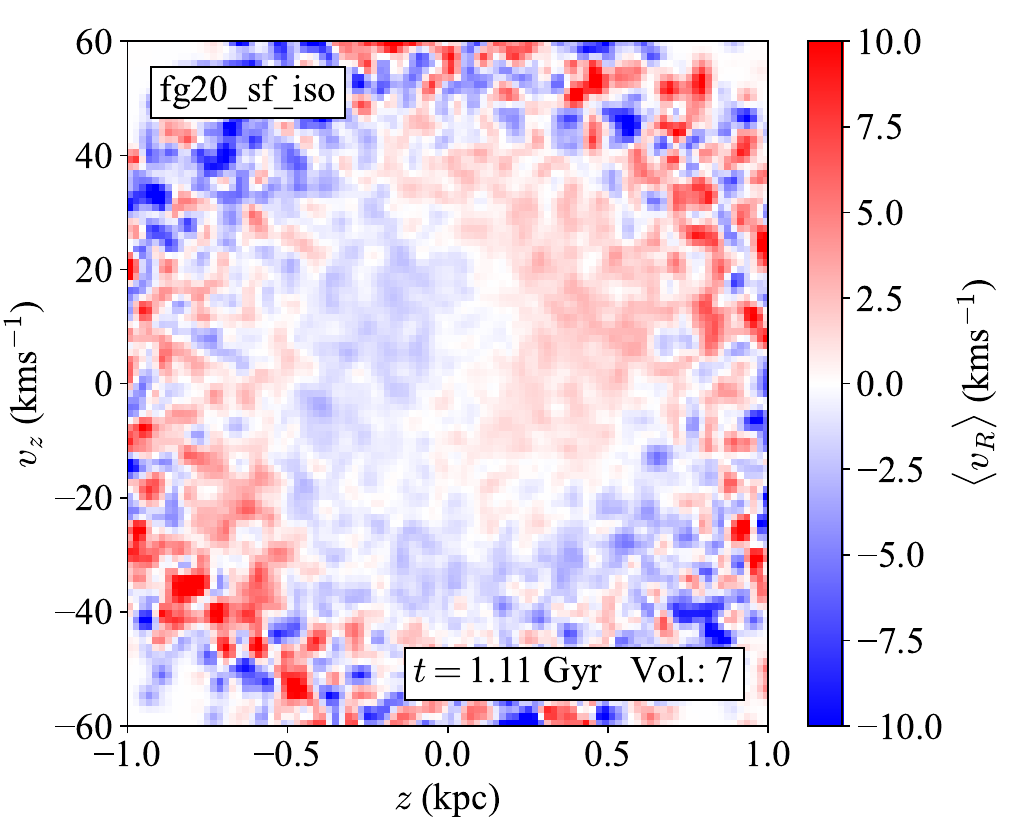}
\includegraphics[width=0.49\columnwidth]{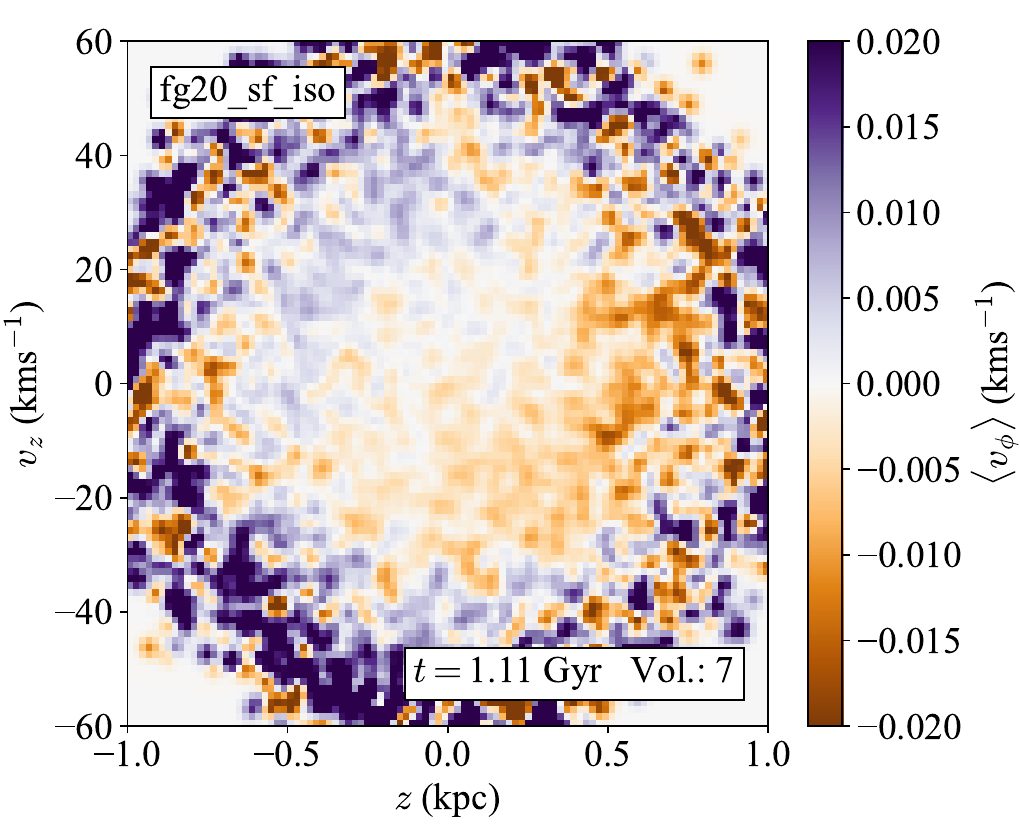}
\caption[  ]{ Examples of phase spirals found in run fg20\_sf (top row, above horizontal line) and in run fg20\_sf\_iso (second through last rows, below horizontal line). The corresponding volume and epoch is indicated on the bottom-right corner of each panel. Left column: Weighted by $\vR$. Righ columnt: Weighted by $\vp$. The panels in the second through last rows are examples of phase spirals likely arising from the `Tremaine-Frankel-Bovy' (TFB) effect, named after \citet*{tre23a}. Note that the PS is cleary visible if weighted either by $\vR$ or $\vp$; i.e. it is kinematically interlocked -- an unexpected result (see Sec.~\ref{sec:kin}).}
\label{fig:ps_tfb}
\end{figure}

We now turn to the discussion of the results shown in Fig.\ref{fig:incidence_map}, focusing on the $\vR$-weighted incidence map (left panel). The first column, which corresponds to fg00, a pure N-body model, clearly shows that the phase spiral emerges some 400~Myr after Sgr's disc crossing (flagged by the horizontal line at the top), but not across all volumes, at least not simultaneously. The signal appears to propagate across a narrow region (a band) in azimuth, until it becomes widespread at $t \approx 750$~Myr (or roughly \mbox{$\Delta t \approx 650$~Myr} after impact). Thereafter, the presence of the phase spiral becomes more stochastic, fading away with time, and virtually disappearing after \mbox{$\Delta t \approx 1150$~Myr}. These results show that, in the absence of additional (external) perturbations, the phase spiral is long-lived but eventually vanishes, probably due to the combined effect of phase-mixing and the settling of the disc \citepalias{tep22x}. The need for multiple interactions to maintain the phase spiral over long periods is consistent with the results by \citet{gar22c}, who found that multiple satellites, even with low mass ($\lesssim 10^9$~\Msun), are capable of triggering and sustaining the phase spiral over several billion years \citep[cf.][]{lap19a,bla19a}.

It is worth stressing that while the choice of an arbitrary threshold value for the slope may affect the earliest epoch at which the PS finder detects a positive signal, it turns out that -- in our simulations -- the fact that the PS signal becomes significant at $t~\gtrsim~500$~Myr (i.e. $\sim400$~Myr after impact) is entirely independent of this choice. This can be clearly seen in fig.~12 of \citetalias{bla21e}, in the case of model fg00. Independent of any automated PS finding, in that figure the PS can be seen to emerge clearly not earlier than $t \approx 500$~Myr, in agreement with the results displayed in  Fig.~\ref{fig:incidence_map}. Given this consistency, we are confident that our choice of slope threshold has not affected the earliest epoch at which the PS is detected by the algorithm.

The dynamic nature of the perturbation is apparent in the form of the azimuthal bands made up of non-empty cells running diagonally down from the right, with a `slope' that appears to decrease as the system evolves. This behaviour has been discussed at length in \citetalias{bla21e}, and has also been reported in other studies \citep{gar22c}. In \citetalias{bla21e}, we argued that the perturbation is supported by the interplay between a bending wave and a kinematic density wave, both triggered by the impulsive interaction. In consequence, its propagation speed is directly related to the difference between the pattern speeds of these waves. As these wrap up with different frequencies, the propagation speed of the perturbation increases, which is reflected in the decreasing slope of the diagonal bands.

Moving on to the results of fg10\_nsf (second column from the left in $\vR$-weighted incidence map), it is immediately apparent that the emergence of the phase spiral is delayed, and it fades away faster, compared to the first column (fg00). We remind the reader that the only initial difference between the fg00 model and the fg10\_nsf model is the presence in the latter of an additional inert gas disc with a mass equivalent to roughly 10\% the total disc mass. Thus, the probable reasons for the difference between their corresponding incidence maps are: 1) the stronger potential close to the plane in fg10\_nsf with respect to fg00, which decreases the disc's responsiveness to an impulsive perturbation; and 2) the dissipative nature of gas (absent in f00), which decreases the timescale for the disc's settling \citepalias[][]{tep22x}.

This interpretation is reinforced by the result shown in the third column (corresponding to fg20\_nsf). In this model, the inert gas disc is twice as massive with respect to fg10\_nsf, and therefore features a stronger in-plane potential and induces a stronger dynamical dissipation. As a result, the phase spiral is mostly suppressed over the entire evolution of the synthetic Galaxy, as indicated by the notably low number of non-emtpy cells on the map.

Remarkably, when the same, massive gas disc is allowed to cool and heat, and to form stars (i.e. when it is `active' as opposed to inert), the phase spiral becomes apparent again, as demonstrated by the results shown in column 4 (from the left), corresponding to our most `realistic' simulation (fg20\_sf). Recall that the Galaxy model in this run is initially identical to the model in fg20\_nsf; the only difference between these is the way the gas is treated during the system's evolution. In model fg20\_sf, the phase spiral is less widespread both in space and time compared to fg00 and fg10\_nsf; in fact, it appears rather intermittently, but its presence peaks around the same time as in these runs ($t \approx 800$~Myr).

We consider the following different hypothesis for this, rather unexpected, result. First, the resurgence of phase spiral may be caused, at least in part, by a decrease in the gas potential relative to fg20\_nsf as a result of the gas depletion due to star production (cf. Fig.~\ref{fig:sfh_fg})  The gas is being replaced by newly formed stars, so there is no net mass loss. But the mass distribution does change, in particular in the direction vertical to the plane. Indeed, the scale height of newly formed stars generally increases with time \citep[][]{bla25a}, which in turn reduces their overall contribution to the potential close to the plane. It is therefore plausible that the overall potential is weaker in fg20\_sf relative to fg20\_nsf (but not as weak as in fg10\_nsf), and thus the disc more susceptible to react to a perturbation.

The second hypothesis is that the phase spiral is triggered (at least partially, i.e. in addition to the interaction) by a mechanism similar to that proposed by \citet*{tre23a}, whereby the cumulative effect of many small, stochastic perturbations imposed onto the stars --- e.g., due to the clumpy nature of the ISM --- results in the formation of spiral-like structures in phase space. We refer to this as the `Tremaine-Frankel-Bovy' (TFB) effect. This hypothesis is supported by the results shown in column 5 (from the left), corresponding to fg20\_sf\_iso. In this model, the gas is forming stars similar to fg20\_sf, but the synthetic galaxy is {\em not} subject to an interaction with Sgr. Nevertheless, the incidence map displays isolated pockets of cells where the phase spiral is detected but which appear to be stochastically distributed along the solar circle over the full evolution of the system. It is worth noting that the incidence rate of the phase spiral in this model is {\em significantly} higher than in fg20\_nsf (see also Fig.~\ref{fig:rate_histo}, discussed in Sec.~
\ref{sec:rate_histo}), but it is not as widespread as in fg20\_sf, probably owing to the absence of an interaction.

In Fig.~\ref{fig:ps_tfb} we show examples of phase spirals found in fg20\_sf in the top row (above the horizontal line) and in fg20\_sf\_iso in the bottom rows. The latter are probably a result of the TFB effect. The left column corresponds to the PS weighted by $\vR$; the right column, by $\vp$. We see clear evidence that the PS is visible on both cases, in general. This was not anticpated by \citet{tre23a}, who modelled the evolution of the phase spiral considering purely the density (counts) of stars in phase space and dit not consider their distribution weighted by $\vR$ or $\vp$.

For the sake of completeness, we contemplate the possibility that the phase spiral may be mainly triggered by numerical noise. To test this, we show in the last column of Fig.~\ref{fig:incidence_map}'s left panel the $\vR$-incidence map corresponding to fg00\_iso, i.e. a pure N-body simulation of a synthetic Galaxy evolved in isolation. Clearly, the phase-spiral incidence is extremely low, comparable to the incidence in fg20\_nsf, and significantly lower than fg20\_sf\_iso, which is also evolved in isolation. Based on this, we conclude that the PS in fg20\_sf (or fg20\_sf\_iso) is not an artefact resulting from numerical noise.\\

The right panel in Fig.~\ref{fig:incidence_map} shows the $\vp$-incidence maps corresponding to the $\vR$ maps shown on the left panel. Qualitatively, we find the same behaviour as in the respective $\vR$-incidence maps on a simulation by simulation basis. But they are by no means identical, and there are instances where the phase spiral is only detected in one but not the other. We return to this point in Sec.~\ref{sec:kin}.

We note a few weak false positives in all runs prior to the disc crossing (indicated by the horizontal dashed line in each column, if applicable). These are either noise-generated signals, or glitches of the phase-spiral finder algorithm applied to noisy data. Their overall incidence rate and strength can be estimated by looking at column 6 (from the left); clearly, they are insignificant and can thus be safely ignored.
Finally, we note the apparent lack of diagonal azimuthal bands -- as seen in the case of fg00 and fg10\_nsf -- in the models fg20\_sf and fg20\_sf\_iso, which is probably a visual bias caused by the low incidence rate in the case of the interaction model, or due to the absence of an impulsive perturbation in the case of the isolated model.

\subsection{Incidence rates over time} \label{sec:rate_histo}

Incidence maps provide a detailed view of when and where the phase spiral appears during the evolution of the disc. 
But given the equivalency\footnote{ Locations along the solar circle are strictlt speaking not entirely equivalent as a result of Sgr's disc crossing, but we currently ignore this circumstance which is of minor importance for our analysis.} in the azimuth along the solar circle -- resulting from the initial axisymmetry of our synthetic Galaxy -- it seems reasonable to consider the incidence rate in time alone (rather than time {\em and} space) by marginalising each incidence map over space. In practice, we bin the incidence rate in time using a bin size of $4 \delta t \approx 40$~Myr (or four rows), then we calculate the fraction of cells with a positive detection at each bin. Thus, if all the cells in a bin had a positive detection, the temporal incidence rate for that bin (i.e. time step) would be 100\%. In this way we calculate for each map the incidence rate of the phase spiral over time.

\begin{figure}
\centering
\vspace{-5pt}
\includegraphics[width=0.89\columnwidth]{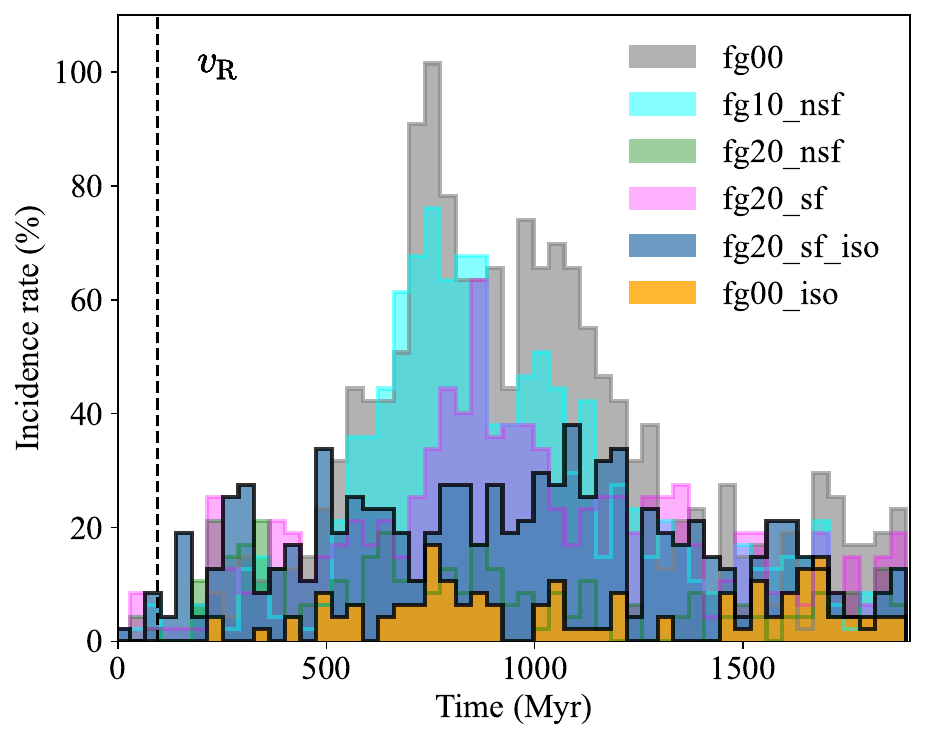}\\
\includegraphics[width=0.89\columnwidth]{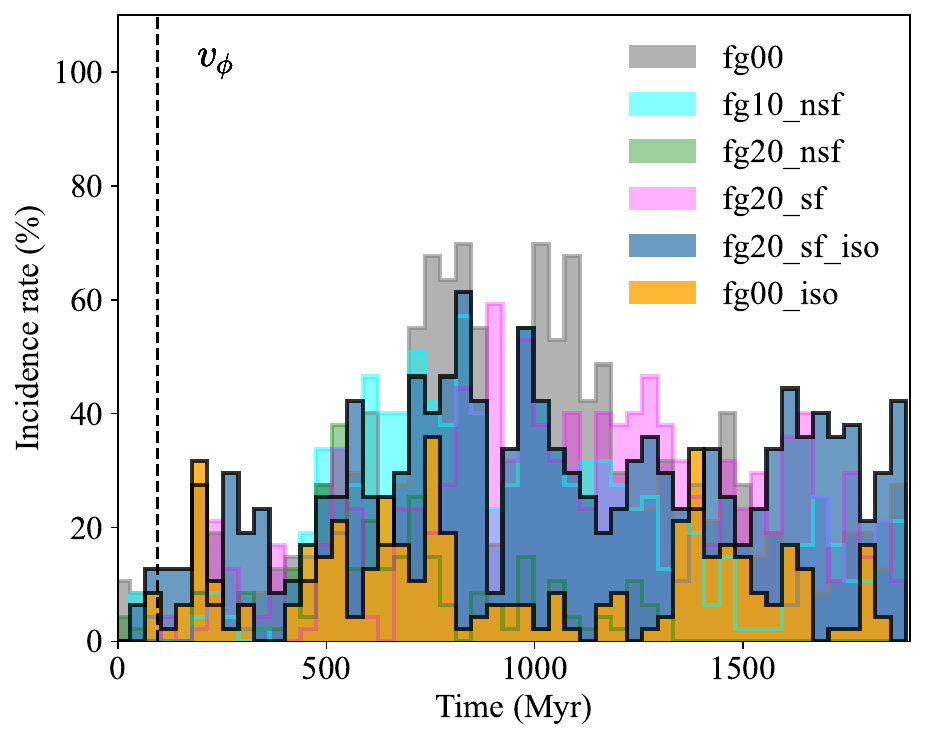}\\
\hspace{5pt}
\includegraphics[width=0.87\columnwidth]{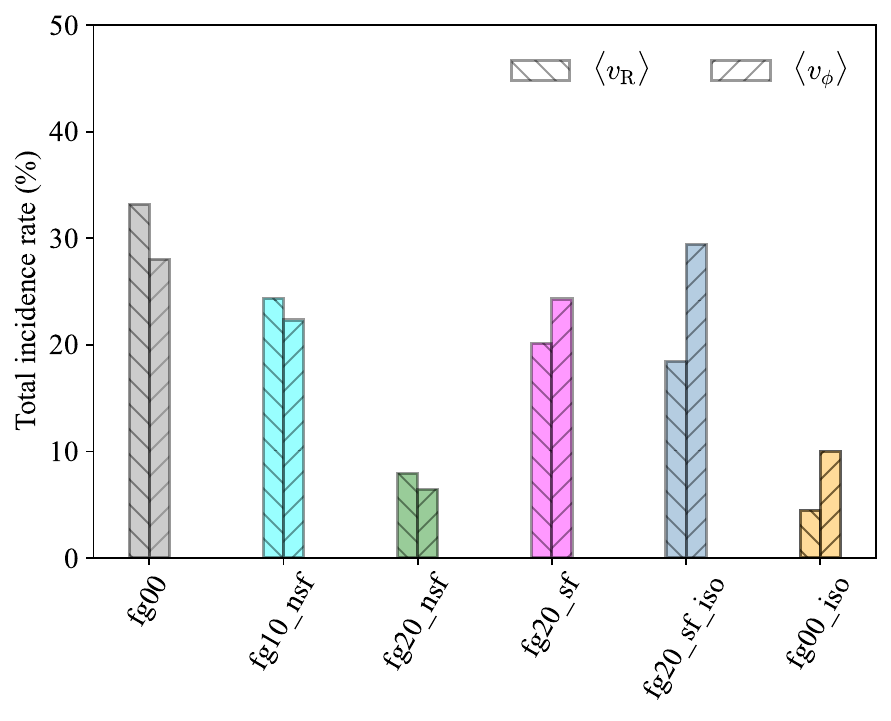}
\caption[]{
Temporal incidence rate, estimated by marginalising the incidence maps over space, adopting a bin size of $\sim40$~Myr. Top: PS weighted by $\vR$. Middle: PS weighted by $\vp$. The vertical dashed line in each panel flags the approximate impact epoch. The distributions of the isolated runs (fg00\_iso, fg20\_sf\_iso) have been added a black border for emphasis.
Bottom: The {\em total} incidence rate, i.e. across all volumes and over all times, for each model. Note the different hatching pattern for each of $\vR$ or $\vp$.
}
\label{fig:rate_histo}
\end{figure}

Fig.~\ref{fig:rate_histo} displays these for each of our runs, for $\vR$ (top) and $\vp$ (middle). The vertical dashed line in each column flags the approximate impact epoch (if applicable). In the case of the interaction models, we see the general trend -- already noted in the previous section -- that the incidence rate increases, first slowly and then rapidly, some \mbox{$\Delta t \approx 400$~Myr} after impact, peaking at around \mbox{$t \approx 700 - 800$~Myr}, and steadily declining after that. The exception to this is model fg20\_nsf, that shows virtually an vanishibly small incidence over the full timespan.

Consistent with the results obtained from the incidence maps, we see a significant difference in the temporal incidence -- both in $\vR$ and $\vp$ -- between the two isolated models (highlighted by a thick black border): the star-forming, multi-phase gas run (fg20\_sf\_iso; blue histogram) features a consistently higher incidence rate at all times compared to its pure N-body counterpart (fg00\_iso; orange histogram).

The bottom panel in Fig.~\ref{fig:rate_histo} shows the {\em total} incidence rate, i.e. the normalised sum of the indicence over all volumes at all times, for each run split by $\vR$ and $\vp$. It is striking that no model displays an incidence rate above 40\%, which suggests that the phase-spiral phenomenon is not widespread neither in space nor in time. It is worth stressing though that the exact rate values are  meaningful only in a relative sense, when comparing simulations to one another. For instance, if the simulations were run for longer, the total rates will all probably uniformily drop, as the phase spiral will weaken in time, unless there was another interaction or more has were supplied to sustain the ongoing star-formation.

In agreement with the above discussion, the lowest total incidence rates are displayed by the inert, 20\% gas fraction run (fg20\_nsf), and the unperturbed, pure N-body run (fg00\_iso), both at less than 10\%. The interacting, pure N-body run (fg00) displays the highest total incidence rates at \mbox{30~--~35\%} (the temporal incidence rate is close to \mbox{60~--~100\%} at its highest; see top and middle panels). The total incidence rate of the interacting, pure N-body model is followed by the inert, 10\% gas fraction run (fg10\_nsf), which has a comparable total incidence rate to the total incidence rate of our more `realistic' run (fg20\_sf) as well as its unperturbed counterpart (fg20\_sf\_iso), all in the range 20 - 25\%. The lowest, total incidence rate is unsurpisingly that of models fg20\_nsf and fg00\_iso, consistent with the results discussed in the previous section.

\subsection{Kinematic interlocking} \label{sec:kin}

We briefly mentioned in Sec.~\ref{sec:chrono} that the $\vp$-incidence maps and $\vR$-incidence maps are qualitatively similar, but not quantitatively. This circumstance is important to consider.

The \gaia\ phase spiral is clearly visible not only in star counts \citep[e.g.,][]{lap19a}, but also when weighing the projected stellar density by $\vp$ or $\vR$. As recognised from the outset \citep{ant18b,bin18a}, this implies that the vertical ($\vz$) and in-plane ($\vp$, $\vR$) velocity components of the stars are 
simultaneously affected by a perturbation in a non-trivial way, or `interlocked.' Thus, we speak of `kinematic interlocking', a feature that {\em must} be reproduced, at least qualitatively, by any model attempting to explain the \gaia\ phase-spiral signature.

\begin{figure}
\includegraphics[width=0.48\textwidth]{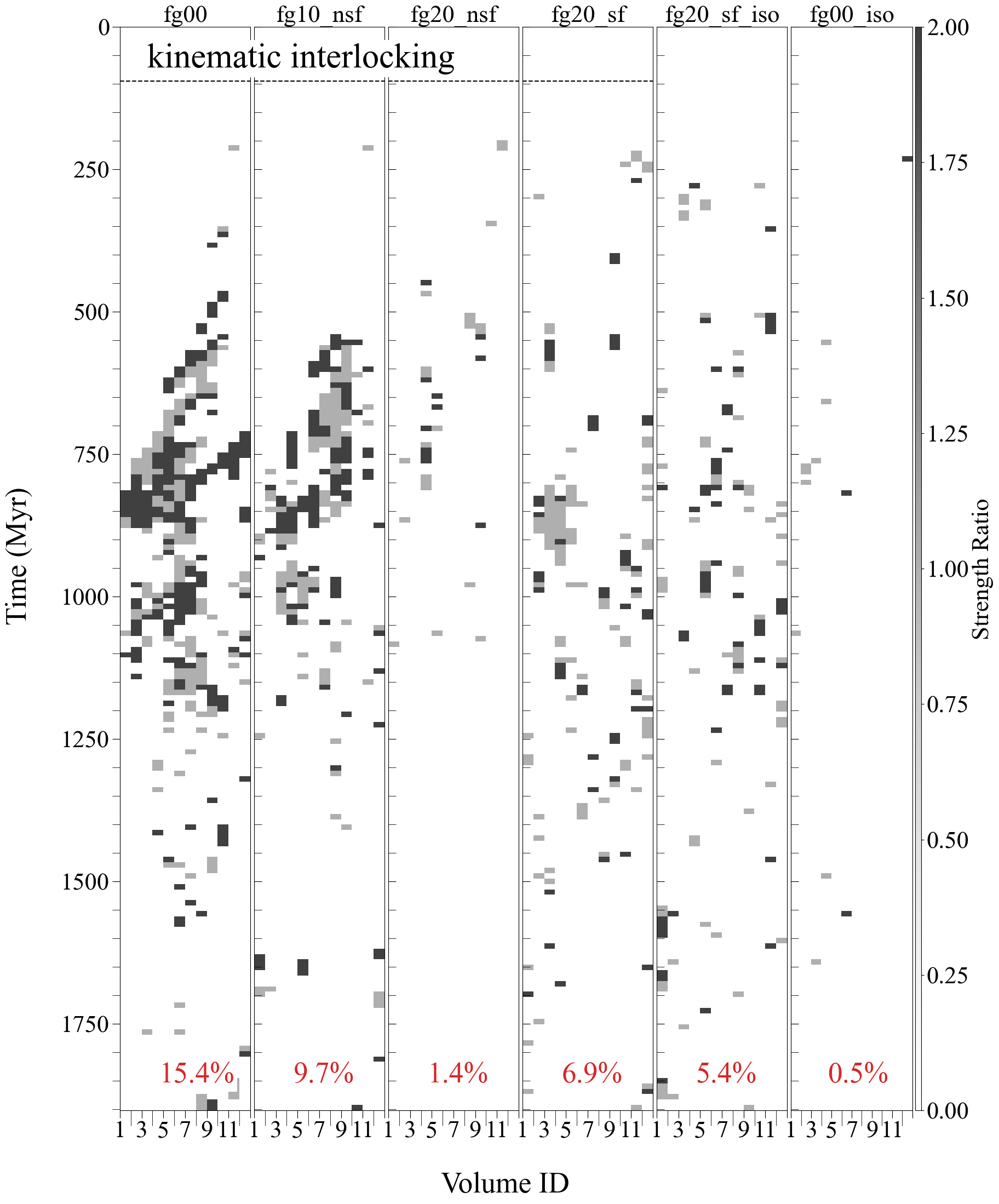}\hfill
\caption[  ]{
Phase spiral `coincidence' map, displaying the simultaneous detection of a positive signal in both $\vR$ and $\vp$, effectively indicating the presence of a {\em kinematic interlocking} between the vertical and in-plane stellar velocity components. The grey scale indicates the ratio of the strength in $\vp$ relative to $\vR$: 2 (1) for a non-zero, finite ratio higher (lower) than 1, and 0 otherwise. The numbers (in red) at the bottom of each column indicate the total coincidence rate, i.e. the ratio of the cells with a positive kinematic interlocking signal, relative to the total number of cells. See text for details.
}
\label{fig:ps_chrono_coinc}
\end{figure}

We assessed the degree of kinematic interlocking in each of our simulations by calculating a `coincidence' map, effectively a cell-by-cell ratio of the $\vp$- over $\vR$-incidence maps, transformed into a ternary signal with a value of 1 for a non-zero, finite ratio lower than 1; a value of 2 for a non-zero, finite ratio higher than 1, and a value of 0 otherwise.
The result of this exercise is displayed, for each simulation, in Fig.~\ref{fig:ps_chrono_coinc}. Black cells indicate cases where the signal in $\vp$ is stronger than in $\vR$; grey cells indicate the opposite. White cells indicate the absence of one or the other in the corresponding incidence map (Fig.~\ref{fig:incidence_map}). The numbers in red at the bottom of each column indicate the total incidence rates.

Qualitatively, we found the same behaviour as seen in the individual maps on a simulation-by-simulation basis (Fig.~\ref{fig:incidence_map}), with the difference that the incidence rate of kinematically interlocked phase spirals is notably lower than the individual $\vR$- and $\vp$ incidence rates. Consistent with the results discussed above (cf. Fig.~\ref{fig:rate_histo}), the pure N-body simulation (fg00) shows the highest coincidence rate, followed by fg10\_nsf. In contrast, fg20\_nsf shows virtually no kinematic interlocking, which is hardly surprising, given the very low $\vp$ and $\vR$ incidence rates. In contrast, and remarkably, the presence of a multi-phase ISM in fg20\_sf\_iso leads to some degree of kinematic interlocking, at a level comparable to fg20\_sf, with the latter having a slightly higher coincidence rate; the latter difference is probably driven by the perturbation induced by the satellite.

While the presence of phase spirals in the $\vp$- and $\vR$-incidence maps in the ISM-bearing simulations could be understood in terms of the TFB effect, the latter did not account for kinematic interlocking, as mentioned earlier.

\begin{figure*}
\centering
\includegraphics[width=0.49\textwidth]{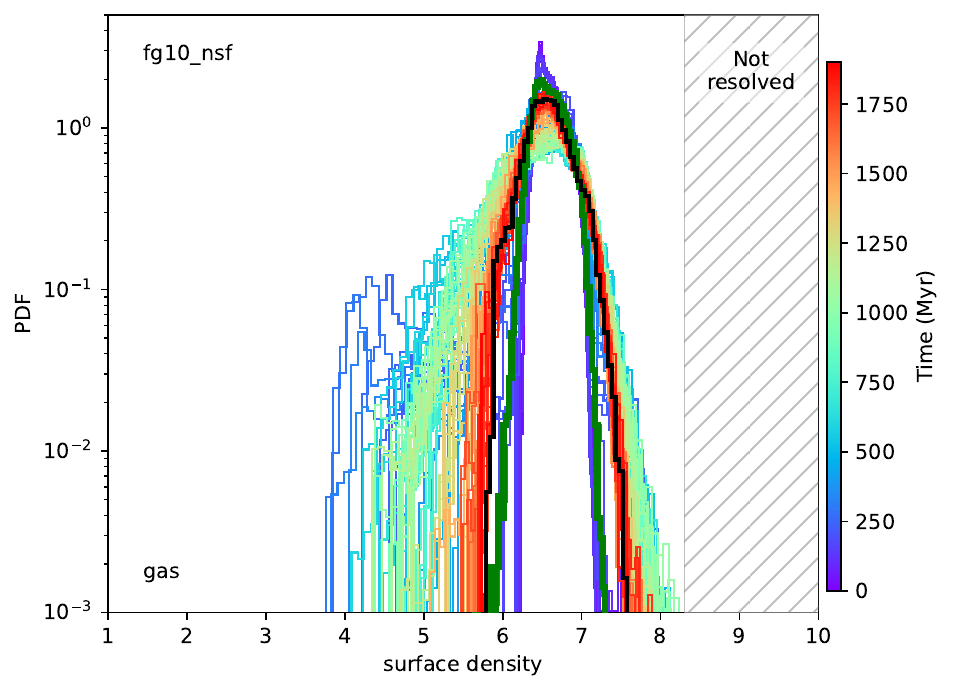}
\includegraphics[width=0.49\textwidth]{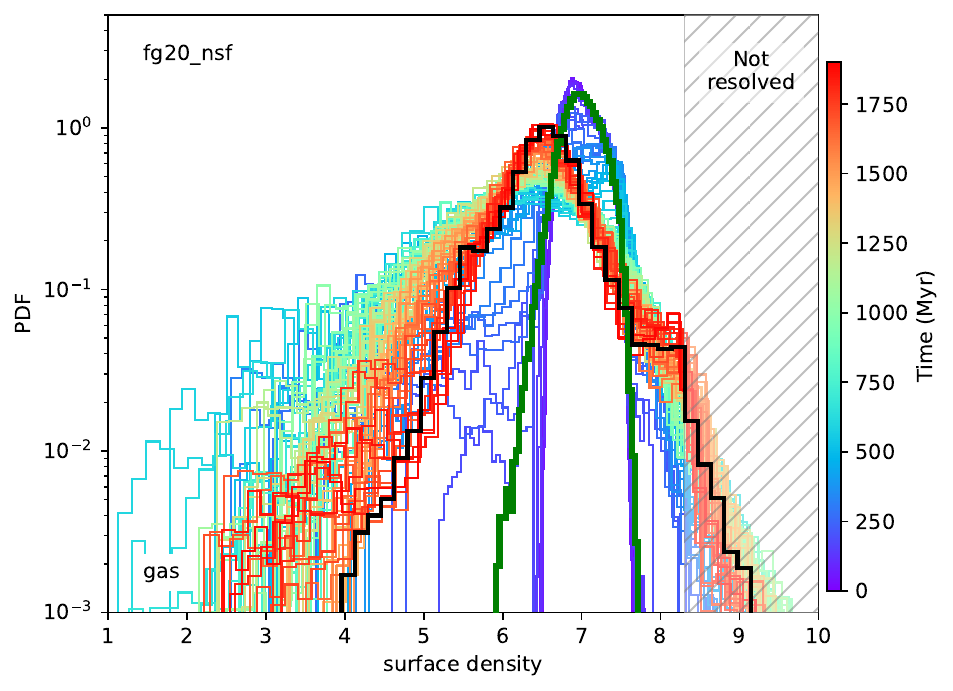}
\includegraphics[width=0.49\textwidth]{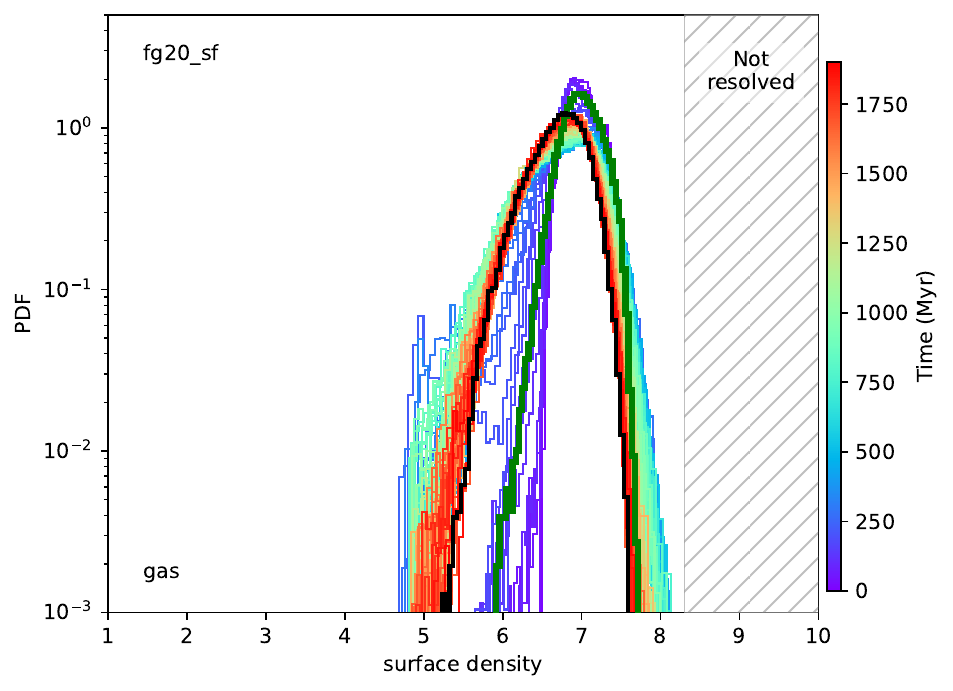}
\includegraphics[width=0.49\textwidth]{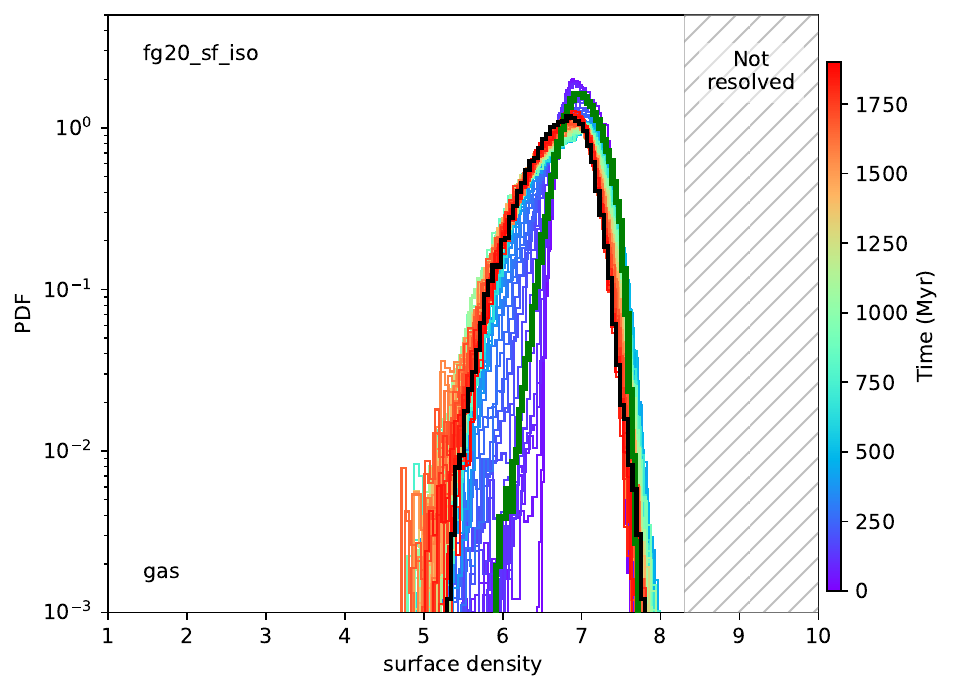}
\caption[  ]{ Distribution of the gas surface density over an area of \mbox{$20\times20~\kpc^2$} centred on the disc in different simulations, one panel for each, as indicated by the top-left label. In each panel, a single histogram corresponds to the gas PDF at a given epoch, as indicated by the colour palette. The initial and final states are highlighted by a thick green and a thick black curve, respectively. The hatched region indicates the densities corresponding to the Jeans density at or below the resolution limit of the simulation grid. }
\label{fig:gas_pdf}
\end{figure*}

\subsection{The effect of the gas structure}

The only difference between the gas bearing models is their initial gas content (disregarding for now the presence of a perturber), and how the gas is treated during the system's evolution. Clearly, this leads to a very particular dynamical response of the disc in each case. It is therefore natural to ask whether the variations in the PS incidence rates between these models can be largely understood in terms of the structural properties of the gas. We address this question in the following.

\subsubsection{Surface density distribution}

In Fig.~\ref{fig:gas_pdf} we show the distribution of gas surface densities (PDF) and its evolution for a each of our hydrodynamic simulations, one panel for each, as indicated on the respective top-left corner. In each panel, a single histogram corresponds to the PDF at a specific epoch, as indicated by its colour and the colour palette to the right, which changes across a rainbow palette from blue hues (`early' times) to red hues (`late' times). For reference, the initial and the final PDFs are highlighted by a thick green histogram and a thick black histogram, respectively. Note that the initial PDF is identical in all the fg20 models (top-right and bottom panels).

The PDFs were obtained for each model and each epoch from face-on projections of the 3D gas distribution centred on the disc over an area of \mbox{$20\times20~\kpc^2$}. Each of these projections, or density maps, was sampled with \mbox{$N_{\rm pix} = 251$} pixels per side, implying a pixel size of \mbox{$\sim80$~pc}, corresponding to roughly twice the limiting, spatial resolution of the AMR grid.

The hatched region in each panel of Fig.~\ref{fig:gas_pdf} indicates the range of \citet{jea15a} densities -- i.e. densities prone to fragmentation -- not resolved in our simulations. The density $\Sigma_{\rm J}$ corresponding to a Jeans length $\lambda_{\rm J}$ equal to the limiting spatial resolution $\delta x \approx 37$~ pc is calculated by \mbox{$\Sigma_{\rm J} = c_s^2 / G~\delta x$},
where $c_s^2 =  k_{\rm B} T / \mu m_{\rm H}$ is the sound speed squared, $k_{\rm B}$ is Boltzmann's constant, $G$ is Newton's constant, $\mu$ is the mean molecular weight and $m_{\rm H}$ is the hydrogen mass. We adopt\footnote{The initial temperature of the fg20 models; the initial temperature of the fg10\_nsf model is lower by a factor 2, and so is the corresponding Jeans density but this is irrelevant, since the model never reaches these densities. } $T = 2\times10^4$ K and $\mu = 1.22$, which results in \mbox{$\Sigma_{\rm J}~\approx~10^8~\Msun~\kpc^{-2}$}, or twice as high over the pixel scale of the maps.

The evolved gas disc structure is distinctly different between synthetic galaxies with different initial gas fractions, but also between models with and without a multiphase ISM. The heavy, inert, isothermal gas disc (fg20\_nsf; top right panel) shows the most dramatic evolution compared to the other models: its surface density (not shown) changes from an initially smooth and featureless mass distribution to a highly structured distribution featuring large-scale, dense filamentary structures; large, dense knots; and large-scale arms on top of a diffuse backround after $t \approx 2$~Gyr. These features presumably are the result of a lack of heating, and they would likely become denser, ultimately resulting in a highly fragmented disc, in the absence of a temperature `floor' imposed by the isothermal EoS. The behaviour of the surface density is reflected in the PDF, where an initially narrow distribution spreads out considerably towards both significantly lower and significantly higher densities, reaching into the Jeans density regime (Fig.~\ref{fig:gas_pdf}, top-right panel).

\begin{figure*}
\centering
\includegraphics[width=0.49\textwidth]{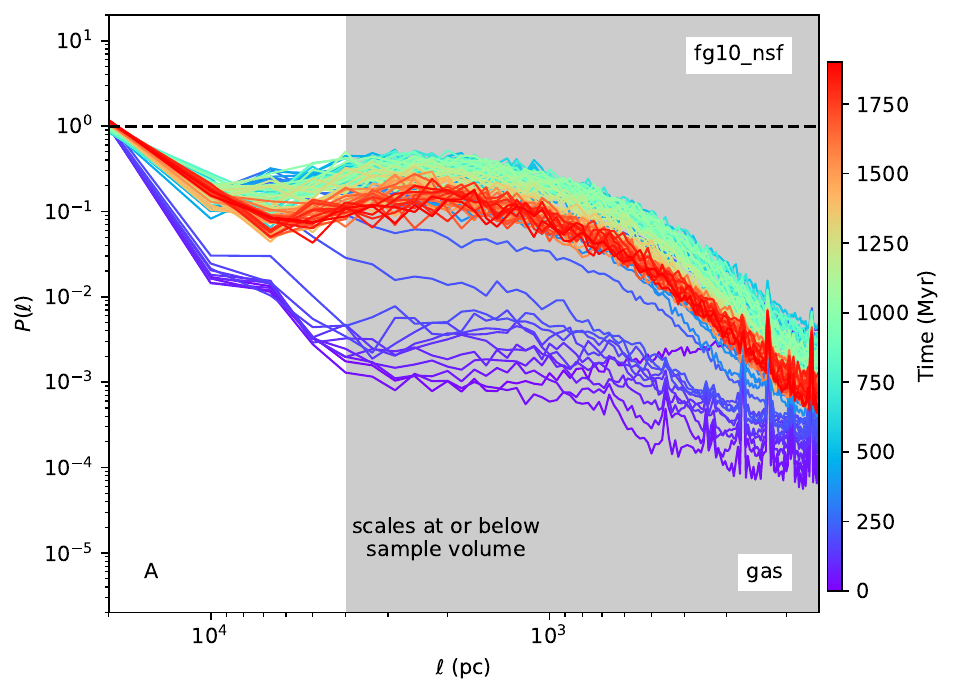}
\includegraphics[width=0.49\textwidth]{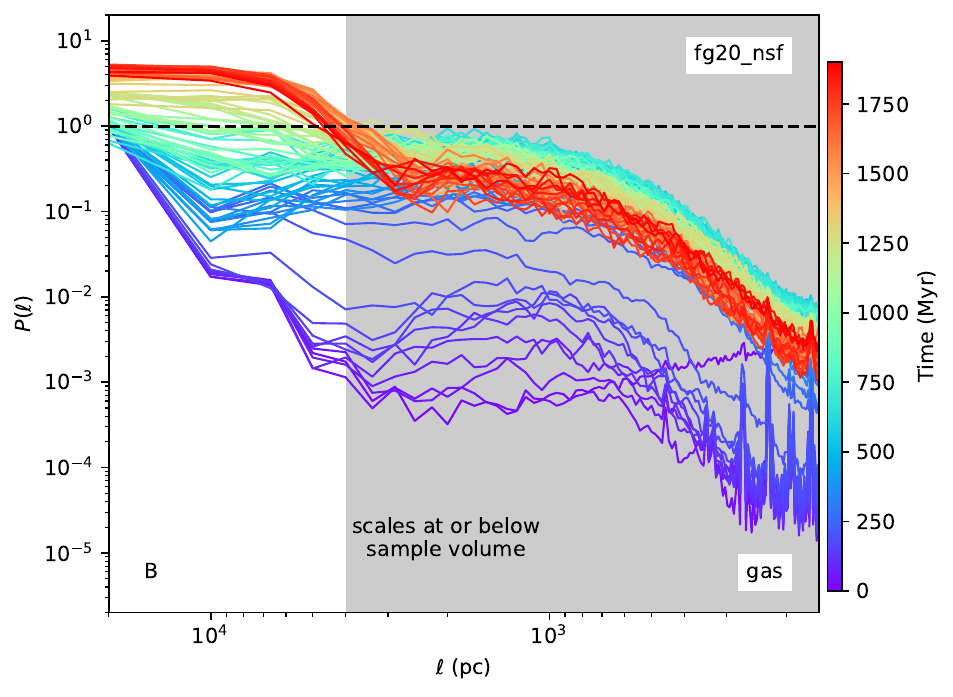}\\
\includegraphics[width=0.49\textwidth]{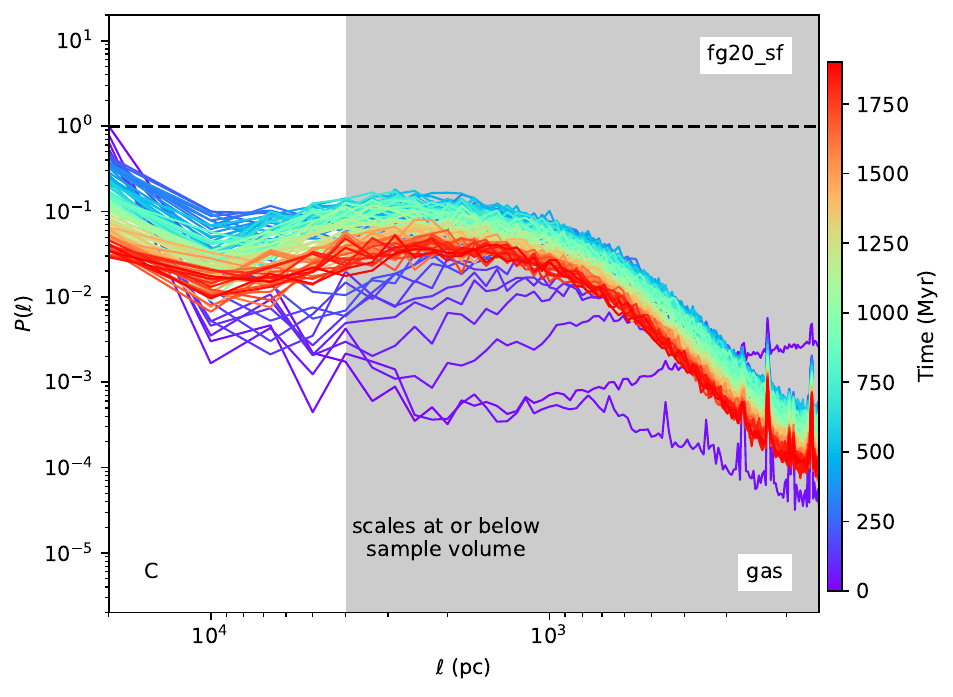}
\includegraphics[width=0.49\textwidth]{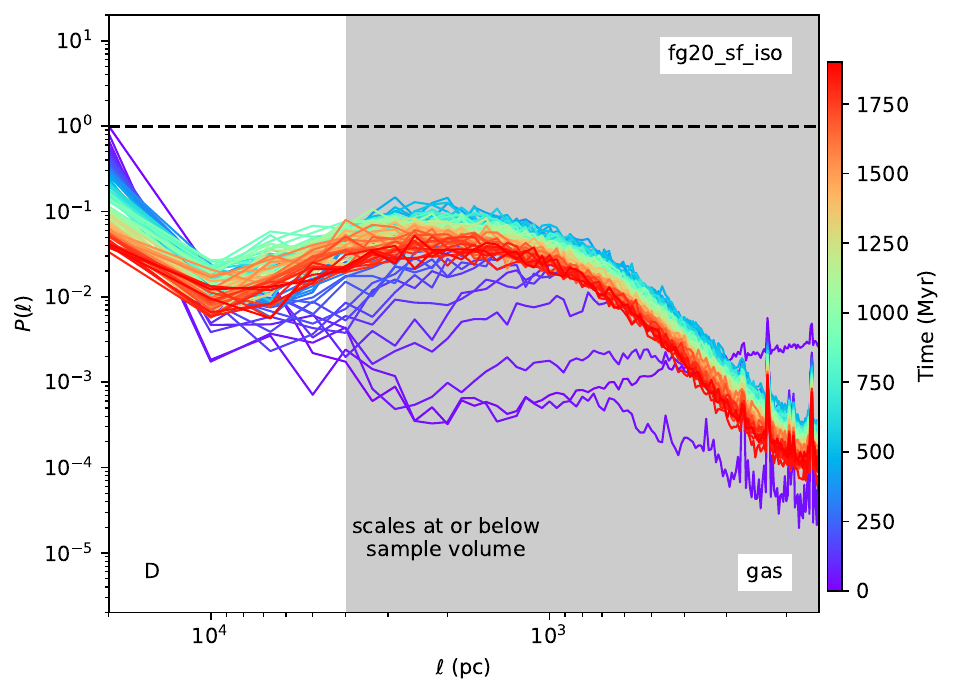}
\caption[  ]{Radially averaged power spectrum (RPS) of the surface density distribution of the gas over an area of \mbox{$20\times20~\kpc^2$}; note that scales decrease from left to right along the $x$-axis. Each panel corresponds to a simulation, as indicated by the corresponding label on the top-left corner. In each panel, each curve corresponds to the RPS at a given epoch, as indicated by the colour palette, normalised to the amplitude of the initial ($t=0$) RPS at the largest scale ($l = 20$~kpc); an horizontal dashed line at this level has been included for reference. The shaded area indicates the scales at or below the size of the sample volume (4 kpc), used to analyse the phase spiral (see Fig.~\ref{fig:vols}). The spikes apparent at scales $\ell \lesssim 500$~pc are likely artefacts of the discreteness imposed onto the gas distribution by the AMR grid, and they are of no significance. The $P(\ell) \propto \ell^2$ behaviour for $\ell \lesssim 1$~kpc displayed by the initial ($t=0$; violet curve) RPS in all models is a reflection of the particle noise inherent to the ICs.}
\label{fig:psd_gas}
\end{figure*}

The surface density of the active, star-forming, multiphase gas disc (fg20\_sf; bottom-left panel) evolves from the same initially featureless distribution as the disc in fg20\_nsf to one dominated by dense spurs embedded within pockets of diffuse, hot material, on top of what appears to be a flocculent spiral structure (not shown). The structure appears to be stationary over hundreds of millions of years, probably sustained by the star-formation activity. As the latter dwindles, the gas distribution becomes somewhat smoother. This evolution results in an extended tail at the lower end of the initial gas PDF. Interestingly, the evolution of the gas structure appears to be largely insensitive to the effect of an external perturbation, as suggested by the similarity between the gas PDF in the fg20\_sf and the fg20\_sf\_iso models (bottom panels).

The surface density of the light, inert isothermal gas disc (fg10\_nsf; top-left panel) displays a similar evolution to that of the star-forming discs, albeit with a better defined spiral structure (not shown). In consequence, its PDF experiences a similar evolution, although the PDF of the fg10\_nsf model displays an enhancement of lower densities at intermediate times, but ultimately becomes narrower at the low end after $t \approx 2$~Gyr (compare black histograms).

\subsubsection{Power spectrum}

For the purpose of a quantitative analysis and comparison of the different gas distributions, we resorted to calculating power spectra of the projected gas density for each of our hydrodynamic models. The use of this mathematical tool is a widely adopted approach to characterise and quantify the structural (and turbulent) properties of the gas distribution in galaxies, not only in observations \citep[e.g.,][]{sta99a,jou06a,zha12a,dut13a}, but also in simulations \citep[e.g.,][]{bou10b,gri17b,ren21a}.

Here, the power spectrum is calculated by performing a discrete Fast Fourier Transform (FFT) over the `raw' (i.e. without windowing, mirroring, or zero-padding) surface density map and averaging the amplitude over the different spatial frequency ($k$) bins, weighing by the width of the frequency annulus. The radially averaged power spectrum (RPS) thus computed is a measure of the variance of the gas distribution at each spatial scale \mbox{$\ell = 2 \pi / k$}. To this end, we adapt the code provided by Bert Vandenbroucke.\footnote{Available at \url{https://bertvandenbroucke.netlify.app/2019/05/24/computing-a-power-spectrum-in-python/}. The results are identical to those obtained with the Python package {\tt cfpack}, written by C.~Federrath, available at \url{https://www.mso.anu.edu.au/~chfeder/codes/cfpack/cfpack.html}.} We have checked that mirroring, windowing, or zero-padding \citep[cf.][]{gri17b} does not noticeably affect the shape of RPS; any of these manipulations only changes its overall amplitude quasi-uniformily across all scales. Therefore, the conclusions based on the comparative analysis of the RPS performed on the raw maps remain unchanged whether mirroring, windowing, or zero-padding are applied or not.

It is worth mentioning that while calculating the RPS we noticed very high density pixels, i.e. those with values beyond the Jeans density (present only in the fg20\_nsf simulation and in relatively low numbers) introduce artefacts in the computed RPS. For this reason, we have reset the density in these pixels to the value of the Jeans density prior to calculating the RPS.

For all gas-bearing models (fg10\_nsf,  fg20\_nsf, fg20\_sf, and fg20\_sf\_iso) we calculate a series of RPS over the full simulation timespan (\mbox{$\sim 2$~Gyr}), one for each time step ($\delta t \approx 10$~Myr). The result is displayed in Fig.~\ref{fig:psd_gas}. Each of the four panels (labelled A, B, C, and D, starting from the top-left) corresponds to one simulation, as indicated by the respective top-left legend. In each panel, each curve corresponds to the RPS at a specific epoch, as indicated by the colour palette to the right, which changes across a rainbow palette from blue hues (`early' times) to red hues (`late' times). For each model, each RPS has been normalised to the amplitude of the initial ($t=0$) RPS at the largest scale ($l = 20$~kpc); an horizontal dashed line at the normalised value of $P(\ell) = 1$ has been included for reference. The grey shaded area indicates the scales at or below the size of the PS sampling volume (see Fig.~\ref{fig:vols}), and it is identical across panels.

In general terms, the amplitude of the RPS in all models increases with time virtually at all scales $\ell \lesssim 10$~kpc, indicating the emergence of substructure as a result of fragmentation. For reference, the theoretically expected value of the fastest growing modes across the isothermal discs is \mbox{$\lambda_{\rm frag} = 2 \lambda_{\rm J} \approx 0.2 - 1.5$~kpc}  (it is of order 1 kpc at the solar radius). The growth of the RPS at early time does appear to conform to this expectation in all models.

The light, inert, isothermal disc (fg10\_nsf, panel A on the top-left corner of the figure) shows little variation at the largest scale, indicating the absence of strong, large-scale coherent structures, although we note that weak spiral arms do appear in the disc (not shown). In contrast, the RPS of the star-forming discs (panels C, D in the bottom row of the figure) display a systematic decrease in amplitude a the largest scale, and a significant increase at smaller scales, indicative of a disc dominated by the appearance of flocculent structure (in accordance with the visual appearance of the surface density maps; not shown). The RPS of the star-forming disc with and without an interaction undergoes a very similar evolution; therefore, the gas structure does not appear to be significantly affected by the perturbation inflicted on the disc by the satellite, consistent with the results shown in Fig.~\ref{fig:gas_pdf}.

It is readily apparent that a heavy, inert isothermal gas disc (fg20\_nsf, panel B on top-right corner of the figure) displays a distinctively different RPS compared to the other fg20 models within the first billion years of evolution (violet through cyan-green curves), in spite of having started from the same state. This is consistent with the visual assessment of the corresponding gas PDFs. Specifically, it displays a rapid increase in the RPS's amplitude at nearly all scales, indicating a strong inhomogeneity in the gas mass distribution.

The difference in the amplitude of the RPS at the relevant scales of $\lesssim 4$~kpc (gray shaded area; corresponding to the size of the sampling volume; see Fig.~\ref{fig:vols}) between the fg20\_nsf model and  the other fg20 models is significant, roughly of an order of magnitude. If the amplitude of the RPS is a measure of the gas clumpiness, then these differences may explain the markedly different emergence and evolution of the PS and its incidence rate. In other words, while the presence of a clumpy medium with the right scale may positively affect the PS (TFB effect), it can also stymie the phenomenon if 'too' clumpy.

As shown in Appendix~\ref{sec:psd_app}, the RPS of the other galaxy components (DM, pre-existing stars, or newly formed stars, if present) either barely evolves, or does in a comparable way across different simulations. This suggests that none of these components significantly impact the onset and evolution of the phase spiral. Thus, {\em the gas and its structure appear to be the key driver} of the differences between models in terms of the associated PS incidence rates.

\section{Summary and conclusions} \label{sec:summ}

The key lessons learned from the the incidence maps in Figs.~\ref{fig:incidence_map} and \ref{fig:ps_chrono_coinc} can be summarised as follows:
\begin{itemize}
    \item Columns 1, 2, 4: A large-scale perturbation, e.g., an external impulse, is needed to generate a semi-coherent phase-spiral (PS) signal on kpc scales persisting with a lifetime of order a few 100 Myr.
    \item Columns 2, 3: Adding inert gas to the disc weakens the PS signature; the suppression appears correlated with the amount of gas in the disc.
    \item Columns 4, 5: Multi-phase, turbulent gas can sustain and even trigger localised, intermitent PS signatures. We associate this with the TFB effect. Unexpectedly, the PS signatures can even be interlocked, which may be a property arising from the (turbulent) gas motions.
    \item Columns 6: A measure of the rate of false positives can be seen in the dry isolated disc. These isolated PS signatures are never interlocked and are noisy artefacts.
    \item Columns 2, 3, 4, 5: The power spectra  corresponding to columns 2, 4 and 5  (panels A, C, and D in Fig.~\ref{fig:psd_gas}, respectively) are qualitatively similar. The power spectrum for column 3 (corresponding to panel B in Fig.~\ref{fig:psd_gas}) shows up to an order magnitude higher amplitude over the relevant scales $\lesssim 4$~kpc; this excessive small-scale structure appears to fully suppress the PS signal (column 3).
\end{itemize}

The upshot is that the phase spiral is influenced by both the quantity and the spatial distribution of gas. Based on our earlier work \citepalias[][]{tep22x}, we know that the amount of gas is key because of the resulting increased potential close to the galactic plane, which affects the stars' dynamics in two ways: 1) by dampening the disc's response to a perturbation; 2) by speeding up the phase-mixing process \citep[q.v.][]{can14c}; both of which negatively impact the PS phenomenon. The role of the gas structure in shaping the phase spiral is indicated the marked differences in the gas PDF and power-spectrum across models. Crucially, the influence of the DM, and of the newly formed and pre-existing stars is of lesser importance compared to the gas.

Our findings are entirely consistent with the result that, without feedback driving, large-scale gas density power spectra have too much power on small scales, a circumstance incompatible with observations \citep{gri17b}. But most importantly, they harmonise with the suggestion that the phase spiral is highly susceptible to scattering by small-scale density fluctuations \citep[TFB effect;][]{tre23a}.

In this context, an unexpected and important result in our study
is that clumpy, turbulent gas is capable of triggering {\em kinematically interlocked} phase-spiral signatures. This in turn suggests that the phase spiral can, in principle, be used to probe the ISM's turbulent structure as well as the interplay between the stars and the ISM on $\sim 100$ Myr timescales, thus meriting further investigation.

\section{Closing remarks} \label{sec:fin}

Our simulations hint that the \gaia\ phase spiral {\it can} coexist with a clumpy, turbulent ISM. {\it But} the signature is highly intermittent {\em both} spatially and temporally: there is not a single epoch at which the phase spiral is observable simultaneously across the solar circle, nor is there a specific location across the solar circle where it is observable at all times; this is in contrast to the findings of billion-particle, {\em pure N-body} models \citep{hun21t,asa25a}.
We should not be surprised by this. Essentially, all phenomena associated with the ISM are intermittent and stochastic by nature \citep{set25a}, including molecular cloud formation and star formation \citep{fed12a}, as well as nuclear activity, disc-halo gas recycling \citep[][]{arm16a}, and transient spiral arms \citep[][]{ros12a}, to name a few. Importantly, the properties (e.g. strength, distribution, interlocking) of the phase-spiral are providing us with information on the strength of small-scale, star-gas coupling in the ISM.

While it may be ambitious to hope to observe a temporal evolution of the phase spiral in the Galaxy, evidence of its spatial variation across the disc is  abundant. \citet{xu20x} established the presence of PS-like signatures beyond the solar neihbouhood out to $R \approx 15$~kpc by direclty mapping the distribution of stars onto $z-\vz$ plane in progressively increasing radial ($R$) bins. \citet{hun22a} took the analysis a step further, and grouping stars by azimuthal action $J_\phi$ (equivalent to angular momentum, $L_z$) and their conjugate angles, $\theta_\phi$, found clear evidence of variations across $J_\phi$ in particular, consistent with the findings by \citet{xu20x}. They also found evidence for radial variations of the \gaia\ PS's morphology, featuring one arm in the outer disc, and two arms in the inner regions. A similar approached was followed by \citet{fra23a} and \citet{dar23a}, both of whom obtained consistent results with earlier work.  
It remains to be seen whether future observations support the degree of intermittency implied by the results from our models or not.

Either way, we need to keep in mind that the models likely are incomplete (i.e. lacking relevant physical ingredients) or inaccurate (e.g., because of an inadequate resolution). Clearly, any simulation may benefit from increased resolution in terms of both the particle number and the spatial scale. An increased particle resolution may reconcile, at least in part, the  different behaviour between our simulations and billion-particle (N-body) simulations, i.e. intermittent vs. continuous. But while an increased spatial resolution is attainable \citep[e.g., on sub-parsec scales;][]{ren13a}, billion-particle N-body/{\em hydrodynamical} simulations are out of reach for the time being.

We anticipate that consideration of additional relevant physical processes such as magnetic fields and highly energetic particle ('cosmic ray') heating, which are expected to reduce the gap between model galaxies and real galaxies, will have a noticeable effect on the interplay between gas and ISM, and by extension on the PS phenomenon. There is a long list of studies that have shown how the presence of an even weak magnetic field can significantly affect the density and turbulence structure of the ISM \citep[q.v.][]{MolinaEtAl2012,ban18a,BeattieEtAl2023}.
Similarly, cosmic rays resulting from stellar activity are know to modify the temperature of gas \citep[e.g.,][]{boo13a}, which in turn affect its density structure.

Despite their caveats, our simulations underline the importance of gas in any study aiming at understanding the dynamical response of stellar discs and the consequent wave phenomena associated with, such as the phase spiral.
It stands to reason that galaxies with gas fractions in excess of 20\% -- the highest gas fraction considered in this work -- may show a {\em weaker} response to an external perturbation. This circumstance is particularly relevant in the context of high-redshift galaxies, which are likely to transit through a gas-rich phase at some point in their evolution, as hinted by early observations \citep[e.g.,][]{gen06h}, and supported by more recent ones \citep[e.g.,][]{riz22w}. This is an important avenue for future research that we intend to pursue.

Conversely, results obtained from the analysis `dry' models, i.e. models that do not include gas,\footnote{We refer to these as `drynamical' models. } may be misleading, because the disc's response appears to be exaggerated $-$ dry models are simply too `reactive.' We have found this to be true in the context of disc corrugations \citepalias{tep22x}, in the context of the MW bar (Davis et al., submitted), and now again in the context of the phase spiral. 
While dry models can still be useful for exploring certain phenomena under specific conditions, incorporating a dissipative component is generally essential for a comprehensive dynamical study of galactic discs.

\section*{Acknowledgments}
We thank the anonymous referee for carefully reading our original manuscript and providing perceptive comments that helped improve the presentation and discussion of our results.
TTG acknowledges financial support from the Australian Research Council (ARC) through Australian Laureate Fellowships awarded to JBH (FL140100278)
and to TRB (FL220100117) from the School of Physics, University of Sydney,
and for partial funding through the James Arthur Pollock memorial fund awarded to the School of Physics, University of Sydney.
We acknowledge the facilities, and the scientific and technical assistance of the National Computational Infrastructure (NCI),  which is supported by the Australian Government, through an NCMAS grant (project IDs: ca64 and ek9).

We made use of {\sc Pynbody}\footnote{\url{https://github.com/pynbody/pynbody}} -- a {\sc Python}\footnote{\url{http://www.python.org} }-based software -- in our analysis for this paper \citep{pon13a}.
This research has made use of NASA's Astrophysics Data System (ADS) Bibliographic Services\footnote{\url{http://adsabs.harvard.edu} }.

\section*{Data availability}

The simulation data and the software underlying this article will be shared on reasonable request to the corresponding author.

\bibliographystyle{mnras} 

\begin{thebibliography}{}
\makeatletter
\relax
\def\mn@urlcharsother{\let\do\@makeother \do\$\do\&\do\#\do\^\do\_\do\%\do\~}
\def\mn@doi{\begingroup\mn@urlcharsother \@ifnextchar [ {\mn@doi@}
  {\mn@doi@[]}}
\def\mn@doi@[#1]#2{\def\@tempa{#1}\ifx\@tempa\@empty \href
  {http://dx.doi.org/#2} {doi:#2}\else \href {http://dx.doi.org/#2} {#1}\fi
  \endgroup}
\def\mn@eprint#1#2{\mn@eprint@#1:#2::\@nil}
\def\mn@eprint@arXiv#1{\href {http://arxiv.org/abs/#1} {{\tt arXiv:#1}}}
\def\mn@eprint@dblp#1{\href {http://dblp.uni-trier.de/rec/bibtex/#1.xml}
  {dblp:#1}}
\def\mn@eprint@#1:#2:#3:#4\@nil{\def\@tempa {#1}\def\@tempb {#2}\def\@tempc
  {#3}\ifx \@tempc \@empty \let \@tempc \@tempb \let \@tempb \@tempa \fi \ifx
  \@tempb \@empty \def\@tempb {arXiv}\fi \@ifundefined
  {mn@eprint@\@tempb}{\@tempb:\@tempc}{\expandafter \expandafter \csname
  mn@eprint@\@tempb\endcsname \expandafter{\@tempc}}}

\bibitem[\protect\citeauthoryear{{Agertz}, {Kravtsov}, {Leitner}  \&
  {Gnedin}}{{Agertz} et~al.}{2013}]{age13a}
{Agertz} O.,  {Kravtsov} A.~V.,  {Leitner} S.~N.,   {Gnedin} N.~Y.,  2013,
  \mn@doi [\apj] {10.1088/0004-637X/770/1/25}, \href
  {http://adsabs.harvard.edu/abs/2013ApJ...770...25A} {770, 25}

\bibitem[\protect\citeauthoryear{{Agertz} et~al.,}{{Agertz}
  et~al.}{2021}]{age21l}
{Agertz} O.,  et~al., 2021, \mn@doi [\mnras] {10.1093/mnras/stab322}, \href
  {https://ui.adsabs.harvard.edu/abs/2021MNRAS.503.5826A} {503, 5826}

\bibitem[\protect\citeauthoryear{{Alinder}, {McMillan}  \& {Bensby}}{{Alinder}
  et~al.}{2023}]{ali23a}
{Alinder} S.,  {McMillan} P.~J.,   {Bensby} T.,  2023, \mn@doi [\aap]
  {10.1051/0004-6361/202346560}, \href
  {https://ui.adsabs.harvard.edu/abs/2023A&A...678A..46A} {678, A46}

\bibitem[\protect\citeauthoryear{{Antoja} et~al.,}{{Antoja}
  et~al.}{2018}]{ant18b}
{Antoja} T.,  et~al., 2018, \mn@doi [\nat] {10.1038/s41586-018-0510-7}, \href
  {http://adsabs.harvard.edu/abs/2018Natur.561..360A} {561, 360}

\bibitem[\protect\citeauthoryear{{Armillotta}, {Fraternali}  \&
  {Marinacci}}{{Armillotta} et~al.}{2016}]{arm16a}
{Armillotta} L.,  {Fraternali} F.,   {Marinacci} F.,  2016, \mn@doi [\mnras]
  {10.1093/mnras/stw1930}, \href
  {http://adsabs.harvard.edu/abs/2016MNRAS.462.4157A} {462, 4157}

\bibitem[\protect\citeauthoryear{{Asano}, {Fujii}, {Baba}, {Portegies Zwart}
  \& {B{\'e}dorf}}{{Asano} et~al.}{2025}]{asa25a}
{Asano} T.,  {Fujii} M.~S.,  {Baba} J.,  {Portegies Zwart} S.,   {B{\'e}dorf}
  J.,  2025, \mn@doi [arXiv e-prints] {10.48550/arXiv.2501.12436}, \href
  {https://ui.adsabs.harvard.edu/abs/2025arXiv250112436A} {p. arXiv:2501.12436}

\bibitem[\protect\citeauthoryear{{Banda-Barrag{\'a}n}, {Federrath}, {Crocker}
  \& {Bicknell}}{{Banda-Barrag{\'a}n} et~al.}{2018}]{ban18a}
{Banda-Barrag{\'a}n} W.~E.,  {Federrath} C.,  {Crocker} R.~M.,   {Bicknell}
  G.~V.,  2018, \mn@doi [\mnras] {10.1093/mnras/stx2541}, \href
  {http://adsabs.harvard.edu/abs/2018MNRAS.473.3454B} {473, 3454}

\bibitem[\protect\citeauthoryear{{Banik}, {Weinberg}  \& {van den
  Bosch}}{{Banik} et~al.}{2022}]{ban22a}
{Banik} U.,  {Weinberg} M.~D.,   {van den Bosch} F.~C.,  2022, \mn@doi [\apj]
  {10.3847/1538-4357/ac7ff9}, \href
  {https://ui.adsabs.harvard.edu/abs/2022ApJ...935..135B} {935, 135}

\bibitem[\protect\citeauthoryear{{Beattie}, {Federrath}, {Kriel}, {Mocz}  \&
  {Seta}}{{Beattie} et~al.}{2023}]{BeattieEtAl2023}
{Beattie} J.~R.,  {Federrath} C.,  {Kriel} N.,  {Mocz} P.,   {Seta} A.,  2023,
  \mn@doi [\mnras] {10.1093/mnras/stad1863}, \href
  {https://ui.adsabs.harvard.edu/abs/2023MNRAS.524.3201B} {524, 3201}

\bibitem[\protect\citeauthoryear{{Binney} \& {Piffl}}{{Binney} \&
  {Piffl}}{2015}]{bin15a}
{Binney} J.,  {Piffl} T.,  2015, \mn@doi [\mnras] {10.1093/mnras/stv2225},
  \href {http://adsabs.harvard.edu/abs/2015MNRAS.454.3653B} {454, 3653}

\bibitem[\protect\citeauthoryear{{Binney} \& {Sch{\"o}nrich}}{{Binney} \&
  {Sch{\"o}nrich}}{2018}]{bin18a}
{Binney} J.,  {Sch{\"o}nrich} R.,  2018, \mn@doi [\mnras]
  {10.1093/mnras/sty2378}, \href
  {http://adsabs.harvard.edu/abs/2018MNRAS.481.1501B} {481, 1501}

\bibitem[\protect\citeauthoryear{Bland-Hawthorn \& Gerhard}{Bland-Hawthorn \&
  Gerhard}{2016}]{bla16a}
Bland-Hawthorn J.,  Gerhard O.,  2016, \mn@doi [Annual Review of Astronomy and
  Astrophysics] {10.1146/annurev-astro-081915-023441}, 54, 529

\bibitem[\protect\citeauthoryear{{Bland-Hawthorn} \&
  {Tepper-Garc{\'\i}a}}{{Bland-Hawthorn} \&
  {Tepper-Garc{\'\i}a}}{2021}]{bla21e}
{Bland-Hawthorn} J.,  {Tepper-Garc{\'\i}a} T.,  2021, \mn@doi [\mnras]
  {10.1093/mnras/stab704}, \href
  {https://ui.adsabs.harvard.edu/abs/2021MNRAS.504.3168B} {504, 3168}

\bibitem[\protect\citeauthoryear{{Bland-Hawthorn} et~al.,}{{Bland-Hawthorn}
  et~al.}{2019}]{bla19a}
{Bland-Hawthorn} J.,  et~al., 2019, \mn@doi [\mnras] {10.1093/mnras/stz217},
  \href {https://ui.adsabs.harvard.edu/abs/2019MNRAS.486.1167B} {486, 1167}

\bibitem[\protect\citeauthoryear{{Bland-Hawthorn}, {Tepper-Garcia}, {Agertz}
  \& {Federrath}}{{Bland-Hawthorn} et~al.}{2024}]{bla24a}
{Bland-Hawthorn} J.,  {Tepper-Garcia} T.,  {Agertz} O.,   {Federrath} C.,
  2024, \mn@doi [\apj] {10.3847/1538-4357/ad4118}, \href
  {https://ui.adsabs.harvard.edu/abs/2024ApJ...968...86B} {968, 86}

\bibitem[\protect\citeauthoryear{{Bland-Hawthorn} et~al.,}{{Bland-Hawthorn}
  et~al.}{2025}]{bla25a}
{Bland-Hawthorn} J.,  et~al., 2025, \mn@doi [arXiv e-prints]
  {10.48550/arXiv.2502.01895}, \href
  {https://ui.adsabs.harvard.edu/abs/2025arXiv250201895B} {p. arXiv:2502.01895}

\bibitem[\protect\citeauthoryear{{Bonnell}}{{Bonnell}}{2025}]{bon25a}
{Bonnell} I.~A.,  2025, \mn@doi [\mnras] {10.1093/mnrasl/slaf024}, \href
  {https://ui.adsabs.harvard.edu/abs/2025MNRAS.540L...1B} {540, L1}

\bibitem[\protect\citeauthoryear{{Booth}, {Agertz}, {Kravtsov}  \&
  {Gnedin}}{{Booth} et~al.}{2013}]{boo13a}
{Booth} C.~M.,  {Agertz} O.,  {Kravtsov} A.~V.,   {Gnedin} N.~Y.,  2013,
  \mn@doi [\apjl] {10.1088/2041-8205/777/1/L16}, \href
  {http://adsabs.harvard.edu/abs/2013ApJ...777L..16B} {777, L16}

\bibitem[\protect\citeauthoryear{{Bournaud}, {Elmegreen}, {Teyssier}, {Block}
  \& {Puerari}}{{Bournaud} et~al.}{2010}]{bou10b}
{Bournaud} F.,  {Elmegreen} B.~G.,  {Teyssier} R.,  {Block} D.~L.,   {Puerari}
  I.,  2010, \mn@doi [\mnras] {10.1111/j.1365-2966.2010.17370.x}, \href
  {http://adsabs.harvard.edu/abs/2010MNRAS.409.1088B} {409, 1088}

\bibitem[\protect\citeauthoryear{{Candlish}, {Smith}, {Fellhauer}, {Gibson},
  {Kroupa}  \& {Assmann}}{{Candlish} et~al.}{2014}]{can14c}
{Candlish} G.~N.,  {Smith} R.,  {Fellhauer} M.,  {Gibson} B.~K.,  {Kroupa} P.,
   {Assmann} P.,  2014, \mn@doi [\mnras] {10.1093/mnras/stt2166}, \href
  {http://adsabs.harvard.edu/abs/2014MNRAS.437.3702C} {437, 3702}

\bibitem[\protect\citeauthoryear{{Darling} \& {Widrow}}{{Darling} \&
  {Widrow}}{2019}]{dar19a}
{Darling} K.,  {Widrow} L.~M.,  2019, \mn@doi [\mnras] {10.1093/mnras/sty3508},
  \href {http://adsabs.harvard.edu/abs/2019MNRAS.484.1050D} {484, 1050}

\bibitem[\protect\citeauthoryear{{Darragh-Ford}, {Hunt}, {Price-Whelan}  \&
  {Johnston}}{{Darragh-Ford} et~al.}{2023}]{dar23a}
{Darragh-Ford} E.,  {Hunt} J. A.~S.,  {Price-Whelan} A.~M.,   {Johnston} K.~V.,
   2023, \mn@doi [\apj] {10.3847/1538-4357/acf1fc}, \href
  {https://ui.adsabs.harvard.edu/abs/2023ApJ...955...74D} {955, 74}

\bibitem[\protect\citeauthoryear{{Dutta} \& {Bharadwaj}}{{Dutta} \&
  {Bharadwaj}}{2013}]{dut13a}
{Dutta} P.,  {Bharadwaj} S.,  2013, \mn@doi [\mnras] {10.1093/mnrasl/slt110},
  \href {https://ui.adsabs.harvard.edu/abs/2013MNRAS.436L..49D} {436, L49}

\bibitem[\protect\citeauthoryear{{Federrath} \& {Klessen}}{{Federrath} \&
  {Klessen}}{2012}]{fed12a}
{Federrath} C.,  {Klessen} R.~S.,  2012, \mn@doi [\apj]
  {10.1088/0004-637X/761/2/156}, \href
  {https://ui.adsabs.harvard.edu/abs/2012ApJ...761..156F} {761, 156}

\bibitem[\protect\citeauthoryear{{Frankel}, {Bovy}, {Tremaine}  \&
  {Hogg}}{{Frankel} et~al.}{2023}]{fra23a}
{Frankel} N.,  {Bovy} J.,  {Tremaine} S.,   {Hogg} D.~W.,  2023, \mn@doi
  [\mnras] {10.1093/mnras/stad908}, \href
  {https://ui.adsabs.harvard.edu/abs/2023MNRAS.521.5917F} {521, 5917}

\bibitem[\protect\citeauthoryear{{GRAVITY Collaboration} et~al.,}{{GRAVITY
  Collaboration} et~al.}{2019}]{gra19a}
{GRAVITY Collaboration} et~al., 2019, \mn@doi [\aap]
  {10.1051/0004-6361/201935656}, \href
  {https://ui.adsabs.harvard.edu/abs/2019A&A...625L..10G} {625, L10}

\bibitem[\protect\citeauthoryear{{Garc{\'\i}a-Conde}, {Roca-F{\`a}brega},
  {Antoja}, {Ramos}  \& {Valenzuela}}{{Garc{\'\i}a-Conde}
  et~al.}{2022}]{gar22c}
{Garc{\'\i}a-Conde} B.,  {Roca-F{\`a}brega} S.,  {Antoja} T.,  {Ramos} P.,
  {Valenzuela} O.,  2022, \mn@doi [\mnras] {10.1093/mnras/stab3417}, \href
  {https://ui.adsabs.harvard.edu/abs/2022MNRAS.510..154G} {510, 154}

\bibitem[\protect\citeauthoryear{{Genzel} et~al.,}{{Genzel}
  et~al.}{2006}]{gen06h}
{Genzel} R.,  et~al., 2006, \mn@doi [\nat] {10.1038/nature05052}, \href
  {https://ui.adsabs.harvard.edu/abs/2006Natur.442..786G} {442, 786}

\bibitem[\protect\citeauthoryear{{Gilman}, {Bovy}, {Frankel}  \&
  {Benson}}{{Gilman} et~al.}{2025}]{gil25a}
{Gilman} D.,  {Bovy} J.,  {Frankel} N.,   {Benson} A.,  2025, \mn@doi [\apj]
  {10.3847/1538-4357/ada963}, \href
  {https://ui.adsabs.harvard.edu/abs/2025ApJ...980...24G} {980, 24}

\bibitem[\protect\citeauthoryear{{Grand}, {Pakmor}, {Fragkoudi}, {G{\'o}mez},
  {Trick}, {Simpson}, {van de Voort}  \& {Bieri}}{{Grand}
  et~al.}{2023}]{gra23a}
{Grand} R. J.~J.,  {Pakmor} R.,  {Fragkoudi} F.,  {G{\'o}mez} F.~A.,  {Trick}
  W.,  {Simpson} C.~M.,  {van de Voort} F.,   {Bieri} R.,  2023, \mn@doi
  [\mnras] {10.1093/mnras/stad1969}, \href
  {https://ui.adsabs.harvard.edu/abs/2023MNRAS.524..801G} {524, 801}

\bibitem[\protect\citeauthoryear{{Grisdale}, {Agertz}, {Romeo}, {Renaud}  \&
  {Read}}{{Grisdale} et~al.}{2017}]{gri17b}
{Grisdale} K.,  {Agertz} O.,  {Romeo} A.~B.,  {Renaud} F.,   {Read} J.~I.,
  2017, \mn@doi [\mnras] {10.1093/mnras/stw3133}, \href
  {https://ui.adsabs.harvard.edu/abs/2017MNRAS.466.1093G} {466, 1093}

\bibitem[\protect\citeauthoryear{{Guo}, {Li}, {Shen}, {Mao}  \& {Liu}}{{Guo}
  et~al.}{2024}]{guo24a}
{Guo} R.,  {Li} Z.-Y.,  {Shen} J.,  {Mao} S.,   {Liu} C.,  2024, \mn@doi [\apj]
  {10.3847/1538-4357/ad037b}, \href
  {https://ui.adsabs.harvard.edu/abs/2024ApJ...960..133G} {960, 133}

\bibitem[\protect\citeauthoryear{{Hernquist}}{{Hernquist}}{1990}]{her90a}
{Hernquist} L.,  1990, \mn@doi [\apj] {10.1086/168845}, \href
  {http://adsabs.harvard.edu/abs/1990ApJ...356..359H} {356, 359}

\bibitem[\protect\citeauthoryear{{Hunt} \& {Vasiliev}}{{Hunt} \&
  {Vasiliev}}{2025}]{hun25a}
{Hunt} J. A.~S.,  {Vasiliev} E.,  2025, \mn@doi [arXiv e-prints]
  {10.48550/arXiv.2501.04075}, \href
  {https://ui.adsabs.harvard.edu/abs/2025arXiv250104075H} {p. arXiv:2501.04075}

\bibitem[\protect\citeauthoryear{{Hunt}, {Stelea}, {Johnston}, {Gandhi},
  {Laporte}  \& {B{\'e}dorf}}{{Hunt} et~al.}{2021}]{hun21t}
{Hunt} J. A.~S.,  {Stelea} I.~A.,  {Johnston} K.~V.,  {Gandhi} S.~S.,
  {Laporte} C. F.~P.,   {B{\'e}dorf} J.,  2021, \mn@doi [\mnras]
  {10.1093/mnras/stab2580}, \href
  {https://ui.adsabs.harvard.edu/abs/2021MNRAS.508.1459H} {508, 1459}

\bibitem[\protect\citeauthoryear{{Hunt}, {Price-Whelan}, {Johnston}  \&
  {Darragh-Ford}}{{Hunt} et~al.}{2022}]{hun22a}
{Hunt} J. A.~S.,  {Price-Whelan} A.~M.,  {Johnston} K.~V.,   {Darragh-Ford} E.,
   2022, \mn@doi [\mnras] {10.1093/mnrasl/slac082}, \href
  {https://ui.adsabs.harvard.edu/abs/2022MNRAS.516L...7H} {516, L7}

\bibitem[\protect\citeauthoryear{{Jeans}}{{Jeans}}{1915}]{jea15a}
{Jeans} J.~H.,  1915, \mnras, \href
  {http://adsabs.harvard.edu/abs/1915MNRAS..76...70J} {76, 70}

\bibitem[\protect\citeauthoryear{{Joung} \& {Mac Low}}{{Joung} \& {Mac
  Low}}{2006}]{jou06a}
{Joung} M.~K.~R.,  {Mac Low} M.-M.,  2006, \mn@doi [\apj] {10.1086/508795},
  \href {http://adsabs.harvard.edu/abs/2006ApJ...653.1266J} {653, 1266}

\bibitem[\protect\citeauthoryear{{Kalberla} \& {Kerp}}{{Kalberla} \&
  {Kerp}}{2009}]{kal09a}
{Kalberla} P.~M.~W.,  {Kerp} J.,  2009, \mn@doi [\araa]
  {10.1146/annurev-astro-082708-101823}, \href
  {http://adsabs.harvard.edu/abs/2009ARA%26A..47...27K} {47, 27}

\bibitem[\protect\citeauthoryear{{Khanna} et~al.,}{{Khanna}
  et~al.}{2019}]{kha19a}
{Khanna} S.,  et~al., 2019, \mn@doi [\mnras] {10.1093/mnras/stz2462}, \href
  {https://ui.adsabs.harvard.edu/abs/2019MNRAS.489.4962K} {489, 4962}

\bibitem[\protect\citeauthoryear{{Khoperskov}, {Di Matteo}, {Gerhard}, {Katz},
  {Haywood}, {Combes}, {Berczik}  \& {Gomez}}{{Khoperskov}
  et~al.}{2019}]{kho19a}
{Khoperskov} S.,  {Di Matteo} P.,  {Gerhard} O.,  {Katz} D.,  {Haywood} M.,
  {Combes} F.,  {Berczik} P.,   {Gomez} A.,  2019, \mn@doi [\aap]
  {10.1051/0004-6361/201834707}, \href
  {https://ui.adsabs.harvard.edu/abs/2019A&A...622L...6K} {622, L6}

\bibitem[\protect\citeauthoryear{{Laporte}, {Minchev}, {Johnston}  \&
  {G{\'o}mez}}{{Laporte} et~al.}{2019}]{lap19a}
{Laporte} C.~F.~P.,  {Minchev} I.,  {Johnston} K.~V.,   {G{\'o}mez} F.~A.,
  2019, \mn@doi [\mnras] {10.1093/mnras/stz583}, \href
  {http://adsabs.harvard.edu/abs/2019MNRAS.485.3134L} {485, 3134}

\bibitem[\protect\citeauthoryear{{Lynden-Bell}}{{Lynden-Bell}}{1967}]{lyn67a}
{Lynden-Bell} D.,  1967, \mn@doi [\mnras] {10.1093/mnras/136.1.101}, \href
  {http://adsabs.harvard.edu/abs/1967MNRAS.136..101L} {136, 101}

\bibitem[\protect\citeauthoryear{{Molina}, {Glover}, {Federrath}  \&
  {Klessen}}{{Molina} et~al.}{2012}]{MolinaEtAl2012}
{Molina} F.~Z.,  {Glover} S.~C.~O.,  {Federrath} C.,   {Klessen} R.~S.,  2012,
  \mn@doi [\mnras] {10.1111/j.1365-2966.2012.21075.x}, \href
  {http://adsabs.harvard.edu/abs/2012MNRAS.423.2680M} {423, 2680}

\bibitem[\protect\citeauthoryear{{Navarro}, {Frenk}  \& {White}}{{Navarro}
  et~al.}{1997}]{nav97a}
{Navarro} J.~F.,  {Frenk} C.~S.,   {White} S.~D.~M.,  1997, \mn@doi [\apj]
  {10.1086/304888}, \href {http://adsabs.harvard.edu/abs/1997ApJ...490..493N}
  {490, 493}

\bibitem[\protect\citeauthoryear{{Pontzen}, {Ro{\v s}kar}, {Stinson}  \&
  {Woods}}{{Pontzen} et~al.}{2013}]{pon13a}
{Pontzen} A.,  {Ro{\v s}kar} R.,  {Stinson} G.,   {Woods} R.,  2013, {pynbody:
  N-Body/SPH analysis for python}, Astrophysics Source Code Library (\mn@eprint
  {ascl} {1305.002})

\bibitem[\protect\citeauthoryear{{Power}, {Navarro}, {Jenkins}, {Frenk},
  {White}, {Springel}, {Stadel}  \& {Quinn}}{{Power} et~al.}{2003}]{pow03o}
{Power} C.,  {Navarro} J.~F.,  {Jenkins} A.,  {Frenk} C.~S.,  {White} S.~D.~M.,
   {Springel} V.,  {Stadel} J.,   {Quinn} T.,  2003, \mn@doi [\mnras]
  {10.1046/j.1365-8711.2003.05925.x}, \href
  {https://ui.adsabs.harvard.edu/abs/2003MNRAS.338...14P} {338, 14}

\bibitem[\protect\citeauthoryear{{Renaud} et~al.,}{{Renaud}
  et~al.}{2013}]{ren13a}
{Renaud} F.,  et~al., 2013, \mn@doi [\mnras] {10.1093/mnras/stt1698}, \href
  {http://adsabs.harvard.edu/abs/2013MNRAS.436.1836R} {436, 1836}

\bibitem[\protect\citeauthoryear{{Renaud}, {Romeo}  \& {Agertz}}{{Renaud}
  et~al.}{2021}]{ren21a}
{Renaud} F.,  {Romeo} A.~B.,   {Agertz} O.,  2021, \mn@doi [\mnras]
  {10.1093/mnras/stab2604}, \href
  {https://ui.adsabs.harvard.edu/abs/2021MNRAS.508..352R} {508, 352}

\bibitem[\protect\citeauthoryear{{Rizzo}, {Kohandel}, {Pallottini}, {Zanella},
  {Ferrara}, {Vallini}  \& {Toft}}{{Rizzo} et~al.}{2022}]{riz22w}
{Rizzo} F.,  {Kohandel} M.,  {Pallottini} A.,  {Zanella} A.,  {Ferrara} A.,
  {Vallini} L.,   {Toft} S.,  2022, arXiv e-prints, \href
  {https://ui.adsabs.harvard.edu/abs/2022arXiv220405325R} {p. arXiv:2204.05325}

\bibitem[\protect\citeauthoryear{{Ro{\v{s}}kar}, {Debattista}, {Quinn}  \&
  {Wadsley}}{{Ro{\v{s}}kar} et~al.}{2012}]{ros12a}
{Ro{\v{s}}kar} R.,  {Debattista} V.~P.,  {Quinn} T.~R.,   {Wadsley} J.,  2012,
  \mn@doi [\mnras] {10.1111/j.1365-2966.2012.21860.x}, \href
  {https://ui.adsabs.harvard.edu/abs/2012MNRAS.426.2089R} {426, 2089}

\bibitem[\protect\citeauthoryear{{Seta} \& {McClure-Griffiths}}{{Seta} \&
  {McClure-Griffiths}}{2025}]{set25a}
{Seta} A.,  {McClure-Griffiths} N.~M.,  2025, \mn@doi [\mnras]
  {10.1093/mnras/staf520}, \href
  {https://ui.adsabs.harvard.edu/abs/2025MNRAS.tmp..512S} {}

\bibitem[\protect\citeauthoryear{{Stanimirovic}, {Staveley-Smith}, {Dickey},
  {Sault}  \& {Snowden}}{{Stanimirovic} et~al.}{1999}]{sta99a}
{Stanimirovic} S.,  {Staveley-Smith} L.,  {Dickey} J.~M.,  {Sault} R.~J.,
  {Snowden} S.~L.,  1999, \mn@doi [\mnras] {10.1046/j.1365-8711.1999.02013.x},
  \href {https://ui.adsabs.harvard.edu/abs/1999MNRAS.302..417S} {302, 417}

\bibitem[\protect\citeauthoryear{{Tepper-Garc{\'\i}a}, {Bland-Hawthorn}  \&
  {Freeman}}{{Tepper-Garc{\'\i}a} et~al.}{2022}]{tep22x}
{Tepper-Garc{\'\i}a} T.,  {Bland-Hawthorn} J.,   {Freeman} K.,  2022, \mn@doi
  [\mnras] {10.1093/mnras/stac1926}, \href
  {https://ui.adsabs.harvard.edu/abs/2022MNRAS.515.5951T} {515, 5951}

\bibitem[\protect\citeauthoryear{{Tepper-Garc{\'\i}a}, {Bland-Hawthorn},
  {Vasiliev}, {Agertz}, {Teyssier}  \& {Federrath}}{{Tepper-Garc{\'\i}a}
  et~al.}{2024}]{tep24a}
{Tepper-Garc{\'\i}a} T.,  {Bland-Hawthorn} J.,  {Vasiliev} E.,  {Agertz} O.,
  {Teyssier} R.,   {Federrath} C.,  2024, \mn@doi [\mnras]
  {10.1093/mnras/stae2372}, \href
  {https://ui.adsabs.harvard.edu/abs/2024MNRAS.tmp.2307T} {}

\bibitem[\protect\citeauthoryear{{Teyssier}}{{Teyssier}}{2002}]{tey02a}
{Teyssier} R.,  2002, \mn@doi [\aap] {10.1051/0004-6361:20011817}, \href
  {http://adsabs.harvard.edu/abs/2002A%26A...385..337T} {385, 337}

\bibitem[\protect\citeauthoryear{{Toomre}}{{Toomre}}{1964}]{too64a}
{Toomre} A.,  1964, \mn@doi [\apj] {10.1086/147861}, \href
  {http://adsabs.harvard.edu/abs/1964ApJ...139.1217T} {139, 1217}

\bibitem[\protect\citeauthoryear{{Tremaine}, {Frankel}  \& {Bovy}}{{Tremaine}
  et~al.}{2023}]{tre23a}
{Tremaine} S.,  {Frankel} N.,   {Bovy} J.,  2023, \mn@doi [\mnras]
  {10.1093/mnras/stad577}, \href
  {https://ui.adsabs.harvard.edu/abs/2023MNRAS.521..114T} {521, 114}

\bibitem[\protect\citeauthoryear{{Vasiliev}}{{Vasiliev}}{2019}]{vas19a}
{Vasiliev} E.,  2019, \mn@doi [\mnras] {10.1093/mnras/sty2672}, \href
  {http://adsabs.harvard.edu/abs/2019MNRAS.482.1525V} {482, 1525}

\bibitem[\protect\citeauthoryear{{Xu} et~al.,}{{Xu} et~al.}{2020}]{xu20x}
{Xu} Y.,  et~al., 2020, \mn@doi [\apj] {10.3847/1538-4357/abc2cb}, \href
  {https://ui.adsabs.harvard.edu/abs/2020ApJ...905....6X} {905, 6}

\bibitem[\protect\citeauthoryear{{Zhang}, {Hunter}  \& {Elmegreen}}{{Zhang}
  et~al.}{2012}]{zha12a}
{Zhang} H.-X.,  {Hunter} D.~A.,   {Elmegreen} B.~G.,  2012, \mn@doi [\apj]
  {10.1088/0004-637X/754/1/29}, \href
  {https://ui.adsabs.harvard.edu/abs/2012ApJ...754...29Z} {754, 29}

\makeatother
\end{thebibliography}
\input{tepper_etal.bbl}

\appendix

\section{Validation of the phase-spiral finder} \label{app:val}

We have conducted an extensive validation of the phase-space finder algorithm described in Sec.~\ref{sec:ana}. To this end, we have created a  synthetic phase spiral model, as follows. We create a rectangular grid of in Cartesian coordinates $(X,Y)$ with $M \times M$ points. On this grid, the amplitude $A(X,Y)$ of the phase spiral of mode $m$ (i.e. $m$-arm), winding $w$, and phase shift $\psi_0$, centred at $(X_c,Y_c)$ is given by the following:
\begin{align}
	&X' = X - X_c\\
	& Y' = Y - Y_c\\
	& \theta = \arctan(Y'/X')\\
	& r^2 = X'^2 + Y'^2\\
	& \psi = \psi_0 + m~\theta + w~r\\
	& A_{\textnormal{PS}} = \mathcal{A}_0 \cos\left[ \psi \right]
\end{align}
For a more realistic model, we optionally add noise described by a Gaussian (normal) distribution $N\left(\mu, \sigma\right)$, with zero mean $\mu \equiv 0$) and standard deviation $\sigma$ equal to the desired noise level. Note that the noise amplitude is chosen relative to the model's. With the above definitions, the full expression for the phase-spiral amplitude is thus
\begin{equation} \label{eq:amp}
	AA_{\textnormal{PS}}(X,Y,\sigma) = \mathcal{A}_0 \left\{ \cos\left[ \psi \right] + N\left(0,\sigma\right) \right\}\, .
\end{equation}
We have tested the finder using a large number of values picked from reasonable ranges for the phase-spiral model parameters: centre, phase shift, and winding, as well as noise levels, focusing for now on one=arm spirals ($m = 1$) with $\mathcal{A}_0 = 1$, and fixing $M = 100$ (the same resolution used in our phase-spiral incidence-map cells). Via a trial-and-error approach we have been able to tune the free parameters of the phase-spiral finder algorithm to yield satisfactory results in the overwhelmingly majority of cases. In fact, the algorithm works flawlessly in idealised case of no noise ($\sigma = 0$). Overall it works remarkably well up to $\sigma \approx 20$ in the overwhelmingly majority of cases we've tested for.

\begin{figure*}
\centering
\includegraphics[width=0.9\textwidth]{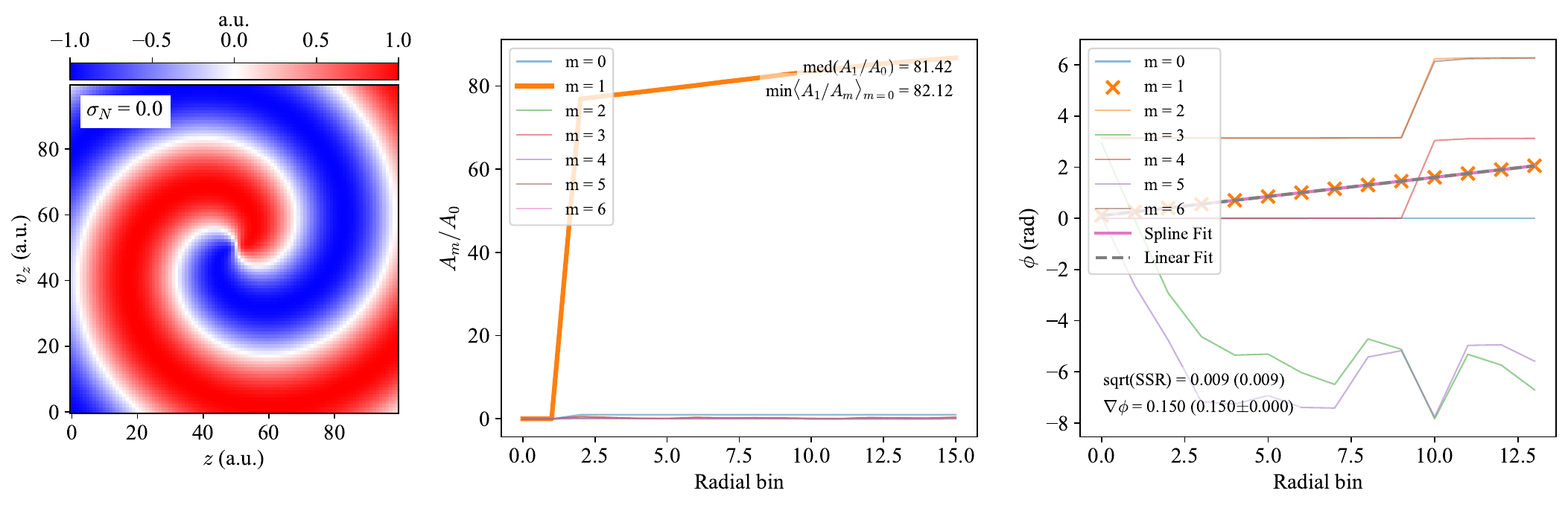}
\includegraphics[width=0.9\textwidth]{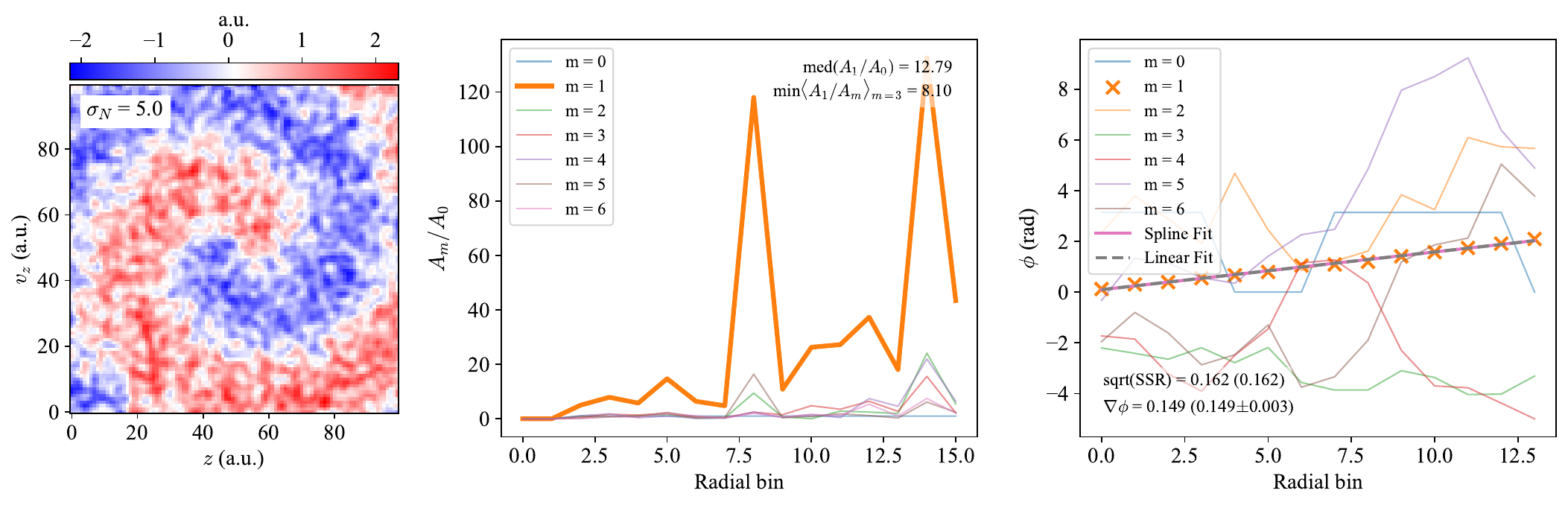}
\includegraphics[width=0.9\textwidth]{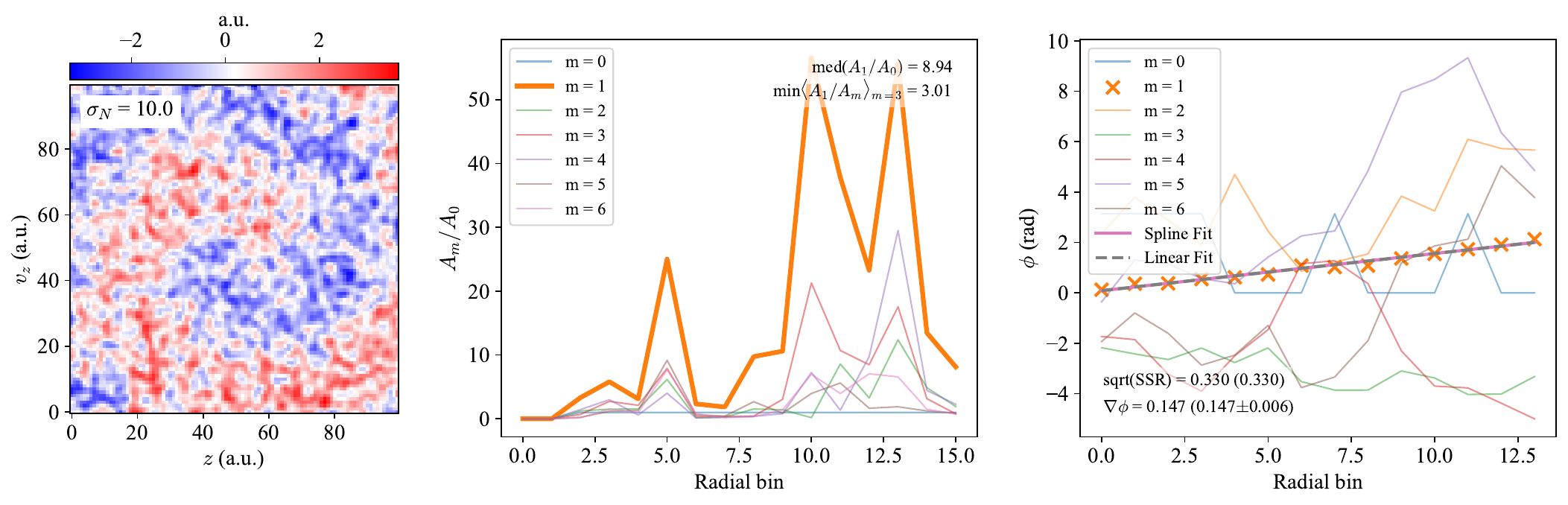}
\includegraphics[width=0.9\textwidth]{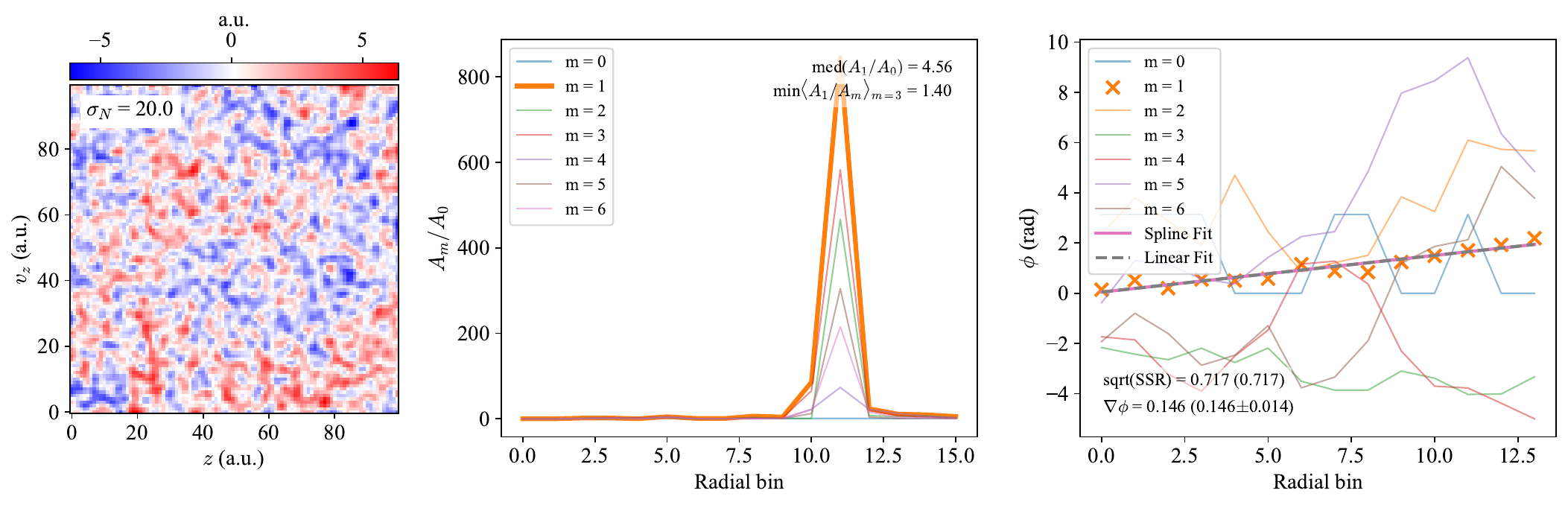}
\caption[  ]{ Synthetic, idealised phase spiral models. Each rows correspond to the same model, but with a progressively increasing amplitude of Gaussian noise. The left column shows a series of examples of synthetic, one-arm ($m = 1$) phase spirals with $w = 1$ and $\psi_0 = 0$. The red-white-blue colour scheme indicates the value of the phase spiral amplitude $A_{\textnormal{PS}}$. Note that the axes units and the units of the amplitude are fully arbitrary. The middle and right columns show, respectively, the amplitude $A_m / A_0$ of the Fourier modes $m = 1, 2, \ldots, 6$ and their corresponding phase $\phi$, as a function of radial bin. The results corresponding to the $m=1$ mode have been highlighted with a thick, orange curve (middle) and crosses (right), for emphasis.
}
\label{fig:ps_toy}
\end{figure*}

An example of such a test case is displayed in Fig.~\ref{fig:ps_toy}. The left column shows a series of examples of synthetic, one-arm ($m = 1$) phase spirals with $w = 1$ and $\psi_0 = 0$, and increasing levels of noise, from $\sigma = 0$ (top) to $\sigma = 20$ (bottom), on a $M = 100$ grid. The middle and right columns show, respectively, the amplitude $A_m / A_0$ of the Fourier modes $m = 1, 2, \ldots, 6$ and their phase $\phi$, as a function of radial bin $r$ (see definitions above).

Regardless of the noise level, it is clear that: 1) the $m = 1$ mode amplitude significantly dominates over all the other modes, but the difference decreases with noise level; 2) the  the $m = 1$ mode's phase increases linearly with radius, but the scatter around a perfect linear behaviour increases with noise level. In stark contrast, the phase of all the other modes varies erratically (with exception perhaps of $m = 3$). It is striking that even in the $\sigma = 20$ case (bottom row), where the phase spiral is nearly invisible to the eye, the phase-spiral finder is able to pick up the signal.

Our extensive test suite demonstrated that a dominant $m = 1$ amplitude throughout and a (ideally) linearly increasing $m = 1$ phase are necessary and sufficient conditions to detect the presence of a structure reminiscent one-arm spiral.

\section{Mass distribution and radially averaged power spectrum (RPS)} \label{sec:psd_app}

\begin{figure*}
\centering
\includegraphics[width=0.33\textwidth]{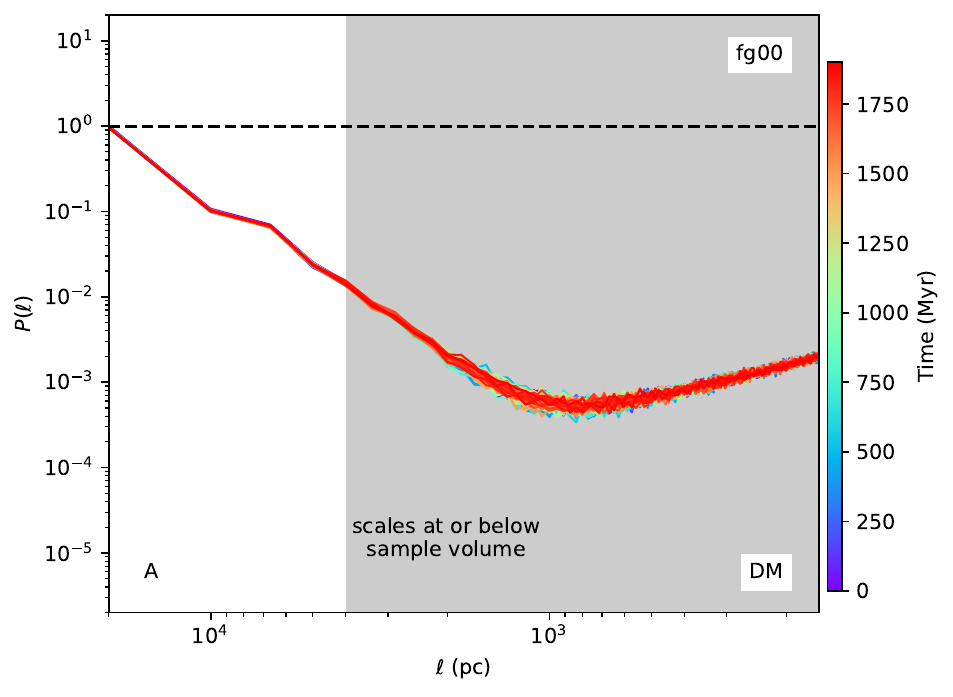}
\includegraphics[width=0.33\textwidth]{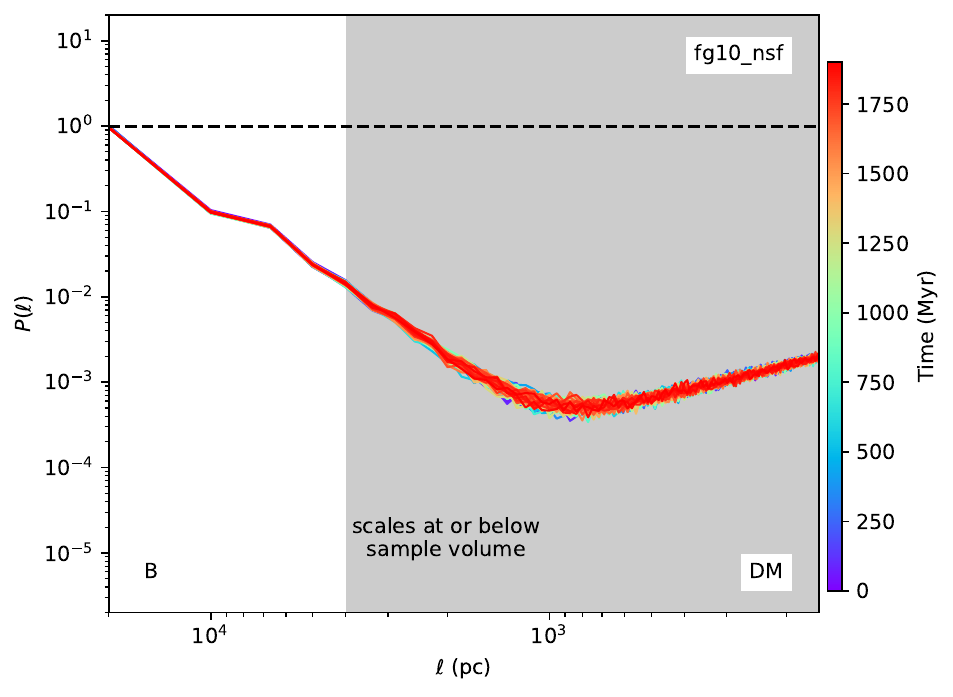}
\includegraphics[width=0.33\textwidth]{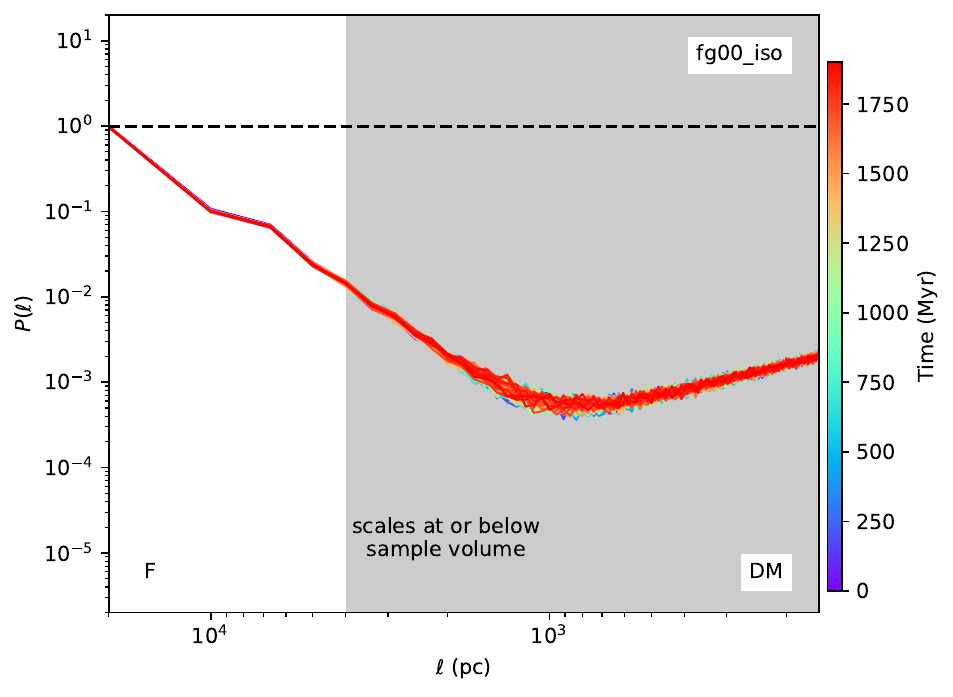}
\includegraphics[width=0.33\textwidth]{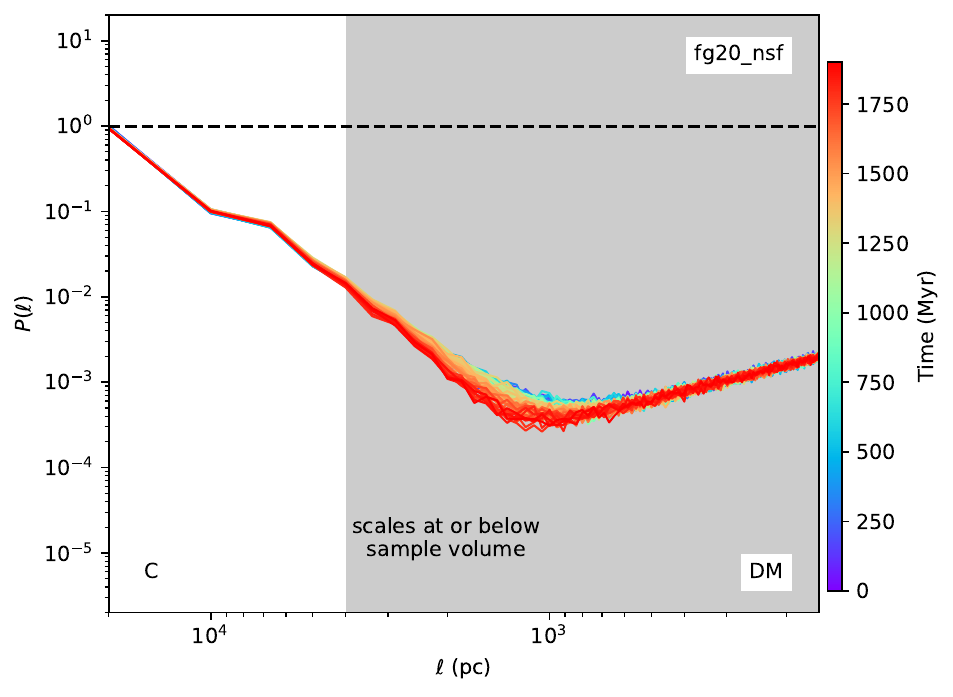}
\includegraphics[width=0.33\textwidth]{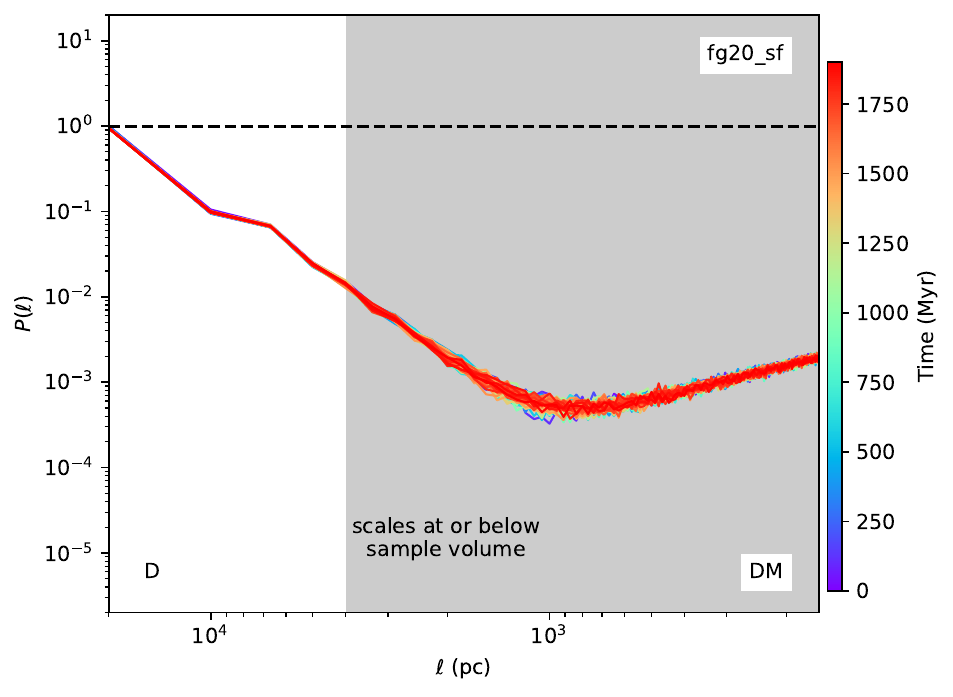}
\includegraphics[width=0.33\textwidth]{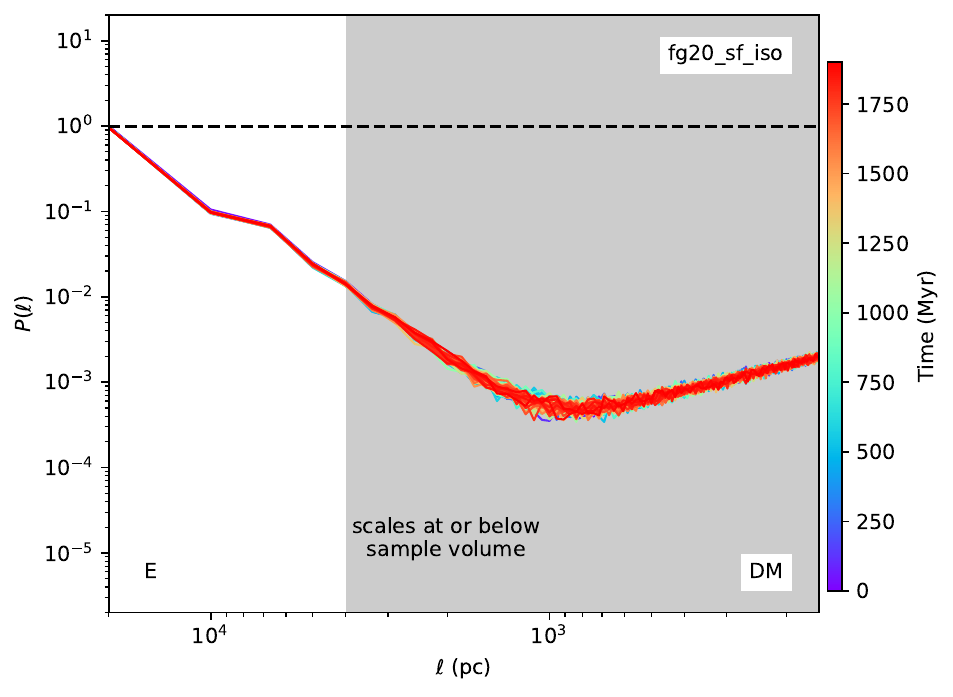}
\caption[  ]{ Similar to Fig.~\ref{fig:psd_gas}, but for the mass distribution of dark matter. Note that there is virtually no change in the initial RPS, and thus the curves for all epochs overlap. The $P(\ell) \propto \ell^2$ behaviour at the smallest scales is a reflection of the particle (discretisation) noise.}
\label{fig:psd_dm}
\end{figure*}

\begin{figure*}
\centering
\includegraphics[width=0.33\textwidth]{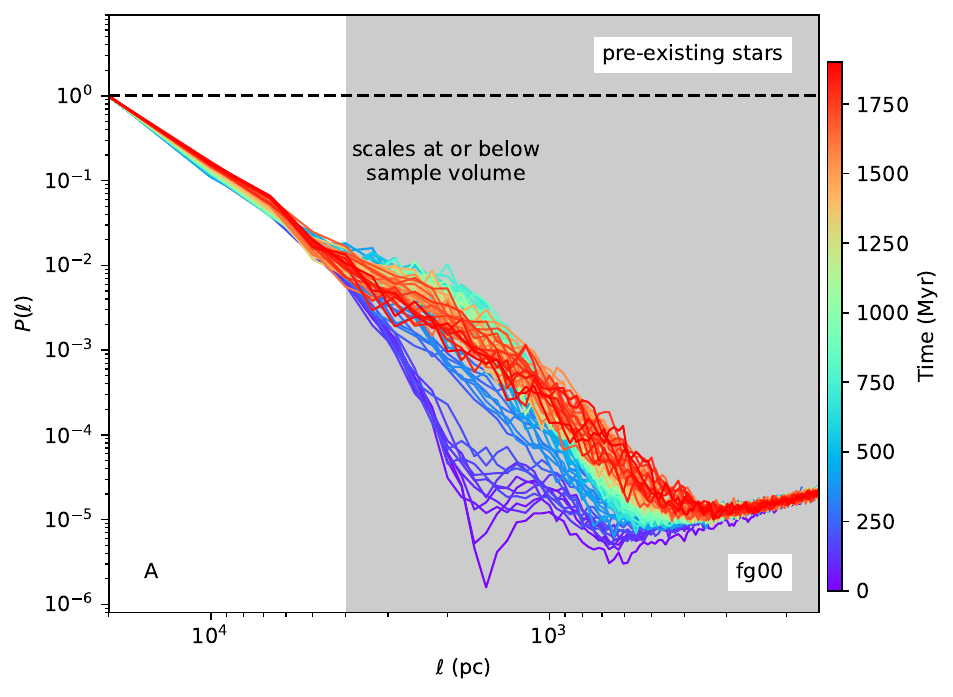}
\includegraphics[width=0.33\textwidth]{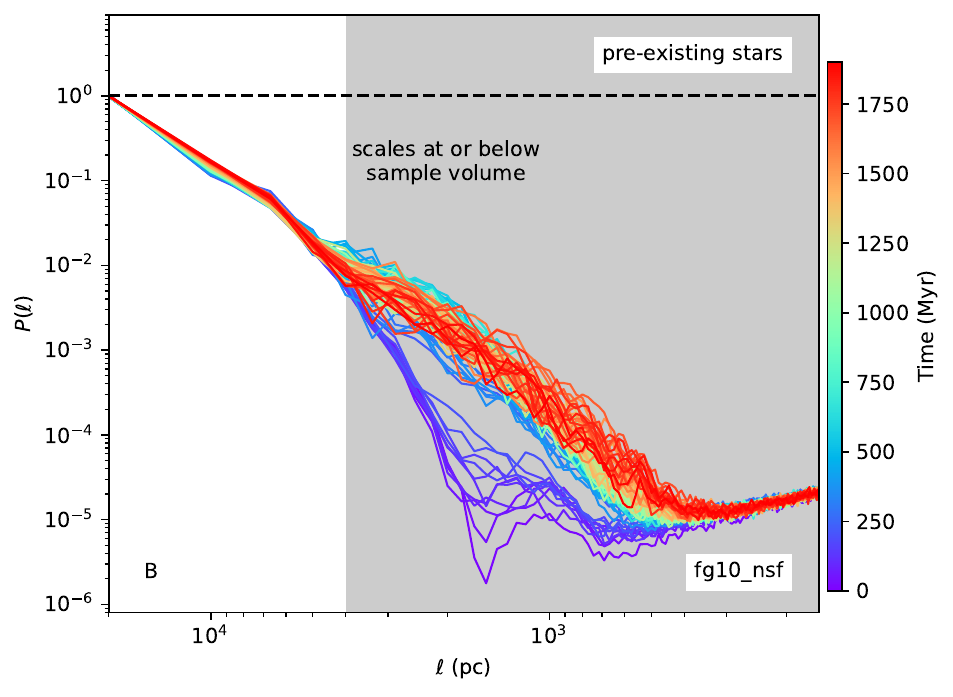}
\includegraphics[width=0.33\textwidth]{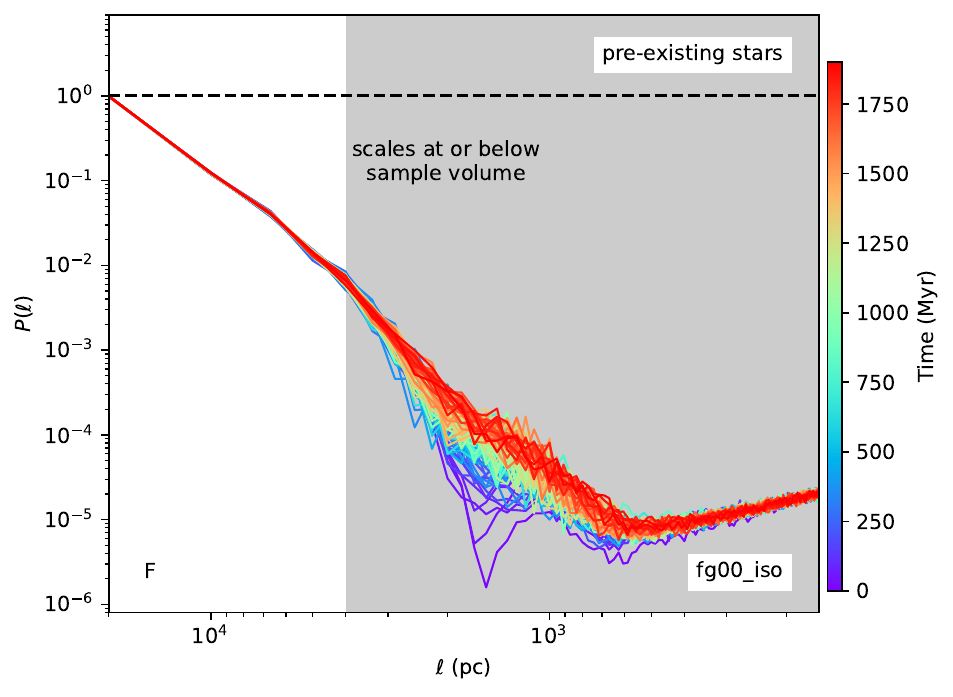}
\includegraphics[width=0.33\textwidth]{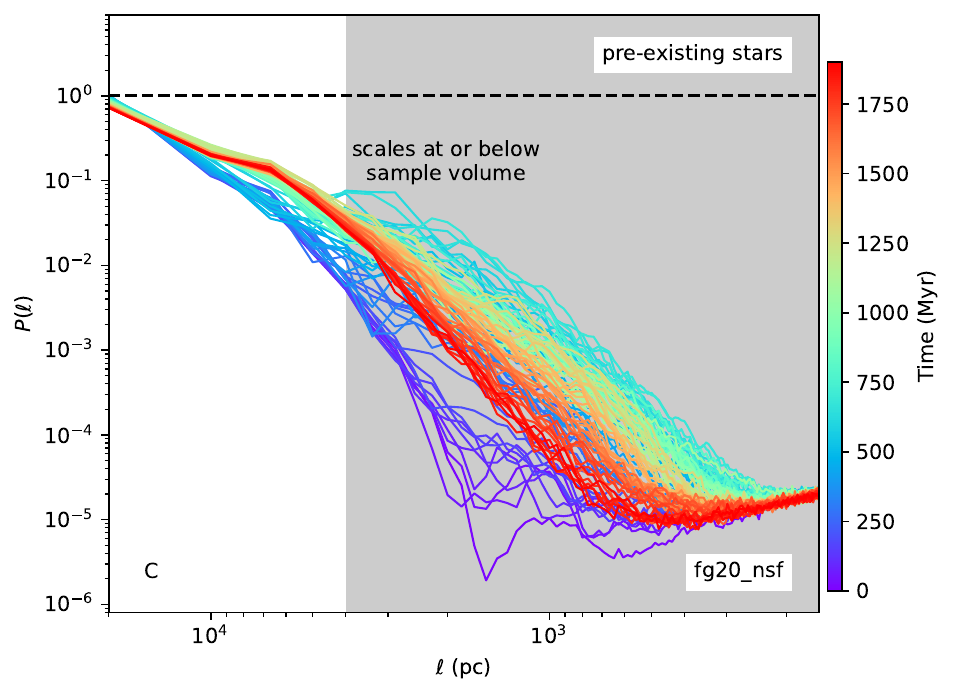}
\includegraphics[width=0.33\textwidth]{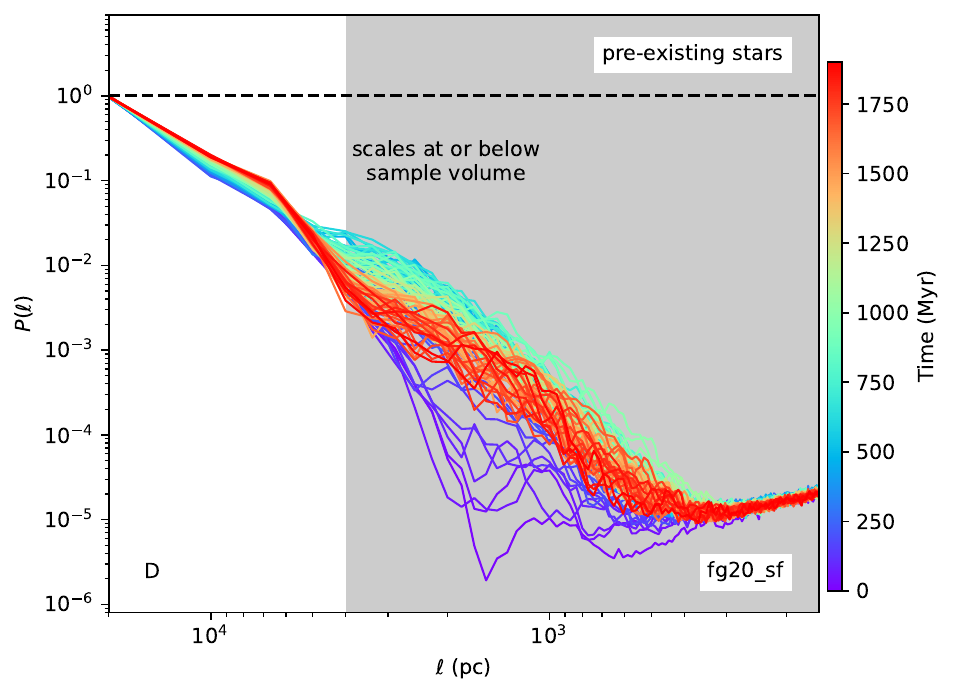}
\includegraphics[width=0.33\textwidth]{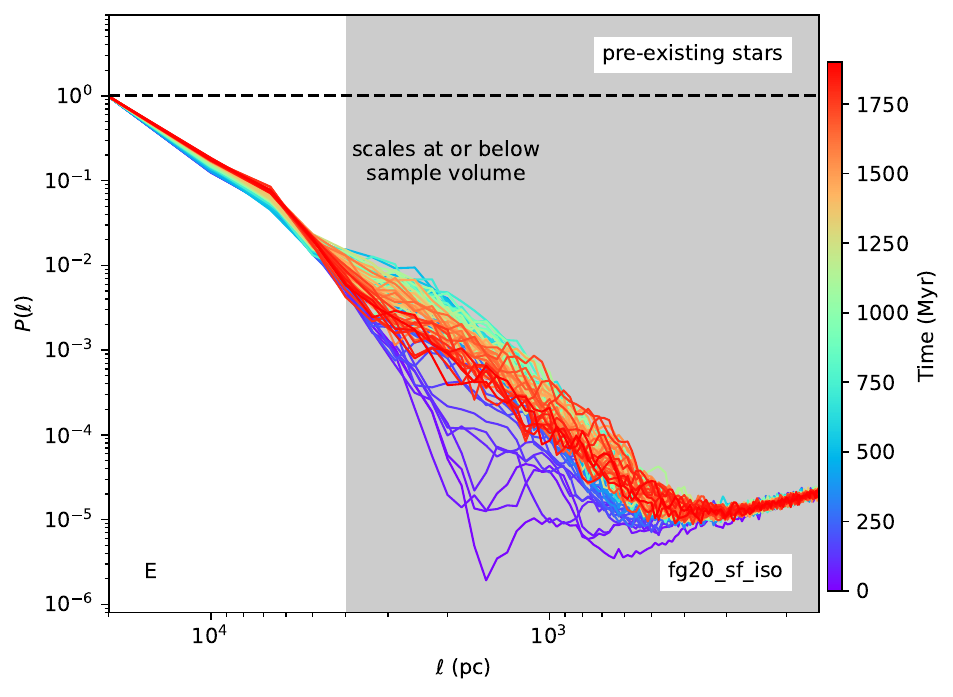}
\caption[  ]{ Similar to Fig.~\ref{fig:psd_dm}, but for the distribution of pre-existing stars. The $P(\ell) \propto \ell^2$ behaviour at the smallest scales is a reflection of the particle (discretisation) noise.}
\label{fig:psd_stars}
\end{figure*}

\begin{figure*}
\centering
\includegraphics[width=0.33\textwidth]{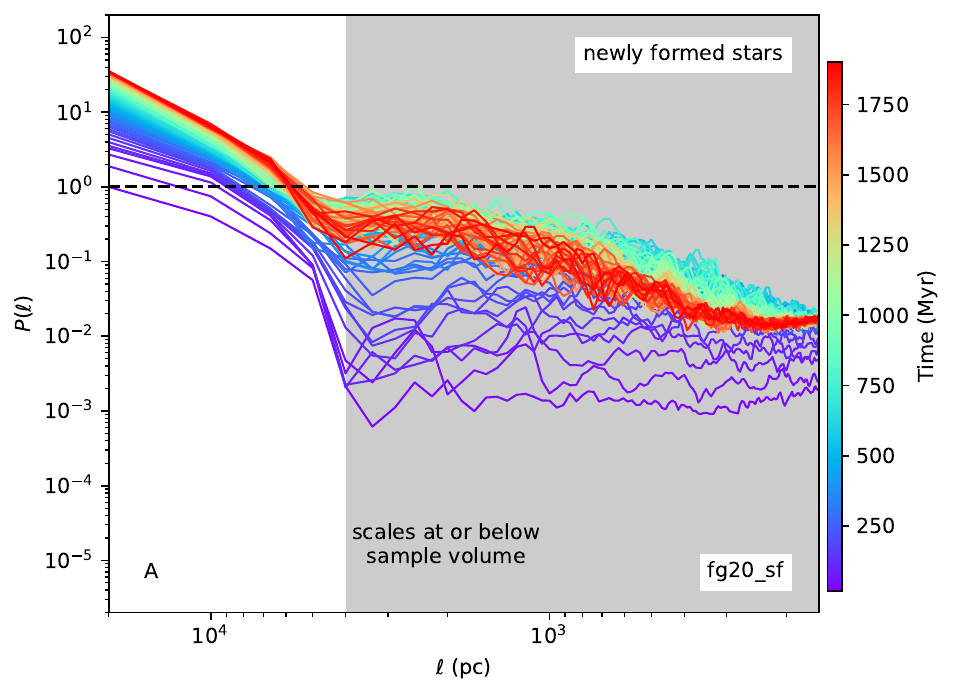}
\includegraphics[width=0.33\textwidth]{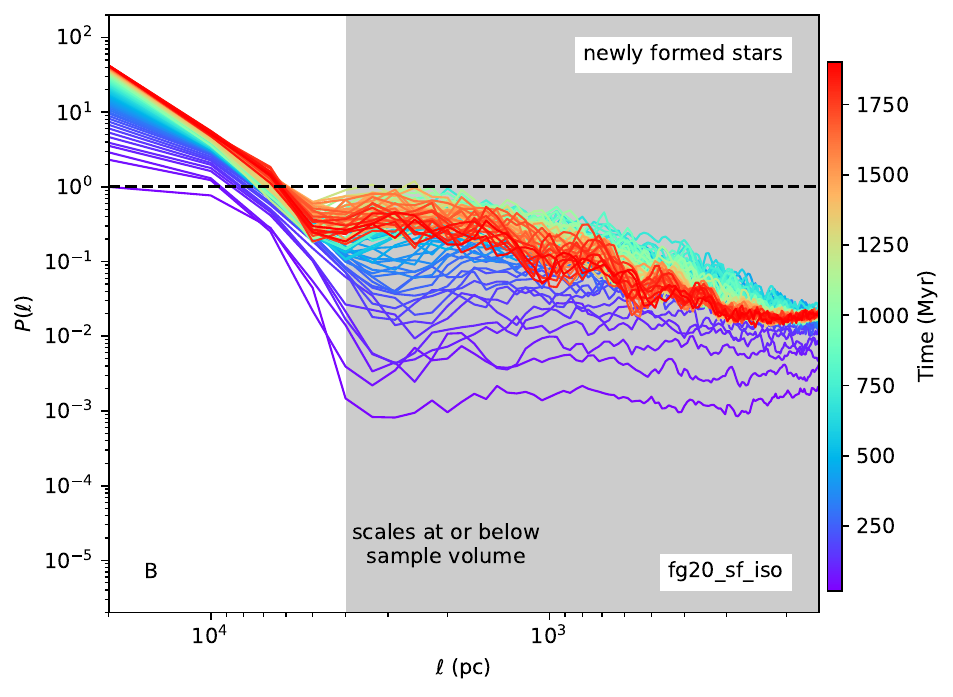}
\caption[  ]{ Similar to Fig.~\ref{fig:psd_stars}, but for the mass distribution of newly formed stars. }
\label{fig:psd_nstars}
\end{figure*}

Here we show the evolution of the RPS of additional galaxy components: dark matter  (Fig.~\ref{fig:psd_dm}); pre-existing stars (Fig.~\ref{fig:psd_stars}); and newly formed stars (Fig.~\ref{fig:psd_nstars}). For details about the RPS, we refer the reader to Sec.~\ref{sec:summ}. The key point to take away from these plots is the lack of power on sub-kiloparsec scales compared to 
Fig.~\ref{fig:psd_gas}. The substructure within these evolving spatial components is neither able to excite nor able to dampen the phase spiral \citep{tre23a,gil25a}.

\bsp	
\label{lastpage}
\end{document}